\documentclass[useAMS,usenatbib,times]{mn2e}

\usepackage{graphicx}
\usepackage{natbib}

\usepackage[usenames]{color}

\definecolor{grey}{rgb}{0.75,0.75,0.75}
\definecolor{Orange}{rgb}{1.0,0.5,0.15}
\definecolor{brown}{rgb}{0.7,0.25,0.0}
\definecolor{pink}{rgb}{1.0,0.5,0.5}
\definecolor{darkerred}{rgb}{0.8,0,0}
\definecolor{darkerblue}{rgb}{0,0,0.8}
\definecolor{Blue}{rgb}{0,0.08,0.65}
\definecolor{Red}{rgb}{0.65,0.08,0.05}
\definecolor{Green}{rgb}{0.15,0.45,0.25}

\def\d{{\mathrm{d}}}  

\def\MG#1{{\mbox{\boldmath $ #1$}}} %bold greek
\def\M#1{{\mathbf #1}}

\def\ie{{\frenchspacing\it i.e. }}

\def\eg{{\frenchspacing\it e.g. }}

\def\R#1{{\mathrm{#1}}}% roman font in math mode
\def\llangle{{\langle\hskip -2.5pt\langle}} 
\def\rrangle{{\rangle\hskip -2.5pt\rangle}} 
\def\Sec#1{{Section~\ref{s:#1}}}
\def\sec#1{{section~\ref{s:#1}}}
\def\seq#1{{~\ref{s:#1}}}
\def\Eq#1{{Equation~(\ref{e:#1})}}% equation reference
\def\eq#1{{equation~(\ref{e:#1})}}% equation reference
\def\Ep#1{{~(\ref{e:#1})}}% equation reference
\def\Eqs#1#2{{equations~(\ref{e:#1})-(\ref{e:#2})}}
\def\EQN#1{\label{e:#1}}        % eqn labelling a la Texsis
\def\Tab#1{{Table~\ref{t:#1}}}        % table reference
\def\Fig#1{{Fig.~\ref{f:#1}}}% figure reference
\def\Fip#1{{~\ref{f:#1}}}% figure reference

\def\M#1{{\mathbf{#1}}}% matrix notation
\def\T#1{{{#1}^{\top}}}% transposition of a matrix
\newcommand{\tmmathbf}[1]{\mathbf{#1}}
\newcommand{\mathe}{\mathrm{e}}
\newcommand{\mathi}{\mathrm{i}}
       \def\bfr {{\bf  r}} \def\bfv {{\bf
v}}

 \def\R#1{{\mathrm{#1}}}  % roman  font  in math mode       
\def\Tab#1{{Table~(\ref{t:#1})}}       %      table       reference
\def\Fig#1{{Fig.~(\ref{f:#1})}}               

\def\Fip#1{{~(\ref{f:#1})}} 
\def\bm{{\bf m}}
\def\ee{\end{equation}} 
\def\ba{\begin{eqnarray}} 
\def\ea{\end{eqnarray}}

\def\d{{\R{d}}}   %  integrant   \def\mdot{\!\cdot\!}   %   matricial  product
\def\MG#1{{\mbox{\boldmath $ #1$}}} %bold greek \def\tmmathbf#1{{\mathbf{#1}}}
\def\bu{{\bf u}} \def\bv{{\bf v}} \def\br{{\bf r}} 
      
    \def\bI{{\bf I}}
 \def\bw{{\bf w}} \def\bu{{\bf u}} \def\bO{{\bf \Omega}} \def\bG{{\bf \Gamma}}
 \def\bV{{\bf  V}}   \def\md{{\mathrm  d}}  \def\bo{{\bf
 \omega}} \def\be{{\bf e}}

\def\T#1{{{#1}^{\top}}}%  transposition   of  a  matrix  \def\mdot{\!\cdot\!}% matricial product

\newcommand{\nicefrac}[2]{\leavevmode\kern.1em
            \raise.5ex\hbox{\the\scriptfont0                      #1}\kern-.1em
            /\kern-.15em\lower.25ex\hbox{\the\scriptfont0 #2}}

\usepackage{times}

\begin{document}
 \title{Dynamical flows  through Dark  Matter Haloes II~: \\ One  and Two points Statistics at the virial radius}

\author[Dominique      Aubert       and      Christophe      Pichon]{Dominique
Aubert$^{1,2,3}$\thanks{E-mail  :aubert@astro.u-strasbg.fr}   and   Christophe
Pichon$^{1,3}$\\$^{1}$ Observatoire Astronomique de Strasbourg,  11 rue de
l'Universite,  67000 Strasbourg, France\\$^{2}$ Service d'Astrophysique, CEA Saclay, 91191 Gif-Sur-Yvette, France\\$^{3}$ Institut  d'Astrophysique de Paris,  98 bis  boulevard d'Arago,  75014  Paris, France  \\ }

   \date{Typeset \today ; Received / Accepted}

   \maketitle

  \begin{abstract}

In a serie of three papers, 
the dynamical interplay between environments and dark matter haloes 
is investigated, 
while focussing on the dynamical flows through the virtual virial sphere.
 It relies on both cosmological simulations, to constrains the environments,
 and an extension to the classical matrix method  to derive the responses  of
 the halo.  A companion paper (\citet{pich}, paper I) showed how perturbation theory allows to propagate the statistical properties of the  environment to an ensemble description of the dynamical response of the embedded halo.  The current paper  focuses   on the statistical characterisation of the environments surrounding haloes, using a set of large scale simulations; 
  the large statistic of environments presented here allows to put quantitative and statistically significant constrains on the properties of flows accreted by haloes.

The description chosen in this paper relies on a ``fluid'' halocentric representation. The interactions between the halo and its environment are investigated in terms of a time dependent external tidal field  and a source term characterizing the infall.  The former accounts for fly bys and interlopers. 
   The latter stands for the distribution function of the matter accreted through the virial sphere.  The method of  separation of variables is used to decouple
   the temporal evolution of these two quantities from their angular and velocity dependence by means of projection on a 5D basis.

It is shown how the flux densities of mass, momentum and energy can provide an alternative description to the 5D projection of the source. {Such a description is well suited to re-generate synthetic time-lines of accretion which are consistent with environments found in simulations as discussed in appendix.}
The method leading to the measurements of these quantities
in simulations is presented  in details and applied to 15000 haloes, with masses between $5\cdot 10^{12} \R{M}_\odot$ and $10^{14} \R{M}_\odot$ evolving between $z=1$ and $z=0$. The influence of resolution, class of mass, and selection biases are investigated with higher resolution simulations. The emphasis is put on the one and two-point statistics of the tidal field,
     and of the flux density of mass, while the full characterization of the other fields { is postponed to  paper III}.

The net accretion at the virial radius is found to decrease with time. This decline results from both an absolute decrease of infall and from a growing contribution of outflows.  Infall is found to be mainly radial and occurring at velocities $\sim 0.75$ times the virial velocity. Outflows are also detected through the virial sphere and occur at lower velocities $\sim 0.6 V_c$ on more circular orbits. The external tidal field is found to be strongly quadrupolar and mostly stationnary, possibly reflecting the distribution of matter in the halo's near environment. The coherence time of the small scale fluctuations of the potential hints a possible anisotropic distribution of accreted satellites. The flux density of mass on the virial sphere appears to be more clustered than the potential while  the shape of its angular power spectrum seems stationnary.
 Most of these results are tabulated with simple fitting laws and are found to be consistent with published work which rely on a  description of accretion in terms of satellites.
\end{abstract}

%\onecolumn
%%%%%%%%%%%%%%%%%%%%%%%%%%%%%%%%%%%%%%%%%%%%%%%%%%
\section{Galaxies in their environment}
%%%%%%%%%%%%%%%%%%%%%%%%%%%%%%%%%%%%%%%%%%%%%%%%%%
%%%%%%%%%%%%%%%%%%%%%%%%%%%%%%%%%%%%%%%%%%%%%%%%%%
%%%%%%%%%%%%%%%%%%%%%%%%%%%%%%%%%%%%%%%%%%%%%%%%%%

Examples of  galaxies interacting with their environments  are numerous.  The
antennae, the  cartwheel galaxy, M51 are  among the most famous  ones.  One of
our  closest neighbor,  M31, exhibits  a giant  stellar stream  which  may be
associated with  its satellites (e.g. \citet{Mac}).  Even the
Milky  Way shows relics  of past  interactions with  material coming  from the
outskirts, such as the Sagittarius dwarf (\citet{Ibata1}).  It appears clearly
that the evolution of galactic systems cannot
be understood only by considering their internal properties but also by taking
into account their environment.  From a dynamical point of view it
is still not  clear  for example if spirals in galaxies are  induced by intrinsic unstable
modes   (e.g.   \citet{DLB},  \citet{Kalnajs2})   or  if   they  are   due  to
gravitational   interactions  with   satellites  or   other   galaxies  (e.g.
\citet{TT}).   Similarly, normal  mode theories  of warps  have  been proposed
(\citet{Sparke}, \citet{Hunter}) but failed to reproduce long-lived warps in a
live halo  for example (e.g. \citet{B98}).  Since warped  galaxies are likely  to have companions
(\citet{Combes}),  it is  natural  to  suggest satellite  tidal  forcing as  a
generating mechanism (e.g.   \citet{W98}, \citet{Tsu}).  Another possibility is
angular momentum misalignment of infalling material (e.g. \citet{OsBin}, \citet{JB99}). The
existence  of the  thick disk  may also be  explained by  past small  mergers (e.g.
\citet{Velazquez}, \citet{Quinn}, \citet{Walker}).  Conversely, very thin disk
put serious  constraints on  the amplitude of  the interactions they  may have
experienced in the past.  

On a larger  scale, dark matter halos are built in a
hierarchical fashion within the CDM model.  Some of the  most serious challenges that these models
are now facing (the over production  of dwarf galaxies in the local group (e.g. \citet{M99}, \citet{Klypin}), the
cuspide  crisis of  NFW-like halos (e.g. \citet{FP}, \citet{M94}),  the over-cooling  problem  and
the momentum crisis  for galactic disks (e.g. \citet{NS97})  occur at these
scales; it  is therefore important to study the  effects of the
cosmological paradigm on the evolution of galaxies in order to address these issues.  

In fact, the properties of galaxies naturally present correlations with their environments.
For example \citet{Tormen} showed that the  shape of halos tends to be aligned
with the  surrounding satellites distribution.  Also, halos  spin is sensitive
to recently accreted angular momentum (e.g.  \citet{Haarlem}, \citet{aubert}).
More generally,  halos inherit  the properties of  their progenitors.  

At this
point, a question naturally arises, namely  `what is the dynamical response of a
galactic system  (halo+disk) to its environment  ?'.  One way  to address this
issue  is to  compute high  resolution simulations  of galaxies  into  a given
environment  (e.g.    \citet{Abadi}, \citet{Knebe}, \citet{Gill}).   However,  if  one   is  interested  in
reproducing  the  variety  of  dynamical  responses  of  galaxies  to  various
environments, the  use of such  simulations becomes rapidly  tedious. 
An alternative way to investigate this topic is presented here, which should complement both high resolution simulations and large cosmological simulations. In a serie of three papers,  an hybrid approach is presented 
 to investigate the interplay between environments and haloes. It relies on both cosmological simulations (to constrains the environments) and
on a straightforward 
extension of the classical tools of  galactic dynamics (to derive the haloes
response). A companion paper (\citet{pich}, Paper I hereafter, in press) describes the analytic theory which allow to assess the dynamics of haloes in the open, secular and non-linear r\'egimes.
The purpose  of the current paper is  to set out  a framework in which to describe statistically the environments of haloes and present
 results on the tidal field and the flux density of matter.
Paper III (\citet{aubert2}, in prep.) will conclude the  complete description of the environments of dark haloes.

\subsection{Galactic infall as a cosmic boundary }
 
Clearly a number of problems  concerning galactic evolution can only be tackled
properly via a detailed statistical investigation. Let  us  briefly  make  an  analogy  to the  cosmological  growth  of  density
fluctuations.   Under certain  assumptions,  one can  solve  the equations  of
evolution   of   those  over-densities   in   an   expanding  universe   (e.g.
\citet{Peebles}, see \citet{Bernardo} for an extensive review). Their  statistical evolution  due  to  gravitational
clustering follows, given the statistical  properties of the
initial  density  field.  For example, the power spectrum, $P(z,k)$, may be computed
for various primordial power  spectra, $P_{\mathrm{prim}}(k)$, and for various
cosmologies.   In  other words,  the  statistical  properties  of the  initial
conditions  are \textit{propagated} to  a given  redshift through  an operator
$\bf \aleph$ given by the non linear dynamical equations of the clustering~:
\begin{equation}
P(z,k)={\bf \aleph}(P_{\mathrm{prim}}(k),z)\,.
\end{equation}

In a similar way, how would  the statistical properties  of environments be propagated  to the
dynamical properties of galactic systems?   This is clearly a daunting task~:
the  previous  analogy  with  the  cosmological growth  of  perturbation  is
restricted to  its principle.  For example  the assumption of  a uniform and
cold  initial state  cannot  be  sustained for  galaxies  and halos.   While
spatial  isotropy is  clearly  not  satisfied by  disks,  and hot,  possibly
triaxial halos,  the velocity  tensor of galaxies  may also  be anisotropic.
Environments  also share  these inhomogeneous and anisotropic  features since they
are  also the  product  of  gravitational clustering  and  cannot be  simply
described  as  Gaussian fields.   These  boundary  conditions  are not  pure
`initial conditions'  since they  evolve with time  and in a  non stationary
manner (e.g. the accretion rate decreases with time).  A whole range of mass
must be taken into account, each with different statistical properties.
Finally, trajectories  cannot be considered  as ballistic (even in  the linear
regime) and must be integrated over long periods.
Notwithstanding  the  above  specificities  of  the  galactic  framework,  two
questions have to be answered:
\begin{enumerate}
\item What is the `galactic'  equivalent of $P_{\mathrm{prim}}(k)$, \ie how to
describe statistically the boundary conditions ?
\item What is  the `galactic' equivalent of $\bf \aleph$,  \ie how to describe
the inner galactic dynamics ?
\end{enumerate}

The  second point  is discussed  extensively  in \citet{pich}  and is  briefly
summarized in \sec{dynamics}.   In that paper, it is  shown how a perturbative
theory can describe the dynamics of haloes which experience both accretion and
tidal  interactions (see  also  \citet{aubert}).  Within  this formalism,  the
environment  is  described  by  the  external gravitational  potential  and  a
\textit{source function}. The former describes  fly-bys and the tidal field of
neighbouring large scale  structures.  The latter describes the  flows of dark
matter, i.e. the  exchanges of material between the  halo and the `inter-halo'
medium.   The  knowledge  of  these  two quantities  fully  characterizes  the
boundary  condition.  The  focus here  is on  well formed  halos which  do not
undergo major merger between $z=1$ and  $z=0$.  This bias is consistent with a
galactocentric description  in which a  perturbative description of  the inner
dynamics  is appropriate  and   equal  mass mergers are explicitly ignored.  As
briefly explained  in \sec{dynamics}, this  formalism provides a  link between
the statistical properties of environments to the statistical distributions of
the  responses of  haloes: this  link  is referred  to as  \textit{statistical
propagation}. In this  manner, the distribution of haloes  dynamical state can
be directly inferred from  the statistical properties of environments, without
relying  on the  follow up  of  individual interacting  haloes.  The  observed
distributions  of  dynamical  features  provides informations  on  the  cosmic
boundaries  which  influence haloes.   This,  together  with the  perturbative
formalism described  in \citet{pich} should allow us  to address statistically
the  recurrent  `nurture or  nature'  problem  of  structure formation  within
galactic systems.

{This statistical formalism is complementary to methods
based on  merger trees (which also  couple environment and  inner properties of
galactic systems see e.g. \citet{Kauff1}, \citet{Roukema} \citet{Somer}). These
'analytic'  or 'semi-analytic' models,  with presciption for the baryons
contained in haloes, angular momentum transfer, cooling and star formation may
predict  properties   of  galaxies  given   their formation history (e.g.  \citet{recipe}).  This  history  may be   provided  analytically  using
extended  Press-Schechter formalism  (see e.g.  \citet{B91},  \citet{LC93}) or
using  simulations  (e.g.  \citet{Kauff99},\citet{Benson}).  Even  tough  this
technic   now  extends   its  field   of  application   to   subhaloes  (see
e.g.   \citet{blaizot}),   it   remains somewhat   limited for  the purpose dynamical
applications. These require a  detailed description of  the geometrical
configuration of  the perturbations,  and of the dynamical  response of the halo. Both of  these are
difficult to reduce  to simple recipes.
Conversely, full analytic theories of the inner dynamics of interacting
haloes    were    developped    in   e.g.    \citet{Tremaine},    \citet{W98},
\citet{Murali}. Relying on the matrix method, these theories do take properly into
account the  resonant processes that  occur when the  halo is perturbed  by an
external potential.  However, they usually do  not 
account for the  perturbations induced  by the
accretion of matter,  while these authors generally considered test-cases where a  halo
responds     to    a     given    configuration     (or     statistics,    see
e.g. \citet{2001MNRAS.328..321W}) of perturbations.
Paper I  extended these  theories to  open  stellar systems  and , while
relying on  numerical simulations  to constrain  the environments,  it reformulated them  in terms  of the 
statistics of the inner dynamics of a  representative population of haloes.}

{Paper I  presents a list of possible applications.  For instance,
gravitational lensing by halos is  affected  by inner density fluctuations, which are induced  by  the halo's  environment~:  hence the statistics of  the
lensing  signal is  be related  to  the statistics  of halo's  perturbations,
therefore to the cosmological growth of  structures. Paper I showed how
this  approach  could  be  extended   to  other  observables,  such  as  X-Ray
temperature maps, SZ  surveys or direct detection of  dark matter. Statistical
propagation allows to relate cosmology  to the inner properties of cluster and
galaxies.   Conversely, it  should be  possible to  show if  the perturbations
measured in simulations are consistent with a secular drift toward a universal
profile of haloes. Closer to us,  the correlation of the numerous artefacts of
past accretion  in the  local group, such  as streams  or tidal tails,  can be
understood  in terms  of statistics  of  environments. }  All processes  which
depend critically on  the geometry of the interactions may  be tackled in this
framework\footnote{{However, all  departure to angular isotropy  on the sphere
will be  ignored here (in contrast  to what was stressed  in \citet{aubert} ),
and   its   implications   will   be   postponed   to   the   discussions   in
\sec{conclusion}.}}.

The statistical propagation relies on the knowledge of the  properties of the environment
 and is stated by the  point (i) mentioned above. This  question is investigated {the current paper} by  using  a  large  set  of simulations, where each halo provides a realization of the environment. From this large ensemble of interacting haloes, the aim is to  extract the global properties of their ``cosmic neighborhood''. Such a task requires an appropriate description of the source and the surrounding tidal field. It is the purpose of this work to implement this  description which should both provide insights on the generic properties of cosmic environments and be useful in a `dynamical' context. 
 Specifically, a method is presented to constrain the exchanges between the halo and its neighborhood, via the properties of accretion and potential measured on \textit{the virial sphere}. The advantages, specificities and caveats (and the methods implemented to overcome them) provided by this \textit{halocentric} approach will be presented in this paper.
% From this point of view, the approach developped in the current paper remains mainly exploratory. 

As shown in the following sections, the source function is given by the phase space distribution function (DF hereafter) of the advected material. As a consequence, its full characterization is a complex task since it involves
sampling a 5 dimensional space and relies on the projection of its DF on a suitable 5D basis. In particular, it is shown how such a description can be used to constrain the kinematic properties of accretion by dark matter haloes in cosmological simulations. The  detailed
 statistical characterization of the higher moments of the  source  is postponed to a paper III. % (\citet{aubert2}, in prep.).
 An alternative description of the source is also  presented; it relies on \textit{ flux densities} through the virial sphere, i.e. the moments of the source DF. Even though it is less suited to the dynamical propagation, this alternative description is easier to achieve numerically and to interpret physically. In particular, it illustrates how the source term may be characterized 
 statistically via its moments. 
The link between these  flux densities and the 5D projection of the source 
is discussed together with the one and two-point statistics of the flux
densities of mass through haloes in simulations. Also,
the mean of 
reprojecting the effect of the external gravitational potential  inside the halo  (through
Gauss's theorem)  while knowing its properties on the virial sphere are discussed. The potential's one and two points statistics are also investigated around simulated haloes and interpreted.

Finally (\Sec{gen}) provide means of  regenerating such  flows ab
initio from its tabulated statistical properties. Such tools yields a way to embed idealized simulations of galaxies into realistics cosmological environments.

  The outline of the paper is the following: \Sec{dynamics}  presents  briefly   the  dynamics  of  open  collisionless
systems and states the principle of statistical propagation.  \Sec{source} presents the procedure  used to compute the source
term, and illustrates  its implementation on a given  halo.  The simulations
and the corresponding  selection biases of our sample  are then described in
\sec{stat}.    \Sec{stat1d}  and   \seq{stat2d}   present  the   statistical
measurements  for one  and two  point statistics  respectively. \Sec{scenario}
draws a global picture of galactic infall on $L^{\star}$ galaxies, while
  a discussion and conclusions follow in \sec{conclusion}.

{Among the different results described in this paper, the reader will find~:
\begin{itemize}
\item a statistical description of the external gravitational field felt by halos :
  the potential is found to be quadrupolar and stationnary.
\item a study of the evolution of accretion : accrection by dark matter haloes
  decrease with time, while the outflows becomes more significant at recent
  times. 
\item constrains on the trajectory of infalling material~: accretion is found to be essentially radial, while outflows are
  found to be more circular.
\item results on the two-point statistics of the external potential measured
  on the viral sphere~: the potential provide hints of an anisotropic perturbation of
  the halo.
\item results on the two-point statistic of the accretion's distribution on
  the virial sphere: accretion is dominated by small scales fluctuations
  and has a shorter coherence than the external gravitational field.
\end{itemize}}

%%%%%%%%%%%%%%%%%%%%%%%%%%%%%%%%%%%%%%%%%%%%%%%%%%%%
\section{Dynamics of open collision-less systems}
\label{s:dynamics}
%%%%%%%%%%%%%%%%%%%%%%%%%%%%%%%%%%%%%%%%%%%%%%%%%%%%
%%%%%%%%%%%%%%%%%%%%%%%%%%%%%%%%%%%%%%%%%%%%%%%%%%%%

The  exchanges   occurring  between  a   halo  and  its  environment   can  be
characterized  in several ways.  One of  the classical  method involves  building a
merger tree  where the whole  history of formation  of a halo is  expressed in
terms  of  global properties of its  progenitors (e.g. \citet{LC93}, \citet{KW93},  \citet{RK99}).
While well  suited to study  the evolution of  those characteristics, it  cannot be
directly applied to predict in details the halos' inner dynamic because of the
lack of spatial information on  these interactions.  One could track the whole
(six dimensional)  phase-space history  of all the  progenitors, but  not only
would it be difficult to store  in practice it would also not give information
on  the  influence  of  large  scale structures  through  their  gravitational
potential.
%We suggest an alternative method to describes the interactions of a given halo
%with its  environment.  
In  the present  paper, following \cite{aubert}, it is suggested  to  measure the  relevant quantities  on a
surface  at \textit{the  interface}  between the  halo  and the  intergalactic
medium.  Accretion is described as a flux of particles through the halos' external
boundaries.

This  section  presents an  extension of  the formalism  developed by e.g. \citet{Tremaine}  and \citet{Murali}  to  open spherical  collisionless
systems. The dynamics  of a dark matter spherical halo  is obtained by solving
the collisionless Boltzmann equation coupled with the Poisson equation~:
\begin{eqnarray}
\partial_t      F+\bv\cdot\partial_\br     F-\nabla\Psi\cdot\partial_\bv     F
&=&0\,,\label{e:bsc}\\ \Delta \Psi&=&4 \pi G \int \d^3 v F ( {\bf v} ).
\end{eqnarray} 
where  $F(\bfr,\bfv,t)$ is  the  system's distribution  function coupled  to
$\Psi(\bfr,t)\equiv   \psi+\psi_e$,   the   total  gravitational   potential
(self-gravitating  +  external  perturbation).   Note that,  in  a  somewhat
unconventional manner, $\psi^e$ refers  here to the external potential, \ie
the  tidal  potential  created   by  the  perturbations  {\sl  outside}  the
boundary. The gravitational field of incoming particles is accounted for by
the  source   term.   \Eq{bsc}  coupled  with  Hamilton's   equations  is  a
conservation equation:
\begin{equation}
\partial_t F+{\tilde \nabla}(\bu F)=0,
\label{e:conserv}
\end{equation}
where     $\bu\equiv(\bv,-\nabla      \Psi)$     and     ${\tilde\nabla}\equiv
(\partial_\br,\partial_\bv)$.   As  a consequence,  considering  a `source  of
material' described by $f_e(\bw)$ located on a surface $S(\bw)$ implies:
\begin{equation}
\partial_t                 F+{\tilde\nabla}(\bu                F)=-\delta_{\rm
D}(S(\bw)-a)\bu\cdot\frac{\nabla S}{|\nabla S|}f_e(\bw),
\label{e:conservopen}
\end{equation}
where  $\bw\equiv(\br,\bv)$ describes  the  phase space  and  $a$ defines  the
surface boundary of  the studied system (here $\delta_{\rm  D}$ stands for the
Dirac delta function).  If this boundary is defined as a spherical surface with
radius $R$, then \eq{conservopen} becomes (e.g. \citet{Appel})~:
\begin{eqnarray}
\partial_t   F+{\tilde\nabla}(\bu   F)&=&-\delta_{\rm  D}(r-R)v_r   f_e(\bw)\\
&\equiv&-\delta_{\rm D}(r-R)s^e(\bw).
\label{e:conservopen2}
\end{eqnarray}
The function  $s^e$ will  be hereafter referred  to as the  `source' function.
Formally the r.h.s.   of \eq{conservopen2} can be seen  as an additional local
rate   of  change   of   the  system's   distribution   function.  Note   that
\eq{conservopen2} involves the external potential, $\psi^e$, via $\bu$.
%%%%%%%%%%%%%%%%%%%%%%
\subsection{Moments of the source term}
%%%%%%%%%%%%%%%%%%%%%%
%%%%%%%%%%%%%%%%%%%%%%
 Integrating \eq{conservopen2} over velocities  leads to the mass conservation
 relation:
\begin{equation}
\partial_t\rho+\nabla(\bv\rho)=-\delta_{\rm                        D}(r-R)(\rho
v_r)_e\equiv-\delta_{\rm D}(r-R) \varpi_\rho, \EQN{defmass}
\end{equation} 
where  the  source  appears as  an  external  flux  density of  matter  $(\rho
\overline{v_r})_e$   or    $\varpi_\rho$.   Taking   the    next   moment   of
\eq{conservopen2} leads to the Euler-Jeans Equation:
\begin{equation}
\partial_t\rho\bv+\nabla\cdot(\rho\bv\bv)+\rho\nabla\Psi=-\delta_{\rm
  D}(r-R)(\rho v_r\bv)_e, \EQN{defEuler}
\end{equation}
where the source  adds a flux density of  momentum, $\varpi_{\rho\bv}$, to the
conventional   Jeans    equation.    Taking   the    successive   moments   of
\eq{conservopen2}  will  generically  include  a  new term  in  the  resulting
equations.  

%%%%%%%%%%%%%%%%%%%%%%
\subsection{Propagating the dynamics}
%%%%%%%%%%%%%%%%%%%%%%
%%%%%%%%%%%%%%%%%%%%%%

 Following  \citet{Tremaine},  \eq{conservopen2}  can  be  solved  along  with
Poisson's  equation in  the  regime  of small  perturbations.  In the spirit of  
paper Iet us define  the
system's  environment by the  external perturbative  potential $\psi^e(\br,t)$,
and the source $s^e(\br,\bv,t)$.   Given this environment, the system's linear
potential  response  $\psi(\br,t)$ can  be  computed.   Writing the  following
expansions:
\begin{eqnarray}
\psi(\br,t)&=&\sum_{\bf n}  a_{\bf n}(t) \psi^{\bf  [n]}(\br),\EQN{defpsie} \\
\psi^e(\br,t)&=&\sum_{\bf    n}    b_{\bf    n}(t)   \psi^{\bf    [n]}(\br),\\
s^e(\br,\bv,t)&=&\sum_{\bf   n}    c_{\bf   n}(t)   \phi^{\bf   [n]}(\br,\bv),
\EQN{defss}
\end{eqnarray}
where $\phi^{\bf  [n]}(\br,\bv)$ and $\psi^{\bf [n]}(\br)$  are suitable basis
functions, and solving the Boltzmann  and Poisson's equation for $a_\M{n}$, one
finds (\citet{aubert})~:
\begin{eqnarray}
{\bf  a}(t)=\int_{-\infty}^\infty   \md  \tau  {\bf  K}(\tau-t)\cdot\left[{\bf
a}(\tau)+{\bf b}(\tau)\right] + {\bf H}(\tau-t)\cdot {\bf c}(\tau).
\label{e:linresp}
\end{eqnarray}
The  kernels  $\bf  K$  and  $\bf  H$ are  functions  
%given  in  the  appendix
%(\Eqs{defK}{defsig})  
of the equilibrium  state distribution  function, $F_0$,
and  of the two  basis, $\phi^{\bf  [n]}(\br,\bv)$, and  $\psi^{\bf [n]}(\br)$
only (see paper~I).  As  a consequence, they may  be computed once  and for all for  a given
equilibrium model.  Since the  basis function, $\psi^{[\M{n}]}$, can be customized
to  the NFW-like  profile of  dark matter  halos, it  solves  consistently and
efficiently the coupled dynamical and  field equations so long as the entering
fluxes of dark matter amounts to  a small perturbation in mass compared to the
underlying equilibrium.

Assuming  linearity and knowledge of  $\bf K$  and $\bf  H$, one  can see  that the
properties  of   the  environments   (through  $\bf  b$   and  $\bf   c$)  are
\textit{propagated} exactly to the inner dynamical properties of collisionless
systems.  Note in  particular that the whole phase space  response of the halo
follow  from the  knowledge  of  $\M{a}$.  For  example,  taking the  temporal
Fourier  transform of  \eq{linresp}, the  cross correlation  matrix  is easily
deduced:
\begin{eqnarray}
\langle  {\tmmathbf{  \hat a}}\cdot  {\tmmathbf{  \hat a}}^{*\T{}}  \rangle&=&
\langle (\bI -{\bf \hat K} )^{-1} \cdot \left[{\bf \hat K} \cdot {\bf \hat b}+
{\bf \hat H}  \cdot {\bf \hat c}\right]\cdot \nonumber  \\&&\left[{\bf \hat K}
\cdot {\bf \hat b}+ {\bf \hat H} \cdot {\bf \hat c}\right]^{\T{} *} \cdot (\bI
-{\bf \hat K} )^{-1 * \T{}} \rangle\,, \EQN{defcor}
\end{eqnarray}
where  ${\bf \hat  x}= {\bf  \hat  x}(\omega)$ is  the Fourier  Transform of  ${\bf
x}(t)$.  The environment's two-points statistic, via $\langle {\tmmathbf{ \hat
b}}\cdot {\tmmathbf{  \hat b}}^{* \T{}}  \rangle $, $\langle  {\tmmathbf{ \hat
c}} \cdot {\tmmathbf{ \hat c}}^{* \T{}} \rangle$ and $\langle {\tmmathbf{ \hat
b}}\cdot {\tmmathbf{\hat  c}}^{* \T{}} \rangle $, modifies  the correlation of
the response  of the inner  halo.

Linear dynamics do not take in account the effects on the perturbation induced
by dynamical  friction.  More  generally the damping  of incoming  fluxes will
ultimately require non linear dynamics  (since the relative temporal phases of the 
infall do matter in
that context).  It  is also assumed  in \eq{linresp}  that the
incoming material  does not modify the  equilibrium state of  the system.  The
secular evolution  of the system should  also be ultimately  taken in account,
through a quasi-linear theory for example (see e.g. paper~I). 
  %The above description is therefore
%a first step towards a complete description of open collisionless systems.

 Let us emphasize that, since the addressed problem is linear, the response,
\eq{linresp},  can  be  recast  into  a  formulation  which  only  involve  an
`external potential', namely  the sum of $\psi^e$ and  the potential created
by the entering  particles described by $s^e$.  While  formally simpler at the
linear deterministic  level, this  alternative formulation does  not translate
well non linearly or statistically (since it would require the full knowledge of 
  the perturbation everywhere in space in a manner which is dependent upon 
  the inner structure of the halo).

In  the   following  sections,   our  aim  is   to describe how  the   two  fields
$\psi^e(\br,t)$ and  $s^e(\br,\bv,t)$ can be extracted from haloes in cosmological simulations. Then it will be  shown how to characterize    their  statistical properties  as  a function  of  time via their  expansion  coefficients ,  $b_{\bf  n}(t)$ and  $c_{\bf n}(t)$.

%%%%%%%%%%%%%%%%%%%%%%%%%%%%%%%%%%%%%%
\subsection{Convention and notations}
%%%%%%%%%%%%%%%%%%%%%%%%%%%%%%%%%%%%%%

In  what follows, let us  characterize the  properties of  2  fields, either
angularly, kinematically, statistically or temporally, or any combination (for
various  classes of  masses).     For  a given  field, $X$, let us
introduce the following notations for clarity~:
\begin{equation}
{\overline X}  \equiv \frac{1}{4 \pi  } \int X(\theta,\phi)\d  \sin(\theta) \d
\phi \,, \EQN{def-temp-avg}
\end{equation}
which represents the  angular average of $X$ over  the sphere.  Alternatively, let us define the temporal average over $\Delta t$ as:
\begin{equation}
{\underline   X}   \equiv\frac{1}{\Delta   T}   \int\limits_0^T  X(t)   \d   t\,.
\EQN{def-temp-avg}
\end{equation}
Finally let us define the ensemble average as
\begin{equation}
{\langle  X   \rangle  }   \equiv  \int \,  X\,  {\cal  F}(X)\,  \d   X=  E   \{  X\}\,,
\EQN{def-temp-avg}
\end{equation}
where $\cal F$ is the density probability distribution of $X$.  Here $E\{X\}$ stands
for the  expectation of $X$.  In  practice, in \sec{stat},  an estimator for
ensemble  average  of  X  measured   for  N  halos  is  given~:\footnote{  An
alternative would be to weight the  sum by the relative number of objects in
each halo, hereby down weighting light  halos. It is found that this alternative
estimator did not significantly affect our measurements }
\begin{equation}
\langle X \rangle_{{}_N} = \frac{1}{N}\sum_i^N X_i \,. \EQN{def-stat-avg}
\end{equation}
The underlying  probability distributions  are
sometimes very  skewed (when,  \eg corresponding to  strong or  weak accretion
event  around massive  or smaller  halos),  which requires  special care  when
attempting  to define statistical  trends.
 Hence let us also  define $\llangle  X \rrangle$ as  the \textit{mode}  (or most
probable value) of the fitting distribution of $\cal F$.
%% In the following sections, we will use the following notations. The average
%% value of $X$ over the different directions on the sphere is defined by:
%% \begin{equation}
%% \overline{X}=\frac{1}{4\pi}\int_{4\pi}X\md\bO.
%% \end{equation}

 All external quantities,  (flux densities, potential etc) will  generally be labeled as
$X^e$.   Let us  introduce  moments  of  the  source over  velocities,  which
correspond to flux densities, noted $\varpi_X$ and their corresponding fluxes,
noted $\Phi_X$. \Tab{varpi}  gives a list of such  flux densities, flux pairs.
Finally  the  harmonic  transform  of  the  field, $X$,  will  be  written  as
$a^X_{\ell,m}$ and its corresponding power spectrum 
$C^X_\ell$, while  the parameters  relative to fitting  the statistics  of the
field will be written as $q_X$.
Note that the contrast of the field, $X$ was also introduced as 
\begin{equation}
\delta_X\equiv  \frac{X -\overline{X
}}{\langle\overline{X}\rangle} \,, \EQN{defcontrast}
\end{equation}
 and its corresponding harmonic transform, ${\tilde a}^X_{\ell m}$.
A summary of all the notations can be found in Appendix \ref{s:notations}.
 
%------------------------------------------------

%%%%%%%%%%%%%%%%%%%%%%%%%%%%%%%%%%%%%%%%%%%%%%%%%%%%
\section{The Source of infall}
%%%%%%%%%%%%%%%%%%%%%%%%%%%%%%%%%%%%%%%%%%%%%%%%%%%%
\label{s:source}

Let  us first  describe our  strategy to  fully characterize  the  source of
cosmic  infall   at  the  virial   radius  via  collisionless   dark  matter
simulations, and enumerate the corresponding biases.  In particular
let us illustrate our procedure on a template halo.

%%%%%%%%%%%%%%%%%%%%%%%%%%%%%%%%%%%%%%%%%%%%%%%%%%%%
\subsection{Describing the source}
\label{s:descsource}
As  argued in  \sec{dynamics}, computing  the response  of an  open  system to
infalling material  requires the knowledge  of the source  function, $s^e({\bf
r},{\bf  v},t)$.  Given  the particles  accreted  by a  halo, one  possibility
involves  storing  those  phase-space  properties for  all  particles.   While
feasible  for  a limited  number  of halos,  this  task  would become  rapidly
intractable  for our large  number of  simulations. In  order to  compress the
information, the accreted  distribution function is projected here  on a basis
of function, following \eq{defss}.

\begin{figure} 
\centering
\resizebox{0.7\columnwidth}{0.9\columnwidth}{\includegraphics{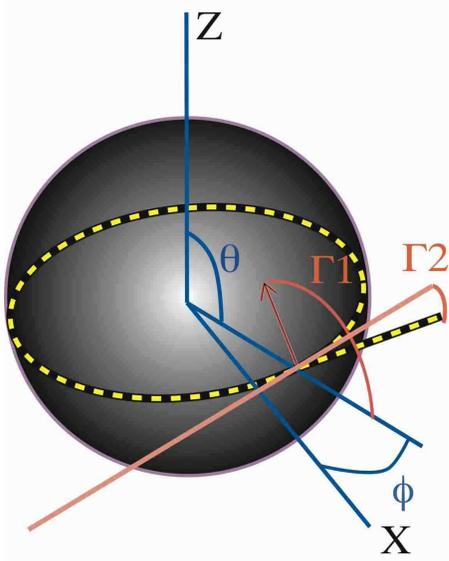}}
\caption{The angles  $\bO$ and  $\bG$. The dot  indicates the position  of the
particle  on  the sphere.   The  dashed  ellipse  represents the  plane  which
contains both  the particle and the  sphere center. $X$ and  $Z$ are arbitrary
directions defined by the  simulation's box.  $(\theta,\phi)={\bf \Omega}$ are
the    particle's    angular     coordinates    on    the    sphere.     ${\bf
\Gamma}=(\Gamma_1,\Gamma_2)$ define the orientation of the particle's velocity
vector (shown as an arrow).}
\label{f:angles} 
\end{figure}

Since the measurement is carried at a fixed radius, the phase space is reduced
from six  to five degrees  of freedom: two for  the angular position  on the
sphere, described  by two angles, $(\theta,\phi)\equiv{\bf \Omega}$,  and three for
the    velocity    space     described    in    spherical    coordinates    by
$(v,\Gamma_1,\Gamma_2)=(v,{\bf  \Gamma})$, where $v$  is the  velocity modulus
and ${\bf \Gamma}$  are the two angles describing  its orientation (see figure
\ref{f:angles}). The  angle $\Gamma_1$ indicates  how radial is  the velocity,
with  $\Gamma_1>\pi/2$  for infalling  dark  matter  and $\Gamma_1<\pi/2$  for
outflows.   $\Gamma_2$ indicates  the orientation  of the  infall's tangential
motion.

Recall that the two fields, $c_\M{n}$ (hence $s^e(\bO,\bG,v,t)$) and $b_\M{n}$
(hence $\psi^{e}(\bO,t)$) are respectively  five dimensional and two dimensional
(as a function of mass and time).   Note also that both $s^{e}$ and $\psi^{e}$ are
statistically  stationary with  respect  to $\bO$,  while  $s^{e}$ is  partially
isotropic and not stationary with respect to $\bG$; neither $\psi^{e}$ and $s^{e}$
are stationary with respect to  cosmic time.

\subsubsection{Harmonic expansion of the incoming fluxes}
The ${\bf \Omega}$ and ${\bf  \Gamma}$ dependence are naturally projected on a
basis    of    spherical    harmonics,    $Y_{\ell,m}({\bf    \Omega})$    and
$Y_{\ell',m'}({\bf \Gamma})$.  The  velocity amplitude dependence is projected
on a basis of Gaussian  functions, $g_\alpha(v)$ with means $\mu_\alpha$ and a
given r.m.s. $\sigma$.  One can write:
\begin{equation}
\phi^{[\M{n}]}(\br,\bv)=Y_{\M{m}      }({\bf      \Omega})Y_{\M{m}'     }({\bf
\Gamma})g_\alpha(v)\,, \EQN{defphin}
\end{equation}
where $\M{n}\equiv (\ell,m,\alpha,\ell',m')=(\M{m},\alpha,\M{m}')$.
The expansion coefficients, $c_{{\bf m'}}^{{\bf m}\alpha}(t)$ are given by~:
\begin{equation}
c^{{\bf       m}}_{{\bf       m      '}\alpha}(t)=({\bf       G}^{-1}\cdot{\bf
s}^{\M{m}}_{\M{m}'})_\alpha. \EQN{defCoef}
\end{equation} 
where:
\begin{equation}
({\bf s}^{\M{m}}_{\M{m}'})_\beta=\int {\mathrm d}\bO{\mathrm d}\bG{\mathrm d}v
v^2   g_\beta(v)   Y^*_{\M{m}   }(\bO)Y_{\M{m}'}(\bO)s^e(v,{\bf   \Omega},{\bf
\Gamma},t),
\end{equation}
given
\begin{equation}
{\bf G}_{\alpha,\beta}=\int {\mathrm d}v v^2 g_\alpha(v)g_\beta(v).
\label{e:defG}
\end{equation}
Note  that  the expansion  defined in  \eq{defss} where the coefficients are given by \eq{defCoef} involves  5  subscripts  spanning the  5
dimensional phase  space, while the  expansion in \eq{defpsie} only  involve 3
subscripts.   This  description  of the  source  term  is  reduced  to a  set  of
coefficients which  depends on time only. Furthermore  this procedure requires
to parse the particles only once,  and all the momenta (\eg mass flux density,
PDF of  impact parameter ...)   of the source  terms can be  computed directly
from these coefficients.   As a consequence, the statistics  of momenta follow
linearly  from  the  {\sl  statistics}  of  coefficients  only,  as  shown  in
\sec{stat}.
%The direct use of particles would have implied several parses of the
%simulations for each momentum to be studied.

\subsubsection{Harmonic expansion of the external potential}
Let us  call $ b'_{\ell m}(t)$  the harmonic coefficients of  the expansion of
the  external potential on  the virial  sphere.  Following  \citet{Murali}, let us
expand    the   potential   over    the   biorthogonal    basis,   $(u_n^{\ell
m},d_n^{\ell,m})$, so that
\begin{eqnarray}
\psi^e(r,\bO,t) &=&\sum_{n,\ell,m}     b'_{\ell    m}(t)\,    Y_{\ell}^m(\bO)\,
 \left(\frac{r}{R_{200}}\right)^\ell\,,  \nonumber \\  &=&\sum_{\M{n}}     b_{\M{n}     }(t)
 \psi^{[\M{n}]}(\br) \,, \label{e:defexp}
\end{eqnarray}
where $ \psi^{[\M{n}]}(\br)  \equiv Y_{\ell}^m(\bO) u_{j}^{\ell m}(r)$.  The
first equality in \eq{defexp} corresponds to the inner solution of the three
dimensional potential  whose boundary  condition is given  by $Y^\ell_m(\bO)
b'_{\ell m}$ on  the sphere of radius $R_{200}$  (defined below).  Since the
basis is biorthogonal, it follows that
\begin{equation}
b_{\M{n}       }(t)      =       \left(\int      d_n^{\ell       m}(r)      \,
\left(\frac{r}{R_{200}}\right)^\ell \d r \right) b'_{\ell m}(t)\,.
\end{equation}
It  is  therefore straightforward  to  recover  the  coefficient of  the  3D
external potential from that of the potential on the sphere.

\subsection{From simulations to expansion coefficients}
\label{s:coeff}
Once  a halo  is  detected, its  outer  `boundary' is  defined  as a  sphere
centered on its  center of mass with a  radius, $R_{200}$ (or \textit{Virial
radius}),  defined  implicitly  by  $3M/(4\pi  R_{200}^3)=200\rho_0$.   This
choice of radius  is the result of a compromise between  being a large distance to
the  halo center, to limit the contribution of halo's inner material to
fluxes,   and being still close enough to the halo's border, to limit  the
simulation's  fraction to  be processed and avoid contributions of fly by
objects. Let us emphasize that several definition of the virial radius can be
found in the litterature, involving e.g. the critical density, or  a different
contrast factor, where the latter may or may not depend on the
cosmology. Hence, one should keep in mind that all the quantitative
results presented in this article depend on our specific choice of a definition.

The time evolution  of accretion is measured backwards  in time by following
the biggest  progenitor of  each halo detected  at redshift $z=0$.   
The  positions and velocities  of particles passing through  the virial
sphere  between  snapshots are then stored.  All  positions  are  measured  relative to  the
biggest progenitor center of mass  while velocities are measured relative to
its  average  velocity, for  each  redshift  z. In one of the simulation described below,  the total comoving drift distance of the center of mass 
was compared to the distance between the halo's position between z=1 and z=0. The haloes were chosen to satisfy the criteria described in $\ref{s:selection}$. It is found that the scattering of the motion of the center of mass represents less than 10 $\%$ of the distance covered in 8 Gyrs. The  center of mass  of  the biggest progenitor seems stable enough to be a reference.    

\begin{figure} 
\centering                 \resizebox{7cm}{6cm}{\includegraphics{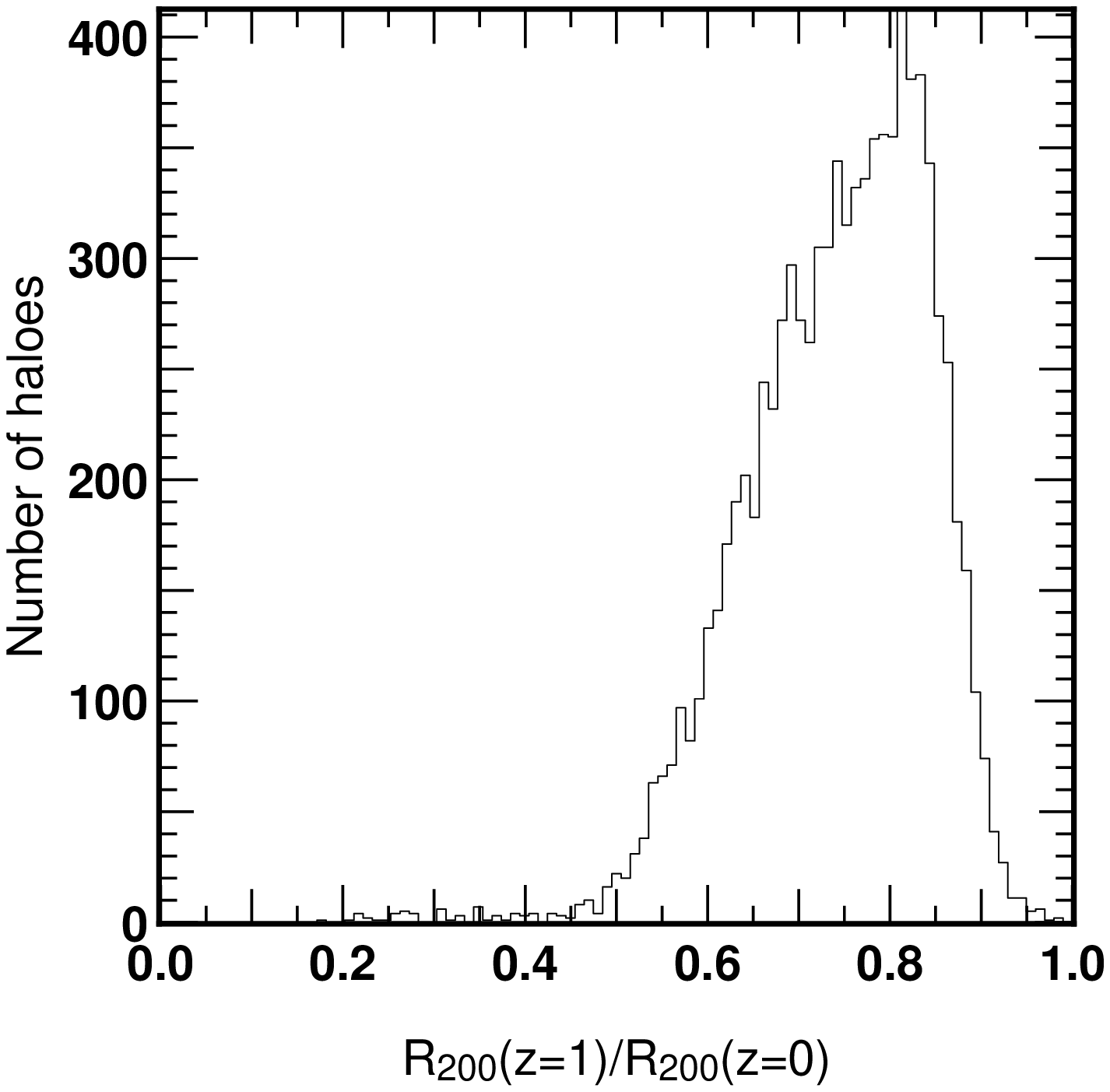}}
\resizebox{7cm}{6cm}{\includegraphics{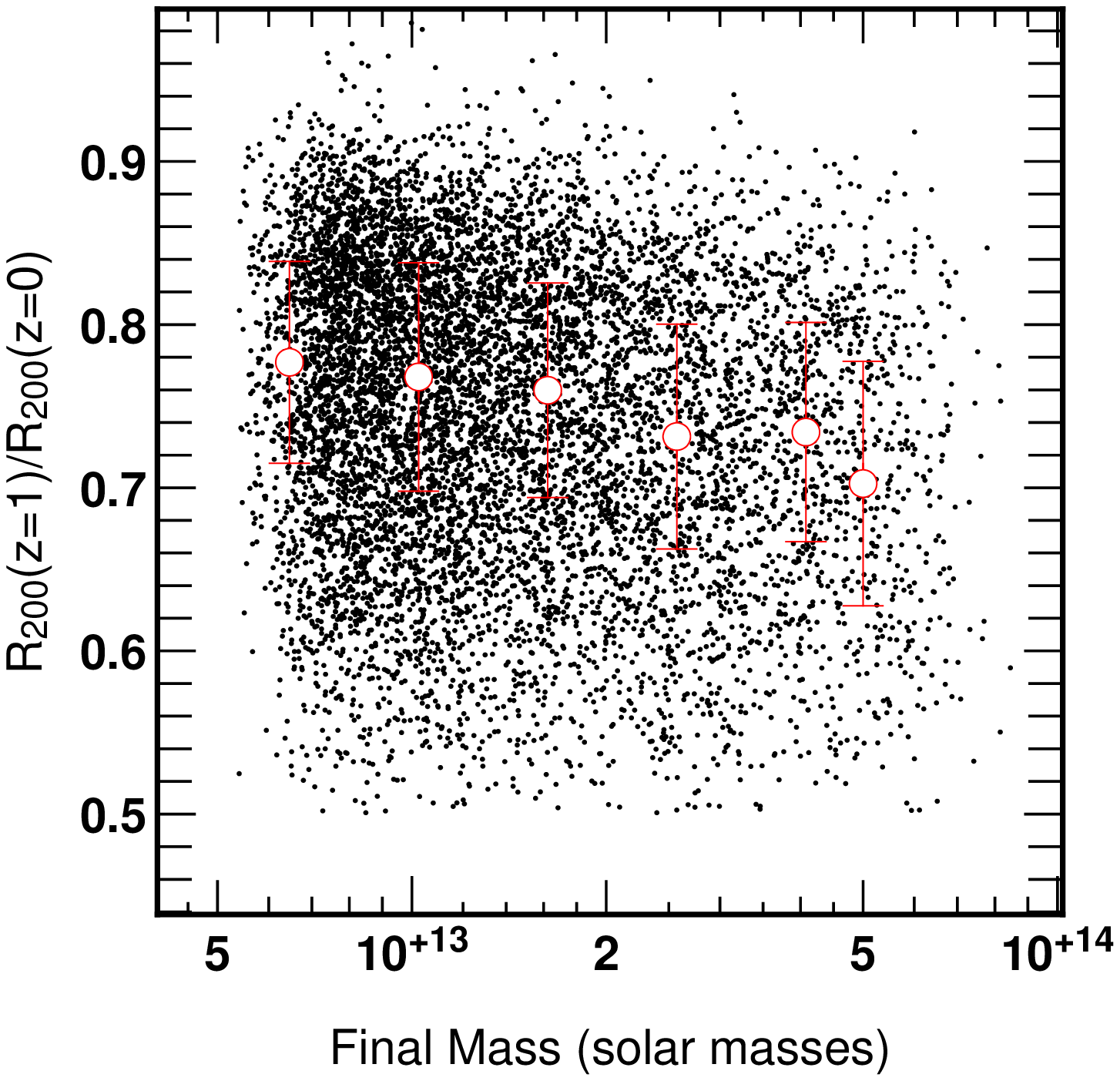}}
\caption{\textit{Top:} the distribution of the ratio between the virial radius
measured  at  $z=1$  and  at  $z=0$.   \textit{Bottom:}  the  distribution  of
$R_{200}(z=1)/R_{200}(z=0)$ as a function of the halo's final mass. Each point
represent   one    halo.    Symbols   stand   for   the    median   value   of
$R_{200}(z=1)/R_{200}(z=0)$ in 6 different  classes of masses.  Bars stand for
the  interquartile. The  two  measurements  were performed  on  9023 haloes  which
satisfy the selection criteria defined in section \ref{s:selection}.}
\label{f:varr200} 
\end{figure}

Sticking to  the  previous
definition  of  $R_{200}$  would   imply  a  changing  outer  boundary and  an
`inertial' flux through a moving surface  would have to be taken in account.
To overcome this effect, the sphere was kept {\sl constant in time }
at a
radius equals to $R_{200}(z=0)$.  {  This choice corresponds to a reasonable
approximation since  the actual virial radius does  not change significantly
with time between $z<2$ for  a  reasonably smooth accretion  history.  As shown  in figure
\ref{f:varr200}  the virial  radius at  $z=1$  is only  $20\%$ smaller  than
$R_{200}$ measured at $z=0$.  Larger  haloes have larger variations but the
median value  of the difference between  the two radii  remains smaller than
$30\%$ for final masses smaller than $10^{14} M_\odot$.}
 Finally, measurements were done  using physical coordinates
(and not comoving coordinates).  These  choices were partly guided by the fact
that they simplify  future applications of these results  to the inner dynamic
of the halos (see paper~I).

\subsubsection{Sampling on the sphere}
%%%%%%%%%%%%%%%%%%%%%%%%%%%%%%%%%%%
As shown in section \ref{s:dynamics},
%$s^e$ is a  phase space flux density
%and formally $s^e=\varpi_f$. We have~:
 the source function $s^e$ reads
\begin{equation}
s^e\equiv   f(\br,\bv,t)  v_r=\sum_i\delta_{\rm  D}^3(\br-\br_i(t))\delta_{\rm
D}^3(\bv-\bv_i(t))v_{r,i}.
\end{equation}
Switching to spherical coordinates leads to:
\begin{eqnarray}
s^e&=&\sum_i^N                                                \frac{\delta_{\rm
D}(R_{200}-r_i(t))}{R_{200}^2}\frac{\delta_{\rm   D}(v-v_i(t))}{v^2}\times\\\nonumber\\
&&\frac{\delta_{\rm                 D}({\bf                 \Omega}-{\bf
\Omega}_i(t))}{\sin{{\Omega_1}}}\frac{\delta_{\rm     D}({\bf     \Gamma}-{\bf
\Gamma_i}(t))}{\sin(\Gamma_1)}v_{r,i}(t),
\label{e:source1}
\end{eqnarray}
where $i$ is the particle index.
Now:
\begin{eqnarray}
v_{r,i}\delta_{\rm   D}(R_{200}-r_i(t))&=&\sum_k   v_{r,k,i}   \left|\frac{\md
t}{\md    r    }\right|\delta_{\rm   D}(t-t_{200,k,i})\,,\nonumber\\    &=&\sum_k
w_{k,i}\delta_{\rm D}(t-t_{200,k,i})\,,
\end{eqnarray}
where $t_{200,k,i}$ corresponds  to the $k$-th passage of  the $i$-th particle
through  the virtual  boundary  $R_{200}$ (and  $v_{r,k,i}$ the  corresponding
radial velocity).  { In our conventions}, the weight function $w_{k,i}$
takes the value 1 if the particle is entering and -1 if its exiting the virial
sphere.  Given  that our  time resolution  is finite, let  us consider  a time
interval $\Delta T$ around $t$ and define the (temporal) average phase space flux density
over $\Delta T$:
\begin{equation}
{\underline  s}^e   (t)\equiv\frac{1}{\Delta  T}  \int_t^{t+\Delta   T}  d\tau
s^{e}(\tau).
\end{equation}
\Eq{source1} becomes:
\begin{equation}
 {\underline s}^e (t)=\sum_{i,k}^N \frac{\delta(v-v_{i,k})}{v^2 \Delta T \cdot
R_{200}^2}\frac{\delta({\bf                                        \Omega}-{\bf
\Omega}_{i,k})}{\sin{{\Omega_1}}}\frac{\delta({\bf                 \Gamma}-{\bf
\Gamma_{i,k}})}{\sin(\Gamma_1)}w_{i,k}.
\label{e:source2}
\end{equation}
The  simulations were  sampled in  time regularly  in $\ln  (z)$  (\ie $\Delta
\ln(z)=\R{constant}$).   From $z=2$ to $z=0.1$, 23  snapshots were taken
(and a $z=0$ snapshot  was added to the sample).  If $\Delta  t$ is small, the
sum over $k$ should mostly involve one passage, \ie:
\begin{equation}
{\underline   s}^e   (t)\sim\frac{1}{\Delta   T   \cdot   R_{200}^2}\sum_{i}^N
\frac{\delta(v-v_{i})}{v^2}\frac{\delta({\bf                       \Omega}-{\bf
\Omega}_{i})}{\sin{{\Omega_1}}}\frac{\delta({\bf                   \Gamma}-{\bf
\Gamma_{i}})}{\sin(\Gamma_1)}w_{i}.
\end{equation}
\begin{figure} 
\centering                                           \resizebox{0.8\columnwidth
}{0.8\columnwidth}{\includegraphics{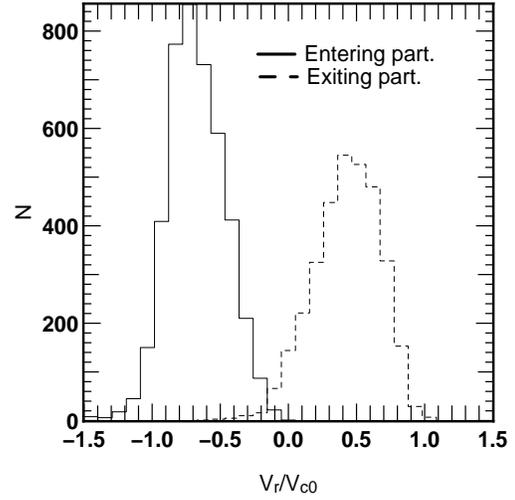}}
\caption{The distribution of interpolated radial velocities $v_r$ of particles
passing through the  virial radius. Those particles were  taken from the whole
history    of   accretion    of   a    typical   halo    ($R_{200}=860$   kpc,
$M(z=0)=3\cdot10^{13}\mathrm{M}_\odot$).   Entering   particles  (solid  line)
$v_r<0$ while exiting particles (dashed line) have $v_r>0$, as it should be.}
\label{f:checkvel} 
\end{figure}
Now these  measurements only give  access to $(v,\bO,\bG)$ at  fixed redshift,
$z$, every  varying $\Delta  z$.  Consequently, these
values   need to   be interpolated at the sought $t_{200,i}$ approximated by~:
\begin{equation}
t_{200,i}=t_{i}(z_n)+\frac{t(z_{n+1})-t(z_{n})}{r_{i}(z_{n+1})-r_{i}(z_{n})}(R_{200}-r_{i}(z_{n})).
\label{e:interp}
\end{equation}
Given these  `crossing' instant, the positions, $\br$,  and velocities, $\bv$,
are also linearly  interpolated. For instance, one gets  for the $x$ component
of the velocity~:
\begin{equation}
v_{x,i}(t_{200})=v_{x,i}(z_n)+\frac{v_{x,i}(z_n+1)-v_{x,i}(z_n)}{t(z_{n+1})-t(z_{n})}(t_{200}-t(z_{n})).
\label{e:interp2}
\end{equation}
Such an interpolation is not strictly self-consistent since a ballistic motion
requires a  constant velocity along  the trajectory.  The  worst-case scenario
would correspond  to particles  which have entered  the virial sphere  with an
outflowing velocity vector  and vice-versa.  
%One could also  use a third order
%more  accurate scheme of  interpolation to  account for  the curvature  of the
%trajectories between the  two snapshots.    
As a
simple but important check,  the distribution of interpolated radial
velocities was plotted (see \Fig{checkvel}).   Those were computed from the whole
history    of   accretion    of   a    typical   halo    ($R_{200}=860$   kpc,
$M_{z=0}=3\cdot10^{13}\mathrm{M}_\odot$).    The   two   types  of   particles
(entering/exiting)  are confined  in their  radial velocity  plane: entering
(resp. exiting)  particles have  negative (resp. positive)  radial velocities.
Velocities are  correctly interpolated. It  also means that our  time steps are
small enough to ensure a small variation of positions/velocities of particles,
validating a posteriori  our assumptions.  A fraction of  exiting particles do
have a negative  radial velocity but represent less than a  few percent of the
total population.  For safety, those particles are rejected from the following
analysis.  

\begin{figure} 
\centering \resizebox{6cm}{9cm}{\includegraphics{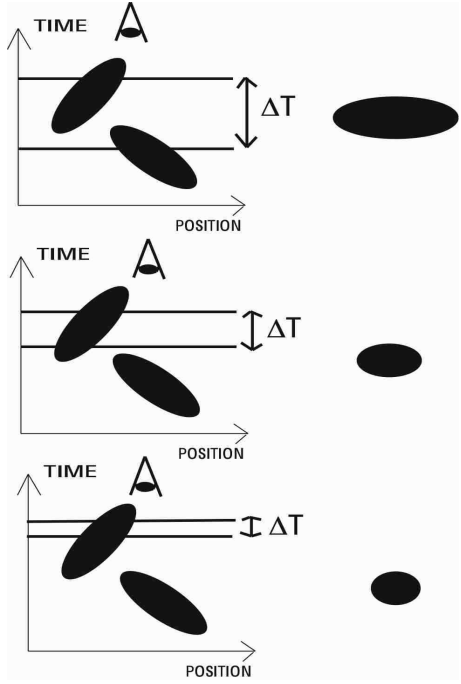}}
\caption{The impact  of time averaging on  the measured scales  of dark matter
passing through the  sphere. On the left, time-position  diagram of dark matter
(black  ellipses)  as  they  pass  through the  sphere.  Time  integration  is
performed  during $\Delta  T$ (the  two horizontal  lines). On  the  right the
accreted  dark  matter  as seen  on  the  sphere.  A longer  integration  time
increases the  length scale  of the  incoming blob.  If  $\Delta T$  gets very
large, different blobs may be seen as one (upper diagram).}
\label{f:scale} 
\end{figure}

One should note  that the  measured angular scales  are sensitive to  the time
sampling (see \Fig{scale}).
%The
%typical length scale  given by a sparse time sampling  would be different from
%the one given  by a high resolution sampling. 
Increasing the sampling time tends to increase the apparent size of objects as
measured  on the  sphere.  Since this  increase  depend on  the  shape or  the
orientation   of  the   objects,   this  effect   \textit{cannot}  be   simply
time-averaged.  As a consequence, a varying time step would induce a variation
of typical spatial scale.  The  interpolation given by \eq{interp} allows also
for a constant  time step resampling of the source term  $\underline s^e $.
 All reference to the time average shall be dropped from now on.

Given  \eq{source2},  computing  the   expansions  coefficients  of  $s_e$  is
straightforward~:
\begin{equation}
c_{\alpha,{\M{m}'}}^{{\M{m}}}=    \left({\bf   G}^{-1}\sum_i^N   w_i\frac{{\bf
g}(v_i)Y_{\M{m}}^*({\bf  \Omega_i})   Y_{\M{m}'}^*({\bf  \Gamma_i})}{\Delta  T
\cdot R_{200}^2}\right)_\alpha.  \EQN{defc}
\end{equation}
It is expected that the above procedure is more accurate than the strategy presented in
\citet{aubert}, where the flux densities  were smoothed over a shell of finite
thickness ($R_{200}/10$).  
%\Xtophe{d'accord?}  \Xtophe{note que  l'on pourrait
%pour alleger  les notation definir  $g' = G^-1  g$ ie orthogonaliser  la base}

The   harmonic   expansion    $b'_{{\M{m}}}$   of   the   external   potential
$\psi^e(R_{200},\bO)$  is computed  direclty  from the  positions of  external
particles (e.g. \citet{Murali2})~:
\begin{equation}
b'_{\ell,m}(t)=-\frac{4\pi            G}{2\ell+1}\sum_j^N            Y^*_{\ell
m}(\bO_j(t))\frac{R_{200}^\ell}{r_j^{\ell+1(t)}},
\label{e:blm}
\end{equation}
where $r_j$  and $\bO_j$ are  the distance and  the two angles  defining the
position  of the  $j$-th  external particles.  The  quantities $r_j(t)$  and
$\bO_j(t)$  at time  $t$ are  obtained by  linear interpolation  between two
snapshots.   Using  \eq{defexp},  $\psi^e(r<R_{200},\bO)$  can  be
reconstructed from the coefficients $b'_{{\M{m}}}$.

\Sec{stat} makes extensive use of  \Eqs{defc}{blm} for {\sl each} halo in our
simulations  to characterise  statistically these two fields.
\subsection{From flux densities to the 5D source}
\label{s:5Ds}
The description of the source term $s^e$ involves time dependent coefficients $c_{\alpha \ell' m'}^{\ell m}(t)$. Their computation from the particles coordinates is quite straightforward and as shown in the previous sections, the different margins can be recovered through
   the manipulation of theses coefficients. 
   Yet a projection of the source on an \textit{a priori} basis is a
complex operation. 
    Here this projection aims at describing 
   a 5D space  for which little is known. 
  As shown in the following sections, the distribution of incidence angles is quite smooth, 
  while  the distribution of velocities appear to be easily parametrized by gaussians.
     The 5D basis presented  in the current paper induces little bias,
    but it is very likely that a more compact basis exists and that the size of the expansions  chosen can be reduced in the future. 

Because of this large amount of information contained in the source, it is not always convenient to relate coefficients or their correlations to physical
quantities, like the mass flux or the flux density of energy. An alternate
description of the source term  was presented  in \citet{aubert}  with the
following ansatz:
\begin{eqnarray}
  && s^e( \bfr,  {\bf  v}  ,t)  =
 \sum_{\bf  m} Y_{\M{m}}(\bO)  \frac{{\hat
  \varpi}_{\rho, \M{ m}} (2 \pi)^{-3/2}} {{\rm det}| {{\hat \varpi}_{\rho
  \M{\sigma} \M{\sigma},\M{m}}}/{{\hat \varpi}_{\rho, \M{m}}}  | \,\,}
\times \\
%\end{equation}
%\begin{equation}
 && \!\!\!\!\! \exp \left[ - \frac{1}{2} \left(  {\bf v} - \frac{ {\hat \varpi}_{\rho {\bf
  v},      \M{m}}}{{\hat      \varpi}_{\rho,      \M{m}}}
  \right)^{\T{}}\left(\frac{  {\hat \MG{\varpi}}_{\rho  \sigma \sigma,\M{m}}
  }{{\hat \varpi}_{\rho, \M{m}} }\right)^{-1}  \left( {\bf v} - \frac{
  {\hat \varpi}_{\rho {\bf v},  \M{m}}}{{\hat \varpi}_{\rho, \M{m}}}
  \right) \right].\nonumber
  \label{e:sourceexpr}
\end{eqnarray}
This representation of the source is by construction consistent with the first two velocity moments:
\begin{equation}
  \int   \d{}^3    \M{   v}s^e(   {\bfr},    {\bf   v}   )   =      {
    \varpi}_{\rho} (\bfr) \,, \quad \int \d{}^3\M{ v} {\bf v} s^e ( {\bfr}, {\bf
    v} )  = { \varpi}_{\rho \M{v}} (\bfr) \,, \label{e:champs}
\end{equation}
while
\begin{eqnarray}
  &&  \int  \d{}^3  \M{v}{( v_{i}  -\frac{  {  \varpi}_{\rho  {\bf v},i}  }{  {
  \varpi}_{\rho}  })}\,{(  v_{j}  -  \frac{{  \varpi}_{\rho {\bf  v},j}  }{  {
  \varpi}_{\rho}}    )}   s^e(    \bfr,    {\bv}   )    = \nonumber  \\&&\!\!\!\!\!\!\!\!\!\!{    \varpi}_{\rho
  \sigma_{i}\sigma_{j}}      (\bfr)     -\frac{{      \varpi}_{\rho     \M{v}}
  (\bfr)^2}{{\varpi}_{\rho}            (\bfr)}+           \sum_{\bf           m}
  Y_{\M{m}}(\bO)\delta(r-R_{200})\frac{ {\hat  \varpi}_{\rho {\bf v}, \M{m}}(t
  )^2}{{\hat \varpi}_{\rho, \M{m}}(t )}\,,\nonumber\\&&\approx{    \varpi}_{\rho
  \sigma_{i}\sigma_{j}}      (\bfr).  \label{e:}
\end{eqnarray}
Obviously, the third moment is not fully recovered from the Ansatz given by  \eq{sourceexpr}. This example should be taken as an illustration and highlights the possibility of building a source term from its moments. It is not unique and more realistic expressions could be found, which satisfies higher moments of the source.
Still, the successive measurements on the sphere of the flux density of mass $\varpi_\rho$, momentum $\varpi_{\rho\bv}$ and velocity dispersion $\varpi_{\rho\sigma\sigma}$ allow a coherent description of the infall of matter. Unlike the coefficients, these flux densities are easier to interpret since they describe physical quantities and are directly involved in specific dynamical processes (see table \ref{t:varpi}). Furthermore, these three flux densities are easily 
expressed in terms of coefficients $c_{\alpha \ell' m'}^{\ell m}(t)$, or more precisely in terms of \textit{a subset} of the source's coefficients, implying a smaller number of computations relative to a complete calculation of  $c_{\alpha \ell' m'}^{\ell m}(t)$. Finally these flux densities are particularly suited to the regeneration of synthetic environments. As shown in appendix~\ref{s:gen}, synthetic spherical maps can be generated from the two-point correlations and cross-correlations of these fields. Such environments would be consistent with the measurements in simulations and will allow us to easily embed simulated galaxies or halos in realistic environments as a function of time.

The expression given in \eq{sourceexpr} has  one important drawback~:  it is
not  of the form of \eq{defss},  i.e. it
would  require a reprojection over  a linear  expansion for a dynamical propagation. Nevertheless, its compacity makes it easier to compute than the full set of coefficients and the associated strategy would be 1) to measure the flux densities from the simulation, 2) build a source term from e.g. the \Eq{sourceexpr} and 3) project over an appropriate 5D basis when needed, i.e. when the source is used as an input to the analytic description of the haloes dynamics.  \\

The following sections will make intensive use of the coefficients described by the \eq{defc} and \eq{blm}. In particular, it will shown how the manipulation of these coefficients allow to recover relevant physical quantities. In the current paper, only the first moment of the source, the flux density of mass $\varpi_\rho$ together with the external potential, will be fully assessed. The kinematical properties of the accreted material will in particular be investigated. 
 The complete characterization of the $\bf c(t)$ coefficients is beyond the scope of the current paper and will be completed in paper III. The full measurements of these 11 fields required by Eq.~\ref{e:sourceexpr} and the comparisons between the two expressions of the source will also be assessed in this future paper as well. Appendix~\ref{s:appen_v} and~\ref{s:appen_disp} 
 describe how the other moments, the flux density of momentum $\varpi_{\rho\bv}$ or the flux density of energy   $\varpi_{\rho\sigma^2}$ may be recovered from the source expansion. 

\begin{table*}
\centering
\begin{minipage}{140mm}
\begin{tabular}{l|lll}
\ & flux density , $\varpi$ & Flux, $\Phi$ & Motivation \\ \hline Mass & $\rho
v_{r}$ & $\d m/\d t$ & heating  \& cooling, \\ Angular momentum & $\rho v_r \,
{\bf r  } \times {\bf v}$ &  $\d L/\d t $  & warp, shape of  halos, \\ Kinetic
energy & $\rho v_r \, \sigma_i \sigma_j  $ & $\d E/\d t$ & virialized objects,
\\ Shear  & $\rho v_{r}(\partial v_j/\partial  x_i 
  + \partial v_i/\partial x_j)$ & 
$\d  c/\d t$ & tidal  field, \\ Vorticity 
& $\rho  v_{r} \nabla \times
{\bf v}$ & $\d \omega/\d t$ & anisotropic accretion.  \\ \hline
\end{tabular}
\caption{Description of the various  flux densities.  The first 10, together
with  the  external  potential  are  sufficient to  characterize  fully  the
environment as shown in Section \ref{s:5Ds}.}
\end{minipage}
 \label{t:varpi}
\end{table*}

%%%%%%%%%%%%%%%%%%%%%%%%%%%%%%%%
\subsection{A template halo}
\label{s:template}
% pour info: halo numero 17 / simu 1946

\begin{figure} 
\centering
\resizebox{0.8\columnwidth}{0.8\columnwidth}{\includegraphics{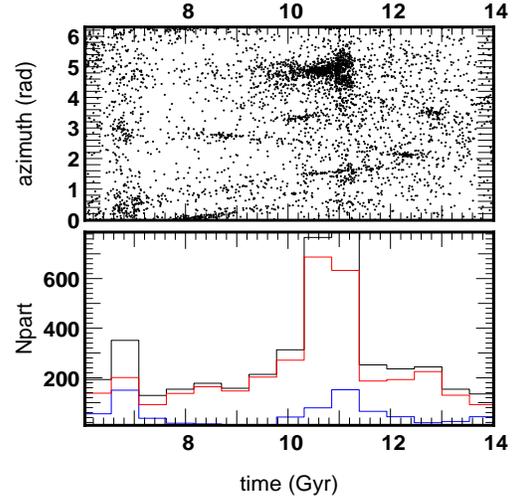}}
\caption{An  example  of   accretion  history.   \textit{Top :}  Azimuth-time
diagram.  Each  point in the  diagram represents one particle  passing through
the virial  sphere at a given azimuth  (y-axis) at a given  time measured from
the  Big-Bang (x-axis).  \textit{Bottom: }  the  distribution of  particles'
crossing  instants  (in black).   Time  increases  from  left to  right.   The
infalling  (resp.    outflowing)  particles  distribution  is   shown  in  red
(resp. blue).}
\label{f:extvsphy} 
\end{figure}

As an illustration, let us first apply the whole machinery to one typical halo.
 At  z=0,  this   `template'  halo  has  a  mass   $M$  of  $3.4\cdot  10^{13}
\mathrm{M}_\odot$, with  a virial radius  $R_{200}$ of 800 $h^{-1}$  kpc.  The
corresponding circular velocity is $V_{c0}=600$ km/s.
Its  accretion history  is shown  in figure  \ref{f:extvsphy} for  $z<1$. Each
point on  the azimuth-time diagram  represents one particle of  the simulation
passing through  the virial  sphere at a  given azimuth  and at a  given time.
Temporal space has  been sampled using 15 equally spaced  bins between z=1 and
z=0 (see  bottom panel in figure  \ref{f:extvsphy}).  For each  time step, the
expansion  coefficients $c_{\alpha,{\M{m}}}^{{\M{m}'}}(t)$  are  computed from
\eq{defc}.  The Gaussian basis  $g_\alpha(v)$ involved 25 functions with means
$\mu_\alpha$ equally distributed from v=0  to v=1.5 in $V_{c0}$ units and with
an r.m.s $\sigma=0.03$.  The harmonic  expansions were carried up to $\ell=50$
in position space and up to $\ell'=25$ in velocity space.
\subsubsection{Advected mass: angular space ($\varpi_\rho$)}
\begin{figure} 
\centering                 \resizebox{7cm}{5cm}{\includegraphics{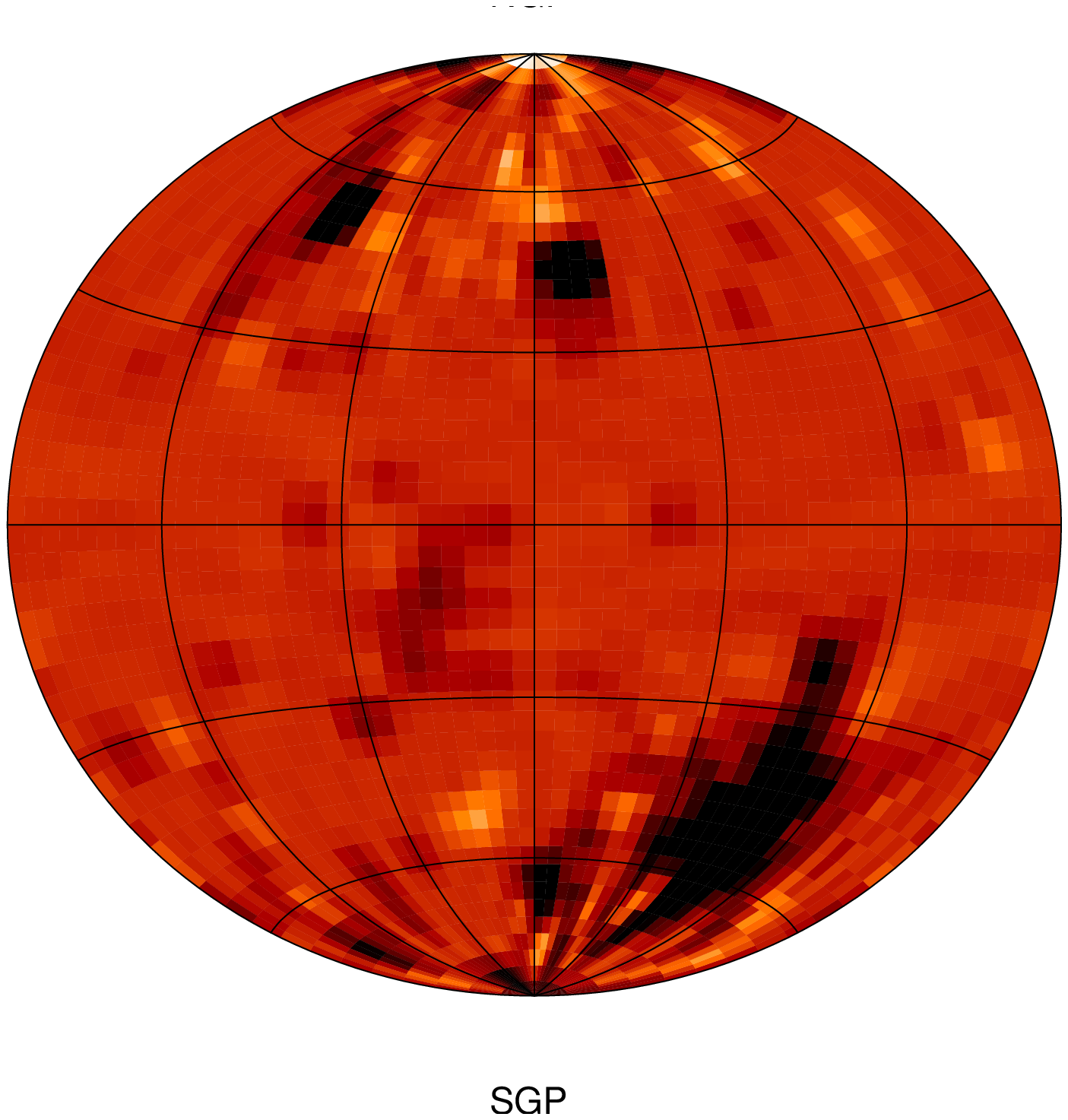}}
\resizebox{7cm}{5cm}{\includegraphics{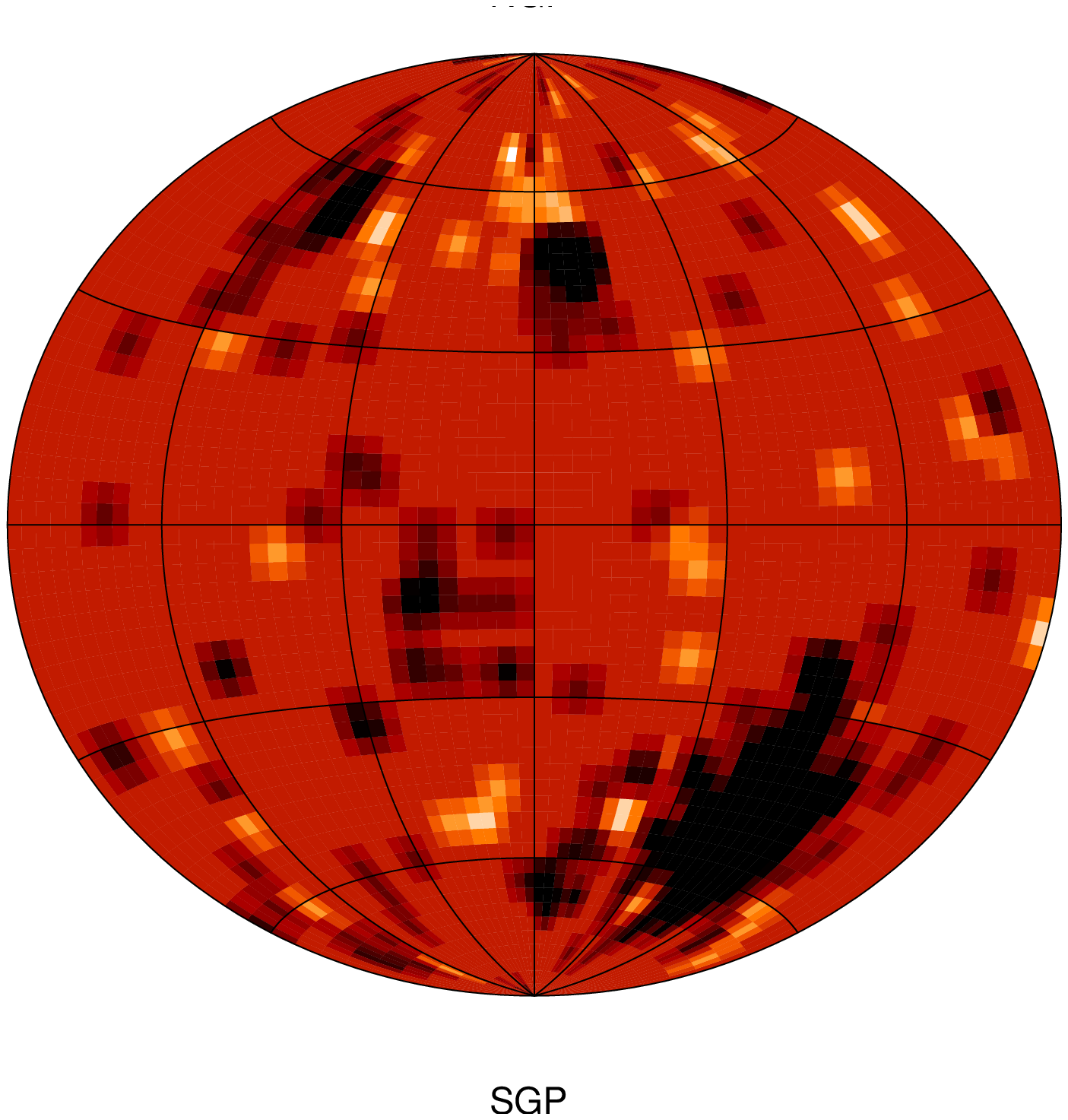}}
\caption{An  example of  a flux  density reconstructed  from  the coefficients
$c_{\alpha,\M{m}}^{\M{m}'}$: the  mass  flux density,  $\rho v_r(\bO)$.   It
represents  the  angular  distribution  of   incoming  mass  as  seen  from  a
halo-centric point of  view. Here $t_s\sim$ 11 Gyr.   Light regions correspond
to strong  infall while darker regions  stand for low  accretion and outflows.
\textit{Top}~:  the   spherical  field  obtained  directly   from  the  spatial
distribution of particles.  \textit{Bottom}: the reconstructed spherical field
from the coefficients, \eq{defc}.}
\label{f:exdens} 
\end{figure}

\begin{figure} 
\centering
\resizebox{0.8\columnwidth}{0.8\columnwidth}{\includegraphics{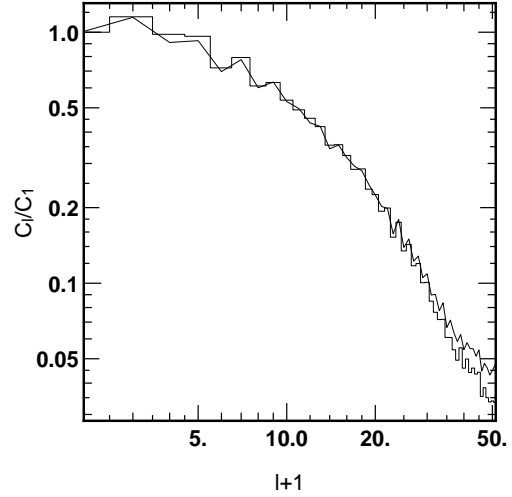}}
\caption{The   angular   power  spectrum,   $C_\ell/C_1$,   (\eq{defcl})  of   the
distribution   of  incoming  matter,   $\varpi_\rho(\bO)$  (shown   in  figure
\ref{f:exdens}) at $t\sim$ 11 Gyr, for our template halo.  For a given $\ell$,
the corresponding  angular scale is $\pi/\ell$.  The  histogram corresponds to
the power spectrum derived  directly from the particles' angular distribution.
The solid line is the power spectrum reconstructed from the coefficients.}
\label{f:exCL} 
\end{figure}

The template  halo accretes an object  at $t_s$=11 Gyr (where  t=14 Gyr stands
for  z=0)  adding  $7.5\cdot10^{12}  \mathrm{M}_\odot$  to  the  system during a $\sim 1$ Gyr interval.   The
corresponding spherical flux density field, $\varpi_\rho(\bO,t_s)$ is shown in
\Fig{exdens}.   It represents the distribution  of accreted particles
as seen from a halo-centric  point of view.  The field $\varpi_\rho (\bO,t_s)$
has been reconstructed from the coefficients (see \eq{defc}). It reads:
\begin{equation}
\varpi_\rho   (\bO,t)=\int  \md   \bG  \md   v  v^2   s^e(v,{\bf  \Omega},{\bf
\Gamma},t)=\sum_\M{m} a_\M{m}(t) Y_\M{m}(\bO).
\label{e:exprhovr}
\end{equation}
Since
\begin{equation}
\int \!\! \md  \bG \, Y_{\ell',m'}=\sqrt{4\pi} \delta_{\ell'0}\delta_{m'0}\,,  \,\,\,
{\rm and} \,\, \int\! \md v v^2 g_\alpha(v)=\mu_\alpha^2+\sigma^2.
\end{equation}
It follows that
\begin{equation}
a_\M{m}(t)=\sqrt{4\pi}\sum_\alpha
(\mu_\alpha^2+\sigma^2)c^{\M{m}}_{\alpha,{\bf 0}}(t),
\label{e:c2a}
\end{equation}
allowing us  to recover $\varpi_\rho(\bO,t_s)$.  Also shown,  the same field
but computed this time using  directly the angular distribution of particles
as described in \citet{aubert} (see below).  All the major features are well
reproduced  by  the  expansion  coefficients, \Eqs{defc}{c2a}.  Clearly,  an
object is  `falling' through  the virial sphere.   It is  straightforward to
obtain the angular power spectrum $C_\ell^{\varpi_\rho}$ from the $c^{\M{m}}_{\alpha,\M{m}
'}$ coefficients via the definition of $a_{\ell m}$ in \eq{c2a}~:
\begin{equation}
C_\ell^{\varpi_\rho}=\frac{1}{4\pi}\frac{1}{2\ell+1}\sum_m |a_{\ell m}|^2.
\label{e:defcl}
\end{equation}
The  angular  power spectrum  of  $\varpi_\rho  (\bO,t_s)$,  derived from  the
expansion, \eq{defc}, is shown in  \Fig{exCL}. From the positions and
velocities  of  particles  it   is  also  possible  to  evaluate  $\varpi_\rho
(\bO,t_s)$  on  an  angular  grid  and  recover  the  angular  power  spectrum
`directly' .  The agreement  between the two  $C_\ell^{\varpi_\rho}$ is
good,  though for  the smallest  scales ($\ell  \ge 30$),  the  power spectrum
computed  from  the coefficients  is  slightly  larger  than the  one  derived
directly from  the particles.  This may be  explained by the fact  that a grid
sampling tend to smooth the  actual $\varpi_\rho$ field.  As a consequence the
amplitude  of small  scales fluctuations  is decreased,  leading to  a smaller
$C_\ell^{\varpi_\rho}$.  A more complete discussion  on harmonic convergence can be found in
appendix  \ref{s:harmconv}.  For  a  given $\ell$,  the corresponding  angular
scale is $\pi/\ell$  in radians.  

Note that the coefficients $a_{00}$ are closely related to the accretion field
averaged over all directions, $\Phi^{M}(t)$, defined by~:
\begin{equation}
\Phi^{M}(t)\equiv{\overline{\varpi_\rho      }}=\frac{1}{4\pi}\int\md\bO      \rho
v_r(\bO,t)=\frac{a_{00}}{\sqrt{4\pi}}.
\label{e:a00phi}
\end{equation}
%Recalling that 
%\begin{eqnarray}
%a_{00}&=&\int \md \bO \rho v_r(\bO,t) Y^*_{00}(\bO)\\
%&=&1/\sqrt{4\pi}\int \md \bO \rho v_r(\bO,t),
%\end{eqnarray} 
%one easily gets:
%\begin{equation}
%\Phi(t)={\overline{\rho v_r}}=\frac{a_{00}}{\sqrt{4\pi}}.
%\end{equation}
Measuring $a_{00}(t)$ amounts to measuring the accretion flux density, \ie the
quantity of dark matter accreted per unit surface and per unit time.

\subsubsection{Advected mass: velocity space}
\begin{figure} 
\centering
%\resizebox{0.8\columnwidth}{0.8\columnwidth}{\includegraphics{exemple_fv2}}
\resizebox{0.75\columnwidth}{0.75\columnwidth}{\includegraphics{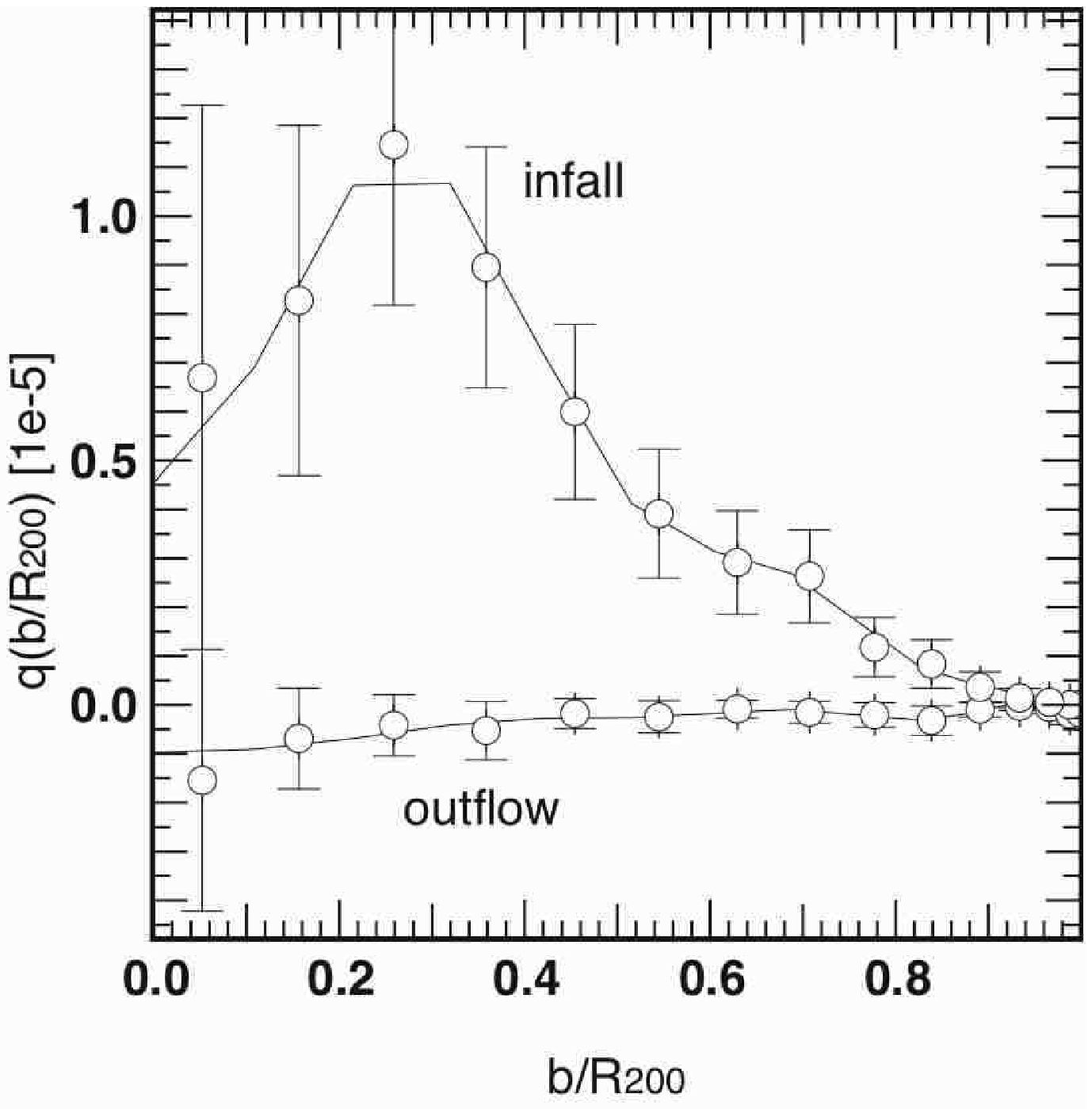}}
%\resizebox{0.8\columnwidth}{0.8\columnwidth}{\includegraphics{exemple_fv}}
\resizebox{0.75\columnwidth}{0.75\columnwidth}{\includegraphics{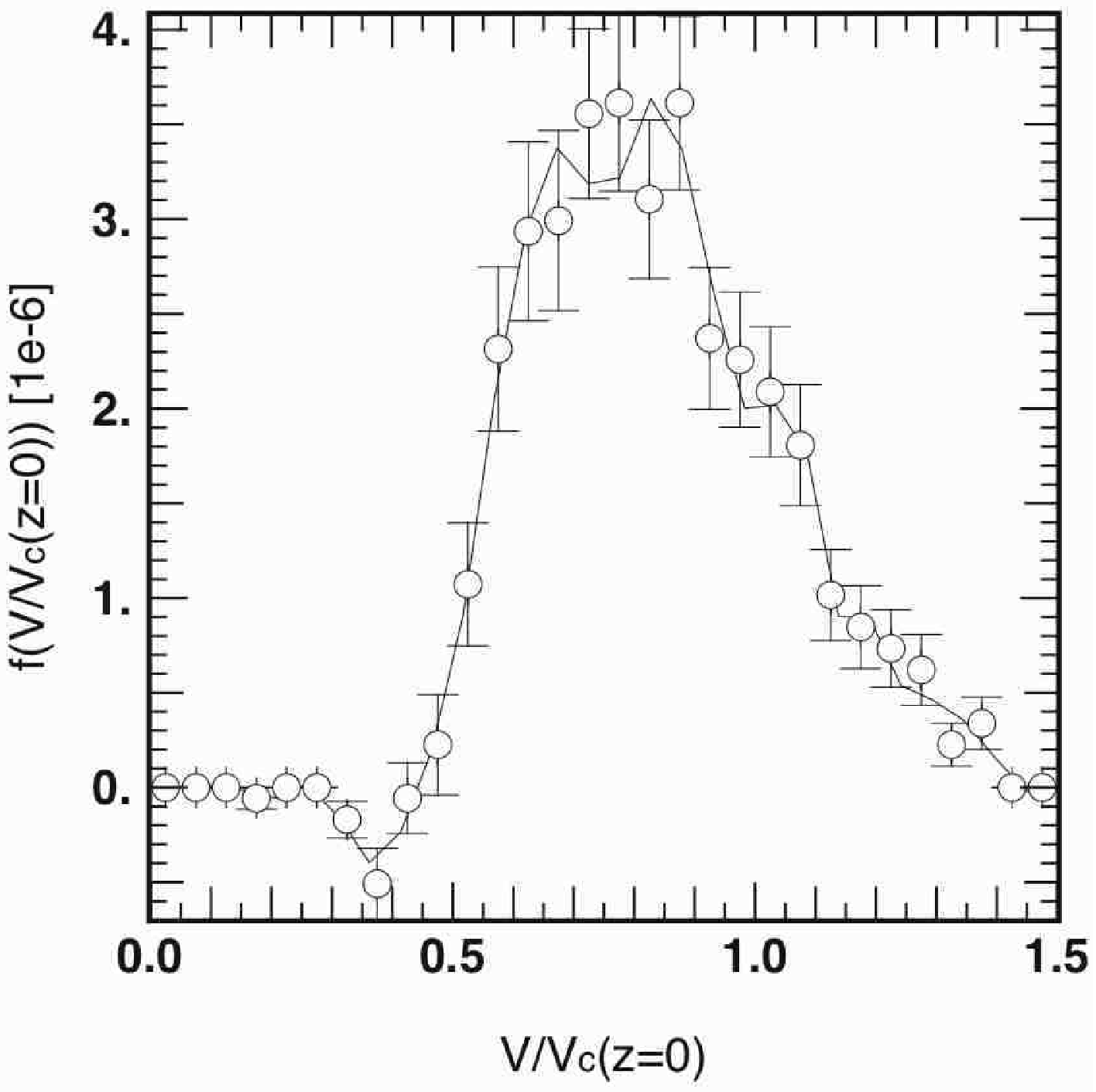}}
\resizebox{0.75\columnwidth}{0.75\columnwidth}{\includegraphics{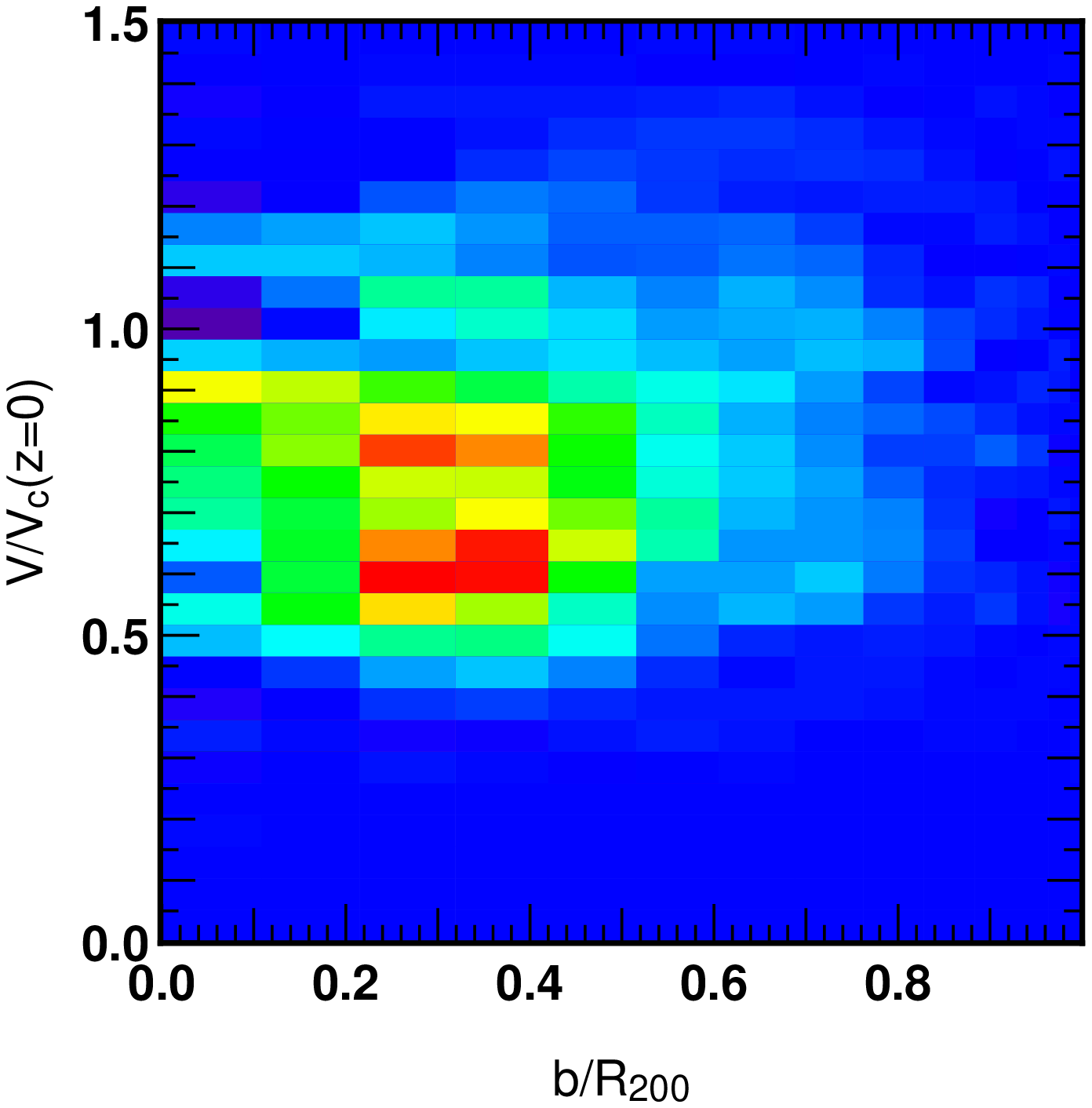}}
%\resizebox{0.8\columnwidth}{0.8\columnwidth}{\includegraphics{templateVimpact}}
\caption{\textit  {Top:} Excess probability  distribution of  impact parameter
$b$, $\vartheta(b)$, derived from the $c_{\alpha,{\M{m}'}}^{\M{m} }(t)$ source
coefficients (\eq{defc}) of our template halo at $t_s\sim$ 11 Gyr (line).  The
histogram  corresponds to  the  same distribution  derived  directly from  the
particles' positions  and velocities.  Error bars stand  for $3\sigma$ errors.
The impact  parameters are given  in units of  $R_{200}$.  The Y-axis  unit is
$5\cdot10^9   \mathrm{M}_\odot$/kpc$^2$/Myr.    Infall   is   mainly   radial.
\textit{Middle:}  The  velocity  distribution  of  particles,  $\varphi(v)$,
accreted at $t_s\sim$ 11 Gyr, for our template halo.  Velocities are expressed
in terms  of the  circular velocity  at z=0.  The  Y-axis unit  is $5\cdot10^9
\mathrm{M}_\odot$/kpc$^2$/Myr.   The  histogram  corresponds to  the  velocity
distribution obtained directly from  the particles velocities.  The solid line
is   the   reconstructed    distribution   from   the   source   coefficients.
\textit{Bottom}:  the probability  distribution, $\wp(b,v)$, of  particles in
the $b-v$ subspace.  Units are the same as above.  Red/blue stand for high/low
densities.   No   correlation  is found  between  $b$ and  $v$ for  this
specific example.}
\label{f:exfv} 
\end{figure}

Integration over  the sphere leads to  the distribution of  accreted matter in
 velocity space~:
\begin{eqnarray}
\rho          v_r(\bG,v,t)&=&\int          \md         \bO          s^e(v,{\bf
\Omega},{\bf\Gamma},t)\,,\\&=&\sqrt{4\pi}\sum_{\alpha,\M{m}'}
c^{0}_{\alpha,\M{m}'} g_\alpha(v)Y_{\M{m}'}(\bG).
\end{eqnarray}
Projections over $\Gamma_2$ and $v$  gives the probability distribution of the
incidence angle $\Gamma_1$, $\vartheta(\Gamma_1,t)$, defined as:
\begin{eqnarray}
\vartheta(\Gamma_1,t)&=&\int     \md    \Gamma_2     \md     v    v^2     \rho
v_r(\bG,v,t)\,,\nonumber\\                                                  &=&
2\pi\sqrt{4\pi}\sum_{\alpha,\ell'}c^0_{\alpha,\{\ell',0\}}(\mu_\alpha^2+\sigma^2)Y_{\ell',0}(\bG).
\label{e:defq}
\end{eqnarray}
The impact  parameter $b$ of  an incoming particle  (measured in units  of the
virial radius) is related to $\Gamma_1$ by
\begin{equation}
\frac{b}{R_{200}}=\sin(\Gamma_1)\,,
\label{e:agaga}
\end{equation}  
therefore the  probability distribution of  impact parameters, $\vartheta(b)$,
is  easily deduced  from  \eq{defq}.  At  $t\sim11$  Gyrs, the  $\vartheta(b)$
computed from  the source  coefficients is compared  to that  derived directly
from the velocities of particles in \Fig{exfv}.
%Error  bars are  computed assuming a  poissonian noise  in each
%impact parameter bin.  
Note that  for pure  geometrical reasons small  impact parameter $b$  are less
likely since  there is  only one trajectory  passing through the  center while
there is a whole cone of  trajectories with $b\ne 0$.  As a consequence errors
are intrinsically larger  for small values of b.   The reconstruction from the
source  coefficients   is  clearly  adequate.   In  this   example,  the  high
probability for infalling particles to have a small impact parameter ($b<0.5$)
imply that velocities are strongly  radial. The object `dives' into the halo's
potential well.

Projection over $\Gamma_1$ leads  to the probability distribution of particles
velocities, $\varphi(v,t_s)$, as they pass  through the virial sphere. The PDF
$\varphi(v,t)$ is defined as:
\begin{equation}
\varphi(v,t)\equiv v^2 \int \md \bO \md \bG s^e(v,{\bf \Omega},{\bf\Gamma},t).
\label{e:dfv}
\end{equation}
Here the $v^2$  weighting accounts for the fact that the
probability distribution  of measuring a  velocity, $v$ within $\d  v$
is of interest here.  Using coefficients,  it follows that:
\begin{equation}
\varphi(v,t)=4\pi\sum_\alpha v^2 g_\alpha(v)c^{\bf 0}_{\alpha,{\bf 0}}.
\end{equation}
The   reconstructed   velocity   distribution   is  also   shown   in   
\Fig{exfv}. It  reproduces well the  actual velocity distribution.  For this
specific  halo,  the satellite  is  being accreted  with  a  velocity of  0.75
$V_{c0}$.

The correlation between the incidence angle $\Gamma_1$ and the velocity's amplitude
$v$ may  be studied by  integrating $\rho v_r(\bG,v,t)$ over  $\Gamma_2$ only.
The distribution function, $\wp(\Gamma_1,v)$, of particles in the $(\Gamma_1,v)$ subspace is
defined by:
\begin{eqnarray}
\wp(\Gamma_1,v)&\equiv  & \int  \md  \Gamma_2  \md  v v^2  \rho
v_r(\bG,v,t)\,,\nonumber\\
&=&2\pi\sqrt{4\pi}\sum_{\alpha,\ell'}c^\M{0}_{\alpha,\{\ell',0\}}
g_\alpha(v)Y_{\ell',0}(\Gamma_1,0).
\end{eqnarray}
Given the relation \eq{agaga}, the correlation $\wp(b,v)$ between the impact parameter and the velocity's amplitude is easily obtained.
 The  $\wp(b,v)$
distribution  is shown  in   \Fig{exfv}.  Again, note that $\wp(b,v)$ represents \textit{an excess} probability of finding an impact parameter $b$ (with a velocity $v$) compared to isotropy. In  this  specific example, no  real correlation may be found between the two quantities.  Finally, the integration
of $\wp(b,v)$, $\varphi(v,t)$  and $\vartheta(\Gamma_1)$ over their respective
space leads to the same quantity, namely the integrated flux $\Phi^{M}(t)$.
%% \begin{eqnarray}
%% \Phi(t)&=&\int\md v \md \Gamma_1 \sin(\Gamma_1) \wp(b,v,t), \\
%% &=&\int \md v \varphi(v,t),\\
%% &=&\int \md\Gamma_1 \sin(\Gamma_1) \vartheta(b,t).
%% \end{eqnarray}

%%%%%%%%%%%%%%%%%%%%%%%%%%%%%%%%%%%1512%%%%%%%25%%%%%%%

%%%%%%%%%%%%%%%%%%%%%%%%%%%%%%%%%%%1512%%%%%%%25%%%%%%%
\subsubsection{External potential}

The final  field needed  on  the virial sphere is  the external
tidal field created by the dark matter distribution around the halo.

\begin{figure}
\centering              
\resizebox{7cm}{5cm}{\includegraphics{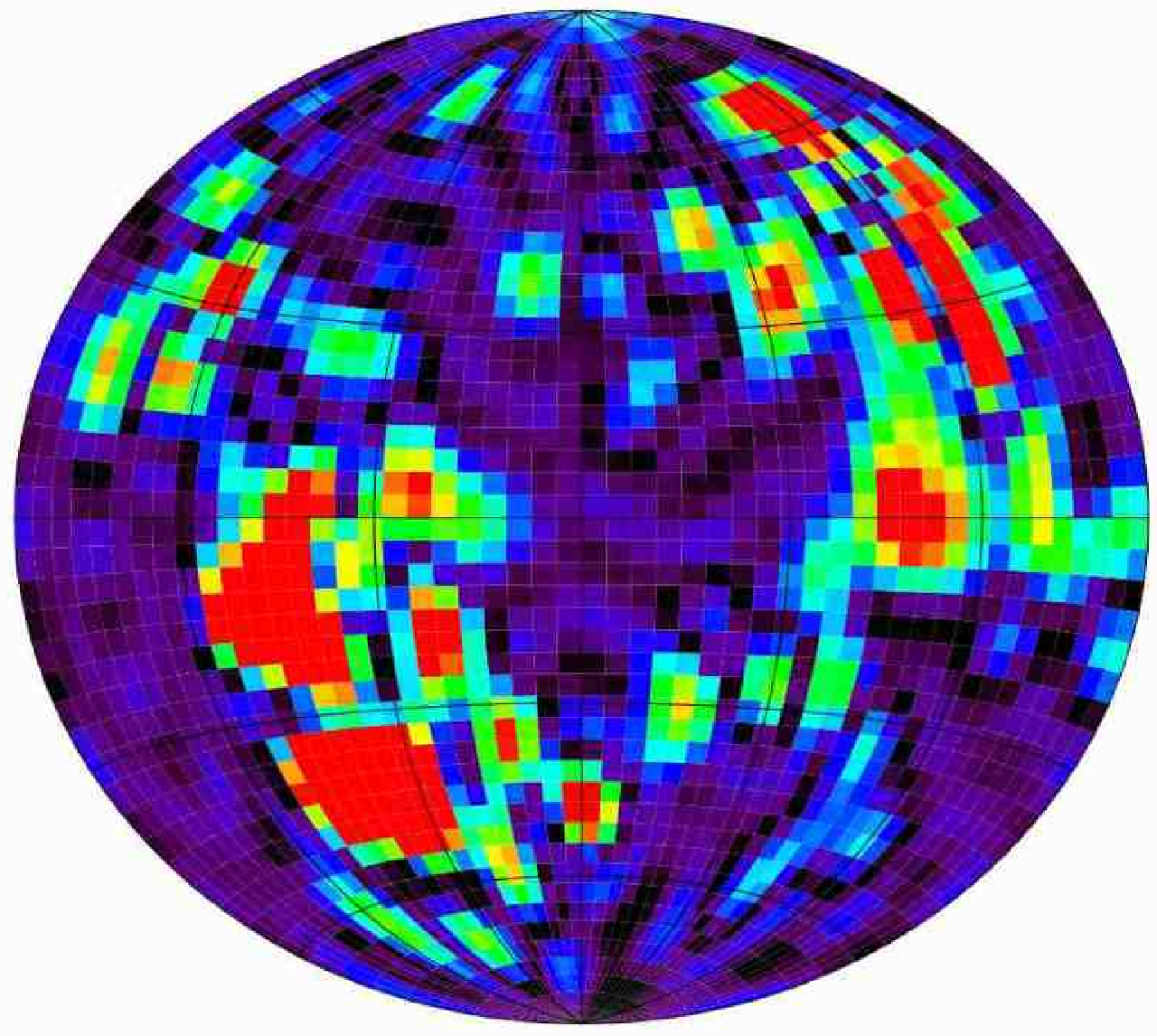}}
\resizebox{7cm}{5cm}{\includegraphics{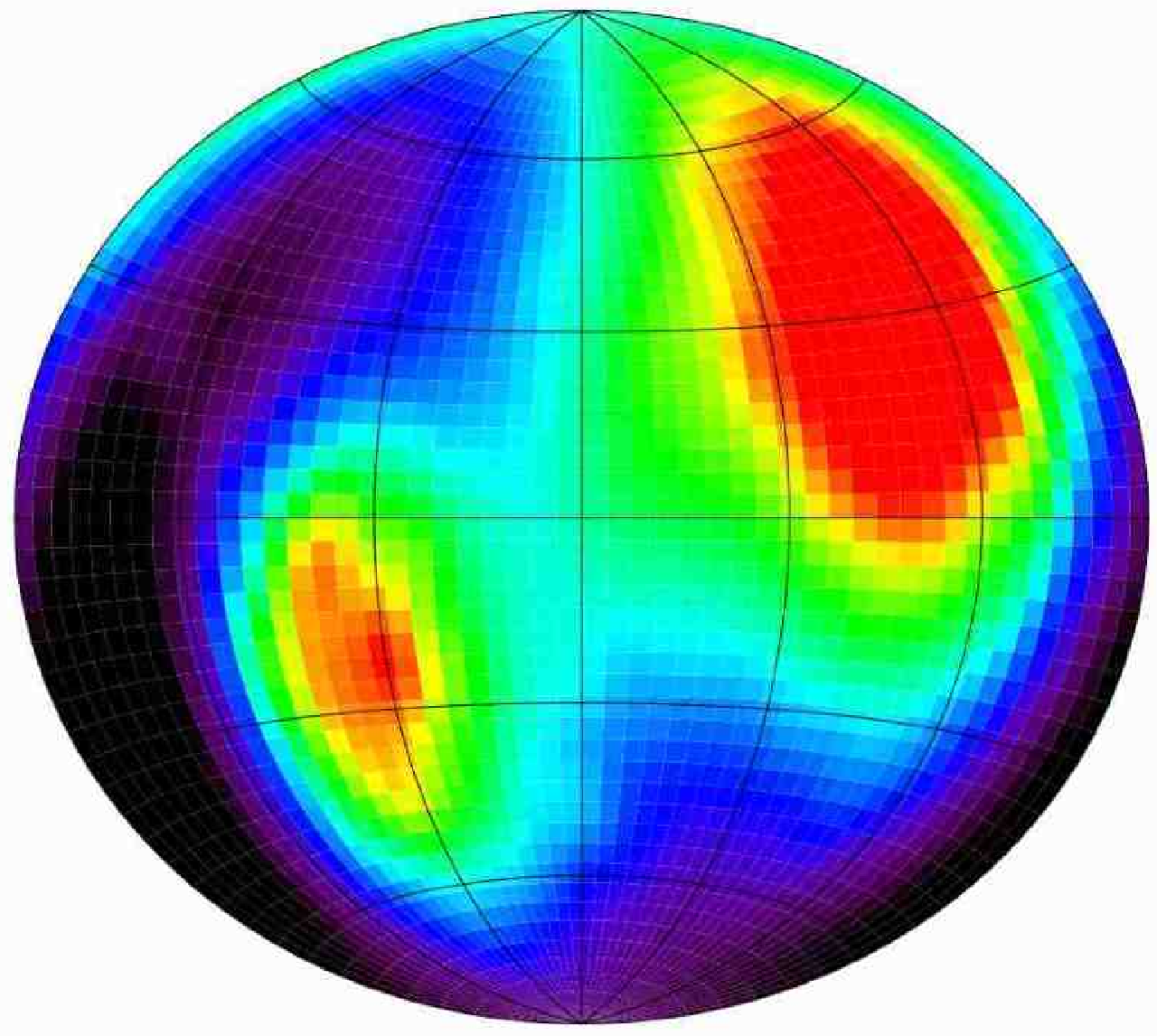}}
\caption{A comparison between the external potential, $\psi^e(\bO)$, and the
modulus of the   flux density of matter, $|\rho  v_r(\bO)|$.  The measurement
is made at $t\sim 7$ Gyrs (measured from the Big Bang) on our template halo.
The  two fields were  respectively reconstructed  from $c_{\alpha,\bm'}^\bm$
and $b_{\bm}$  coefficients with $\ell_{\mathrm{max}} \le  20$.  Even though
the  two fields  are similar  and exhibits  a strong  quadrupolar component,
$\psi^e(\bO)$ is smoother than $\rho v_r(\bO)$.  It is  expected that the corresponding
expansion coefficients be statistically correlated.  }
\label{f:compotdens} 
\end{figure}

\begin{figure} 
\centering
\resizebox{0.8\columnwidth}{0.8\columnwidth}{\includegraphics{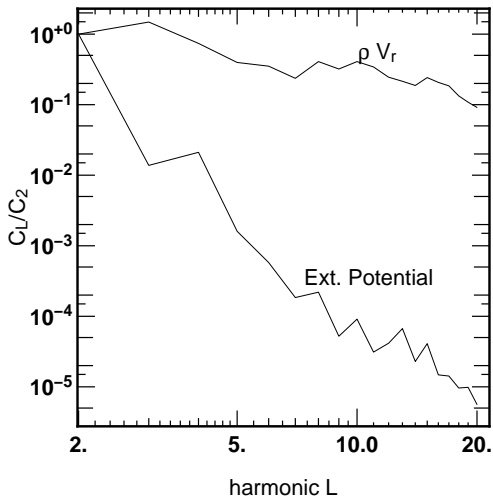}}
\caption{A comparison  between the angular  power spectrum of  $\rho v_r(\bO)$
and $\psi^e(\bO)$  for our template halo  (the two fields are  shown in 
\Fig{compotdens}). The  two power spectra $C_\ell$ are  normalized by $C_2$,
i.e the quadrupole  contribution. The slope of the  potential's power spectrum
is clearly  stronger.  Large scales  (i.e.  small $\ell$ values)  dominate the
angular distribution of $\psi^e(\bO)$, as expected.}
\label{f:CLcompotdens} 
\end{figure}

Using \eq{blm},  the external  potential $\psi^e(\bO,t)$ is  easily computed
from  the  positions  of  external  particles, having  restricted the sampling  to
particles  within a  4 Mpc (physical)  sphere centered  on the  halo.  The  position of
external particles  are linearly interpolated  at a given  measurement time.
  The $b_\bm$ coefficients for  the
 template halo are computed  at $t_s\sim 7$
Gyrs (measured  from the Big Bang). The  reconstructed $\psi^e(\bO,t)$ field
is shown  in \Fig{compotdens}  along with the  modulus of the  advected mass
$|\rho  v_r(\bO)|$. The  two  reconstructions were  restricted to  harmonics
$\ell\le20$.

The two  spherical fields show the  same main features.   However, almost no
small scales  features is seen  in the map  of the external  potential even
though they have the same  resolution.  Since the gravitational potential is
known to  be smoother than  the associated density  and is dominated  by the
global  tidal  field, it  is  not  surprising  that $\psi^e(\bO,t)$  appears
smoother than the advected mass field $|\varpi_\rho(\bO)|$.

The potential's  angular power  spectrum may also  be computed  by replacing
$a_{\ell m}$  by $b_{\ell m}$  in \eq{defcl} (see  \Fig{CLcompotdens}).  The
power spectrum  of the  potential, $C_\ell^{\varpi_\rho}$, sharply  decreases with
$\ell$,  while $C_\ell^{\varpi_\rho}$ has  a gentler  slope. Large  scales are
clearly  more  important for  the  potential  than  for the  advected  mass.
Furthermore,  $C_\ell^{\psi}$ systematically peaks  for even  $\ell$ values,
reflecting the `even' symmetry of the potential measured on the sphere.

%%%%%%%%%%%%%%%%%%%%%%%%%%%%%%%%%%%%%%%%%%%%%%%%%%%%%%%%%%%%%%%%%%
%\section{From a large set of simulations to statistics of the source }
\section{Simulation sample \& statistical biases}
%%%%%%%%%%%%%%%%%%%%%%%%%%%%%%%%%%%%%%%%%%%%%%%%%%%%%%%%%%%%%%%%%%
\label{s:stat}
%%%%%%%%%%%%%%%%%%%%%%%%%%%%%%%%%%%%%%%%%%%%%%%%%%%%%%%%%%%%%%%%%%
%%%%%%%%%%%%%%%%%%%%%%%%%%%%%%%%%%%%%%%%%%%%%%%%%%%%%%%%%%%%%%%%%%
 \Sec{template}  details  the  measurement strategy for a  given typical
halo.  It is now possible   to reproduce the above  measurements for all
the  halos of  the  simulation sample.   Let  us first  describe  in turn  the
construction of our sample, and  the corresponding biases, which constrain our
ability  to convert  a large  set of  simulations into  the statistics  of the
source.

%%%%%%%%%%%%%%%%%%%%%%%%%%%%%%%%%%%%%%%%%%%%%%%%%%%%%%%%%%%%%%%%%%
\subsection{Simulations}
%%%%%%%%%%%%%%%%%%%%%%%%%%%%%%%%%%%%%%%%%%%%%%%%%%%%%%%%%%%%%%%%%%
In  order to  achieve a  sufficient  sample and  ensure a  convergence of  the
measurements,    a set  of  $\sim  500  $ simulations was produced
 as 
discussed in \citet{aubert}.   Each of them consists of  a 50 $h^{-1}$ Mpc$^3$
box  containing  $128^3$  particles.   The mass  resolution  is  $5\cdot10^{9}
M_\odot$.   A  $\Lambda$CDM  cosmogony  ($\Omega_m=0.3$,  $\Omega_\Lambda=0.7$,
$h=0.7$   and  $\sigma_8=0.928$)   is  implemented   with   different  initial
conditions.    These   initial    conditions   were   produced   with   GRAFIC
(\citet{Grafic})  where  a BBKS  (\citet{BBKS}) transfer  function was chosen to
compute  the initial  power spectrum.   The  initial conditions  were used  as
inputs to the parallel version of the tree code GADGET (\citet{Gadget}).
The softening length was set to 19 $h^{-1}$ kpc.\footnote{
A second set of simulations with a resolution increase by a factor $2^{3}$ (resp. $2^{6}$) was carried in order to investigate the convergence of some measurements 
(see Sec.~\ref{s:reso}).}
The halo  detection was performed  using the halo finder  HOP (\citet{Hop}).
   The   density    thresholds   suggested   by   the   authors
($\delta_{\mathrm{outer}}=80,
\delta_{\mathrm{saddle}}=2.5\delta_{\mathrm{outer}},
\delta_{\mathrm{peak}}=3.\delta_{\mathrm{outer}}$) were used.

%%%%%%%%%%%%%%%%%%%%%%%%
\subsection{Selection criteria}
\label{s:stat}
\begin{figure} 
\centering
\resizebox{0.8\columnwidth}{0.8\columnwidth}{\includegraphics{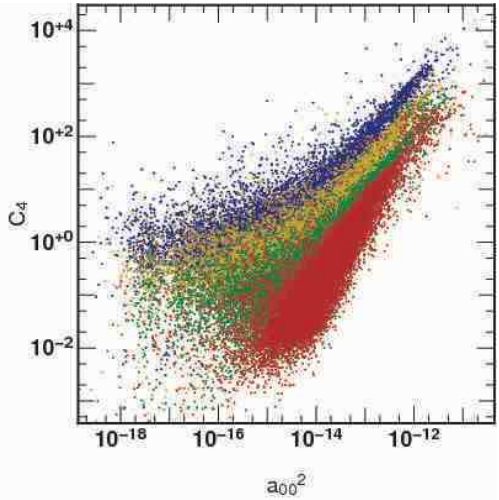}}
\caption{Scatter plot of $C^{\varpi_\rho}_4$ versus $a_{00}^2$ measured at lookback times t= 7.8 (red), 5.6 (green), 4. (yellow) and 2.9 (blue) Gyrs. The quantity $a_{00}$ scales as the average accretion rate of the haloes while  $C^{\varpi_\rho}_4$ scales as the contribution of $\ell=4$ structures in the  flux density of mass measured on the sphere. This plot illustrates how a threshold on the accretion rate affects in a non trivial way the typical clustering measured for $\varpi_\rho$. In particular, one should note how $C^{\varpi_\rho}_4$ remains constant at recent times for low accretion rates.}
\label{f:C4} 
\end{figure}

%%%%%%%%%%%%%%%%%%%%%%%%
\label{s:selection}
%The following  measurements are  concerned with a  galactic context. 
As shown in \citet{aubert} the completion range in mass of 
the simulations spans from $3\cdot 10^{12} M_\odot$ to $3\cdot 10^{14} M_\odot$.  
Since the  emphasis is  on $L_\star$ galaxies, the 
survey is  focused mainly on galactic
 halos and  light clusters,  only  haloes with a  mass smaller
 than $10^{14}  {\rm M}_{{}_\odot}$ at z=0 were considered.  The
  interest is for halos already
 `formed', \ie which  will not experience major fusions  anymore. To satisfy
 these  requirements the  focus is on  the last  8 Gyrs  (redshifts $z<1$  in a
 $\Lambda$CDM cosmogony).   Since  the  history of a  given halo is followed by
 finding its  most massive  progenitor, it is  required not to  accrete more
 than half its  mass in a two-body  fusion. As a final safeguard  a halo is
 rejected  if it  accretes more than  $5\cdot10^{12} \mathrm{M}_\odot$
 between two timesteps (i.e. per 500 Myrs, see next subsection).  This mass  corresponds approximatively  to the smallest haloes considered at $z=0$. The final range of mass of haloes which satisfy these criteria is $\sim 5\cdot10^{12} \mathrm{M}_\odot - 10^{14} \R{M}_\odot$, the fraction of rejected haloes being $\sim 20 \%$. Clearly, such \textit{a priori} selection criteria will modify the distributions of measured values and the related biases may be difficult to predict. For instance, the \Fig{C4} shows the scatter plot of the contribution of $\pi/(\ell=4)=45$ degrees fluctuations to $\varpi_\rho$ field vs $a_{00}$, i.e. the accretion rate. It appears from this plot that modifying the treshold for the accretion will modify the average angular scale of $\varpi_\rho$ in a non trivial way. Since only a small fraction of haloes is rejected, the biases are expected to be moderate, but as for now their impact cannot be estimated accurately  on the average source or its moments.

Let us  emphasize that the above  selection criteria should be  added to those
corresponding to  the simulations themselves. Aside from the fact that a $50 \R{Mpc}^3 h^{-1}$ box size implies a limited range of mass, the universe described in these simulations is more homogeneous than it should be, since each box must satisfy a given mean density. In other words, the probability of rare events is reduced. This effect should not influence the number of haloes with high accretion rate, since strong accretions are rejected a priori. On the other hand, it should influence the number of objects which experience low accretion history, which are probably less numerous in our simulations than in larger simulated volumes since voids are less likely. Furthermore, the intrinsic mass resolution sets a minimum accretion rate equals to one particle mass ($5\cdot10^9 M_\odot$) per time interval. One could imagine an object with a mass smaller than the particle's mass which would not be included in the simulation at the current resolution. Furthermore, an object with a mass equals to a few times the minimum mass is be considered as diffuse accretion. Finally this mass resolution is related to the spatial resolution, which limits intrinsically the angular description of fluxes on the virial sphere. For a given type of simulation, all these effects cannot be avoided and reduce the representability of the following measurements. 

In short, this strategy  involves a bias in mass, in redshift, in resolution and in
strength of merging event. However these biases should only influence somewhat extreme realisations (related to e.g. very low accretion or equal mass mergers) of the source or the external potential and since the focus is on the typical scales, presumably related to moderate interactions, they should hopefully not significantly affect the measurements.

\subsection{Reduction procedure}
\begin{figure} 
\centering
\resizebox{0.8\columnwidth}{0.8\columnwidth}{\includegraphics{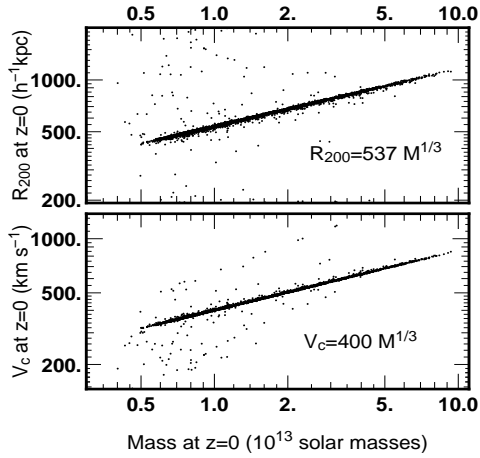}}
\caption{Virial  radii  ($R_{200}$)   and  circular  velocities  ($V_{c}$)  as
functions  of  halos final  masses.   The  quantities  have been  measured  at
redshift $z=0$.   Scaling relations between  $R_{200}$ or $V_{c}$ and  the final
mass are also given.}
\label{f:RVCvsM} 
\end{figure}

{ In the following discussion  most of  the distances  (resp.  velocities)  will be  expressed as
functions of the virial radius  $R_{200}$ (resp.  the circular velocity $V_c$)
measured at $z=0$.  These quantities are related to the halo's virial mass by
\begin{equation}
V_c=\sqrt{\frac{GM_{200}}{R_{200}}} \,.
\end{equation}
Here $M_{200}=M(r<R_{200})$. The mass-dependence of $R_{200}$ and $V_c$ are given in \Fig{RVCvsM} and may
be fitted by:
\begin{equation}
R_{200}=537M^{{1}/{3}}\,\,,\quad V_c=400M^{{1}/{
3}}, \EQN{RV200}
\end{equation}
where  $R_{200}$ is expressed  in ($h^{-1}$kpc),  $V_c$ in  km/s and  $M$ in
units  of $10^{13}  \mathrm{M}_\odot$. Here $M$ stands for the total mass of
the halo, returned by the halo finder HOP. In \Eq{RV200}, $R_{200}$ and  $V_c$  appear to  be
strongly  correlated to  the final  masses of  haloes, the few outliers being
related to external subhalos or peculiar halo geometries.  Since  the selection
criteria are quite restrictive, most of the halos experience the same relatively
quiet history of accretion and account for the lack of  scatter.}
  
The simulations in that redshift  interval involved 15 snapshots sampled with
$\Delta(\log  z)={\mathrm{cst}}$  for $z  \le  1$  down  to $z=0.1$  plus  a
snapshot at $z=0$.  The gap between  the last snapshot and the second to the
last  is nearly  1.4 Gyr.   As a  consequence, the  assumption  of ballistic
trajectories  is  not  valid  anymore  (see appendix)  .   Simulations  were
re-sampled in  15 bins  distributed regularly in  \textit{time} (i.e  not in
redshift) using the procedure described  in \Sec{coeff}: the corresponding time step is $\sim 500$ Myrs.  To take in account
the last gap,  results obtained from the last three  `new' bins (which cover
the last 1.4 Gyr) were averaged into a single bin centered on 0.8 Gyr.

The  source  coefficients,  $c_{\alpha,{\M{m}'}}^{\M{m} }(t)$,  were  computed
following the procedure described  in \sec{template}.  Maximum harmonic orders
were set to $\ell_{\mathrm{max}}=50$  for the position-angular description.
 For a typical halo with $R_{200}=500$ kpc, $\ell_{\mathrm{max}}=50$ corresponds to a spatial scale of $30$ kpc, i.e. equals to 1.5 times the spatial resolution of the simulation.
The harmonic description of the velocities angular dependence is restricted to $\ell'_{\mathrm{max}}=15$. The velocity amplitude is projected on a gaussian
basis  which involves 25 functions  regularly spaced  from $v/v_c=0$  to $v/v_c=1.5$ with an r.m.s of 0.03. These parameters allow a satisfying reproduction of distributions computed from particles. 

The external  potential coefficients, $b_\bm(t)$, were  computed following the
procedure described  in \sec{template}. Only  particles within a 4 Mpc physical sphere centered on the halo are taken in account. Maximum harmonic orders were set to $\ell_{\mathrm{max}}=20$. 
%100 simulations have  been reduced up to now allowing to compute for 15000 halos.  

A set of 100 simulations have been fully reduced 
 allowing us to compute $c^{\M{m}}_{\alpha,,\M{m} '}(t)$ and  $b_\bm(t)$ for 15000 halos.
Since   a  well defined,  (if  only  biased) sample  of
histories of haloes was constructed in our simulations,  it may  be projected  on our basis, to
compute  the  external potential and the flux density of mass,  following  \Sec{template}.   Let us now
characterize the corresponding coefficients, via one point (\Sec{stat1d}) and
two points (\Sec{stat2d}) statistics.

%%%%%%%%%%%%%%%%%%%%%%%%%%%%%%%%%%%%%%%%%%%%%%%%%%%%%%%%%%%%%%%%%%
\section{One point statistics}
%%%%%%%%%%%%%%%%%%%%%%%%%%%%%%%%%%%%%%%%%%%%%%%%%%%%%%%%%%%%%%%%%%
\label{s:stat1d}
%%%%%%%%%%%%%%%%%%%%%%%%%%%%%%%%%%%%%%%%%%%%%%%%%%%%%%%%%%%%%%%%%%
In this section, let us first describe the evolution and the statistical distributions of the global properties (i.e. integrated over the sphere) of the source and the potential. Let us discuss the evolution of the mean potential, of the mass flux $\Phi^{M}(t)$ and the kinematical properties of $s^e$ via the velocity distribution $\phi(v)$ and the impact parameter distribution $\vartheta(b)$.

Let  us then describe  in turn  the  statistical PDF,  mean and  variance of  the
integrated  fluxes, their  corresponding flux  densities, and  finally their
mean kinematical features, following the steps of \sec{template}.

%%%%%%%%%%%%%%%%%%%%%%%%%%%%%%%%
\subsection{Mean External potential}
%%%%%%%%%%%%%%%%%%%%%%%%%%%%%%%%
The mean external potential on the sphere is actually somewhat meaningless but is being used as a normalization value for potential fluctuations (see section \ref{s:statpot}). Because of isotropy, the mean potential is seen as a monopole and only the $b_{00}(t)$ coefficient is statistically different from zero. Furthermore,  following \eq{defexp}, the three dimensional potential component induced by the monopole is a constant potential throughout the sphere volume. As a consequence, it influences the halo's dynamics only through its temporal variation. 

\begin{figure} 
\centering           
\resizebox{0.8\columnwidth}{0.8\columnwidth}{\includegraphics{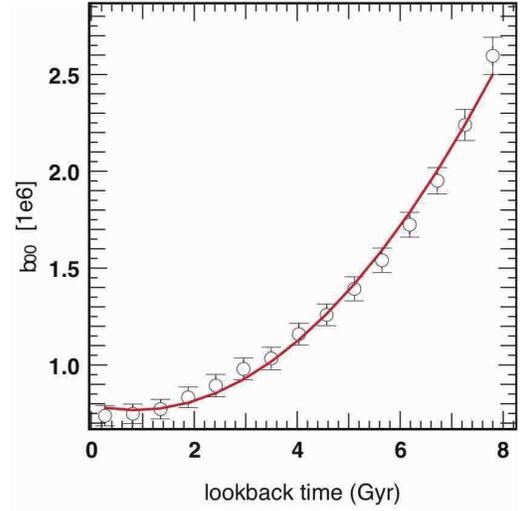}}
\caption{The  time  evolution of  the  monopole  component  of the  external
potential,  $b_{00}(t)$.  The  time evolution  is  fitted by  a second  order
polynomial $-b_{00}(t)=35948*t^2-61480.7*t+793067  $.  Look back time  $t$ is
expressed in Gyrs while $b_{00}$ coefficients are expressed in units of $GM/R$. M is expressed in $10^{10} M_\odot$, R in kpc$h^{-1}$ and G=43007 in internal units.}
\label{f:b00t} 
\end{figure}

The time evolution of the $\llangle b_{00} \rrangle$ coefficient is given in \Fig{b00t}. The $b_{00}(t)$ distribution exhibits a tail due to  large $-b_{00}$ values and is better fitted with a log-normal distribution than with a Normal distribution (see Fig. \ref{f:b001}). Hence $\llangle b_{00} \rrangle$ stands for the most probable value of the log-normal fitting distribution. \footnote{Since the measured distribution is quite peaked, fits made with a Normal distribution (not shown here) return a very similar time evolution of the mode position.} The evolution shown in \Fig{b00t} reflects the measurement procedure. Given that the potential is computed from all the particles contained within a fixed physical volume, the overall expansion implies that particles tend to exit the measurement volume with time. In other words, the average density in the measurement volume decreases with time. This effect leads naturally to the decline of the average potential  within the virial sphere due to external material. 

\subsection{Mass flux~: $\Phi^M(t)$}

\begin{figure} 
\centering
\resizebox{0.8\columnwidth}{0.8\columnwidth}{\includegraphics{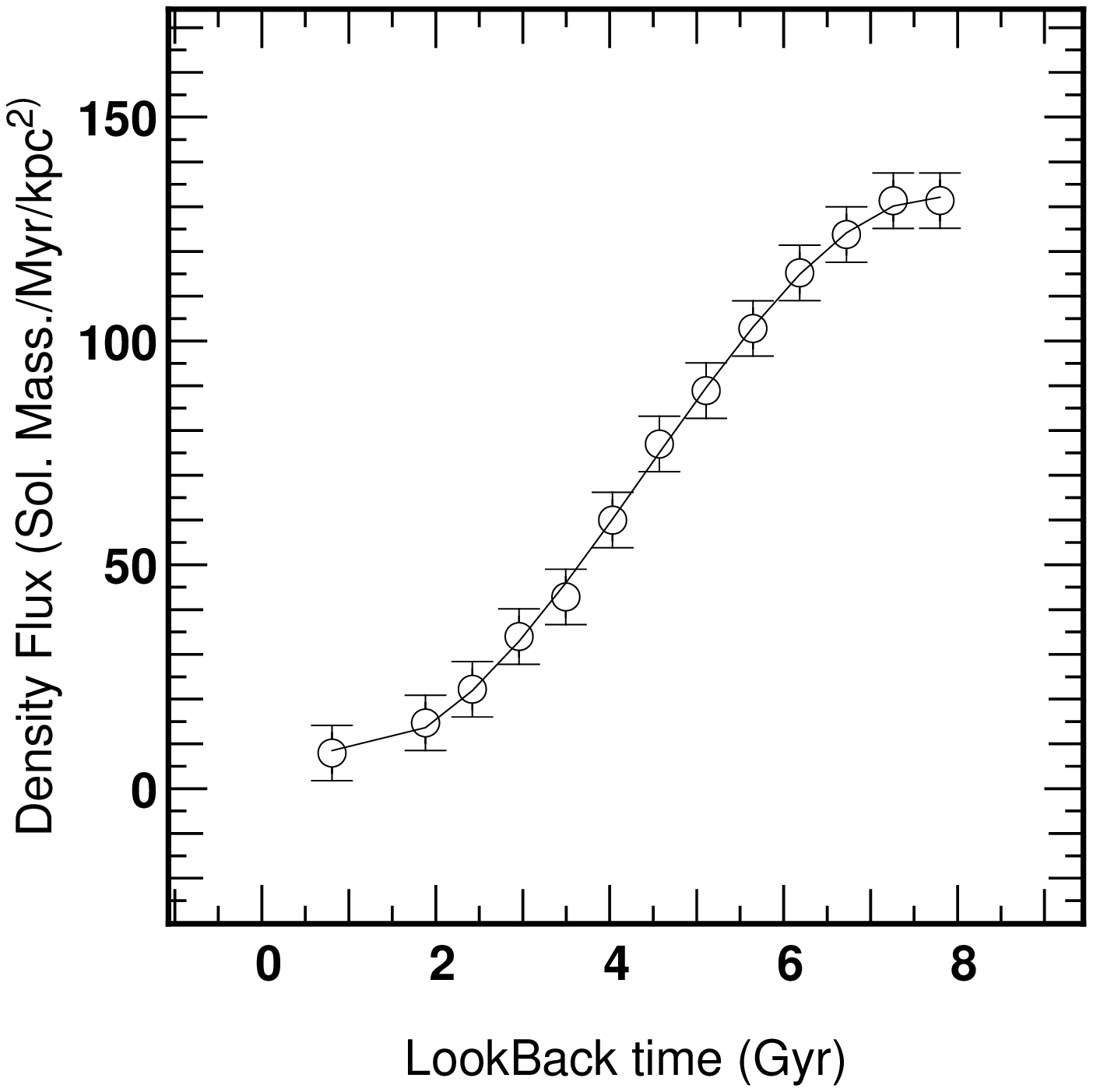}}
\resizebox{0.8\columnwidth}{0.8\columnwidth}{\includegraphics{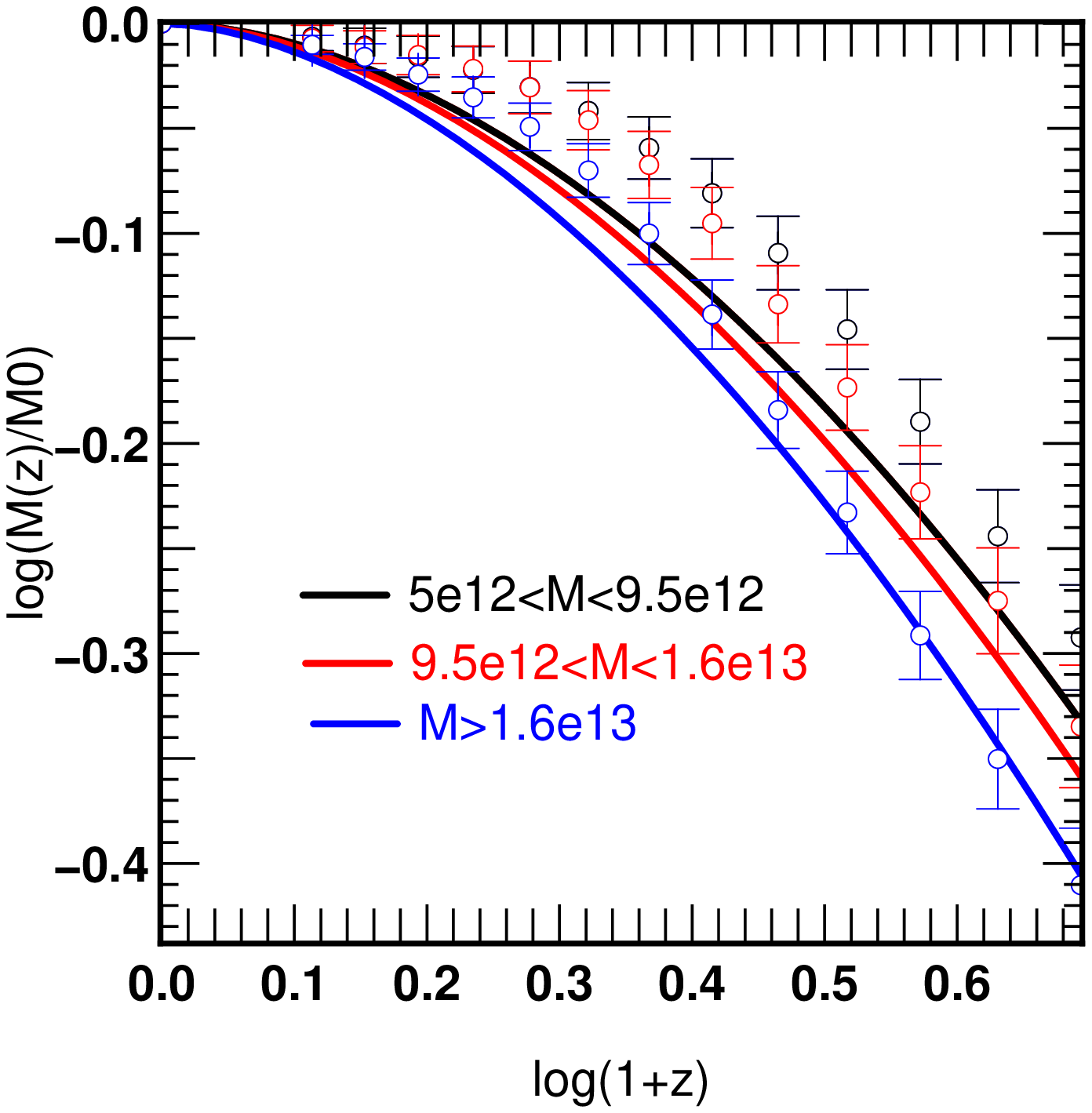}}
\caption{\textit{Top:} Time  evolution of the  average flux density  of matter
through the  virial sphere, $\langle  \Phi^M(t) \rangle=\langle \overline{\varpi_\rho} \rangle (t)$ (\textit{symbols}).  Bars stand for $3\sigma$ errors. Here $
\Phi^{M}(t)$   is  computed  directly   from  $a_{00}$   coefficients  following
\eq{masst}. Its  time evolution is  fitted by a 3rd  order polynomial(\textit{solid
line}).
  \textit{Bottom:} the  mass accretion history $\log{M(z)/M(z=0)}$ for
three different class of masses. Masses are expressed in solar masses.  Symbol
represent the median value  of $\log{M(z)/M(z=0)}$ within each classes.  Lines
represent  the fitting  function suggested  by \citet{VDB}.   Even  though the
global  behavior  is  reproduced  by  the  fitting  functions,  the  measured
accretion rate  is systematically smaller.  This discrepancy  has already been
noted by \citet{VDB}.}
\label{f:rhovrt} 
\end{figure}

   At each  time-step, the  $a_{0,0}$ distribution is  fitted by  a Gaussian
function  with mean  $\langle a_{0,0}  \rangle  $ (see also \Fig{a001}). This gaussian hypothesis is clearly verified for low redshifts while strong accretions events give rise to a tail in the $a_{0,0}$ distribution at high $z$. At these epochs (lookback time $t>7$ Gyr) the gaussian fit tends to slightly overestimate the mode position. Yet the gaussian hypothesis remains a good approximation of the distribution while the time evolution of the gaussian mean value $\langle a_{0,0}  \rangle (t) $ represents well the evolution of the mode of $a_{0,0}$.   

The  time
evolution of the average flux  of matter through the sphere, $\Phi^{M}(t)=\overline{\varpi_\rho}$  is directly derived from the evolution of the monopole (see \eq{a00phi}) and  is shown in \Fig{rhovrt}. It can be fitted by:
\begin{equation}
\Phi^{M}(t)\equiv\langle\overline{\varpi_\rho}\rangle(t)=-0.81t^3+10.7t^2-19.3t+17.57,
\label{e:fita00}
\end{equation}
where $\Phi^{M}(t)$ is  in units of $\mathrm{M}_\odot/\mathrm{Myr}/\mathrm{kpc}^2$
and look-back time $t$ is expressed in Gyr.  As expected, the average quantity
of material accreted  by halos decreases with time. For  $z<1$, a large fraction of the objects of interest in are  already `formed'  and only  gain matter  through  the accretion of small
objects  or diffuse material.  In  a hierarchical  scenario, such  source of
matter   become  scarcer,  inducing a   decrease in the  accretion   rate. Furthermore, recall
  that $\Phi^{M}(t)$ is measured as a net flux, i.e. the outflowing material may cancel a fraction of the infalling flux. Therefore, the decrease with time may also be the consequence of an increasing contribution of outflows: the measurement radius $R_{200}(z=0) $ becomes the actual virial radius of the halo as time goes by, i.e. the radius where the inner material is re-processed and where outflows are susceptible to be detected. 

As a  check the  average mass accretion history  (MAH hereafter) of
the haloes was computed. The MAH $\Psi (t)$ is defined as:
\begin{equation}
\Psi (t)\equiv\frac{M(t)}{M(z=0)},
\label{e:MAH}
\end{equation}
where  $M(z)$ is  the halo's  mass  at a  given instant.   Using the  extended
Press-Schechter formalism, \citet{VDB} showed that halos have an universal
MAH, fitted by the following formula:
\begin{equation}
\log(\Psi(M(z=0),t))=-0.301\left[\frac{\log(1+z)}{\log(1+z_f)}\right]^\nu,
\end{equation}
where $z_f$ and $\nu$ are two  parameters which depend on the considered class
of mass only. These two parameters are found to be correlated and for instance
\citet{wechsler} found a similar relation using a single
parameter.   For each halo, its mass  evolution $M(t)$ was computed from its
final mass $M(z=0)$ and its integrated flux of matter $\Phi^{M}(t)$:
\begin{equation}
M(t)=M(t=0)-4\pi R_{200}^2\int_{t=0}^t \md t \Phi^{M}(t).
\label{e:masst}
\end{equation}
From  \eq{MAH}    $\Psi  (t)$ was computed  for  each  halo.   
The
\textit{median} value  of $\Psi (t)$  for three classes  of mass 
was compared to  the fit
suggested  by \citet{VDB}  (see  \Fig{rhovrt}). For  the  three classes, the
agreement with the  fitting formula is qualitatively satisfying: the three
measurements evolve  in the expected  manner while their  relative positions
are  the same  as the  relative  positions of  the three  fits. However  our
measurements are  quantitatively inconsistent with the three  curves. At low
redshift, $\Psi (t)$  is systematically larger than the  expected value (\ie
the accretion  rate is \textit{smaller}).  The median mass at $z=1$ is well
recovered even though the two method disagree slightly  quantitatively. In
other words, our measurements overestimates the accretion at high redshift and
underestimates accretion at low redshift. This may be related to the
measurement procedure through the sphere~: at higher redshift, accreted
material is assumed  to be added to the biggest progenitor, even though it has
not  yet reached the central object and its mass is overestimated. Still, this material end up in the most massive progenitor and the final mass is recovered.
Note that since  specific selection  criteria
were applied, these halos may    not   be    completely    representative   of    the   whole    halos population. Finally
recall that the \textit{median} value of $\Psi
(t)$  was represented here 
 because of  strong outliers while the fitting formula  is given for the
\textit{average} MAH  (extracted from  merger  trees).   
A similar discrepancy  between the extended Press-Schechter theory and the results obtained
from numerical simulations  had  already been  noticed by  \citet{VDB} and \citet{wechsler}. In particular, \citet{VDB} found that the Press-Schechter models tend to underestimates the formation time haloes compared to simulations. Clearly,  our measurements seem to confirm this discrepancy.  Since the  global behavior of MAHs is  recovered and since the median mass at $z=1$ is recovered, it is concluded that the measure  of $\Phi^{M}(t)$ through  the virial sphere  reproduces the accretion history of halos.

%%%%%%%%%%%%%%%%%%%%%%%%%%%%%%%%
\subsection{Mean kinematics}
%%%%%%%%%%%%%%%%%%%%%%%%%%%%%%%%

Let us now turn to the kinematical properties of the flow, while averaging 
the source over the virial sphere.

%%%%%%%%%%%%%%%%%%%%%%%%%%%%%%%%
\subsubsection{probability distribution of the  modulus of velocities}
\label{s:kinematic}
\begin{figure} 
\centering           
\resizebox{0.8\columnwidth}{0.8\columnwidth}{\includegraphics{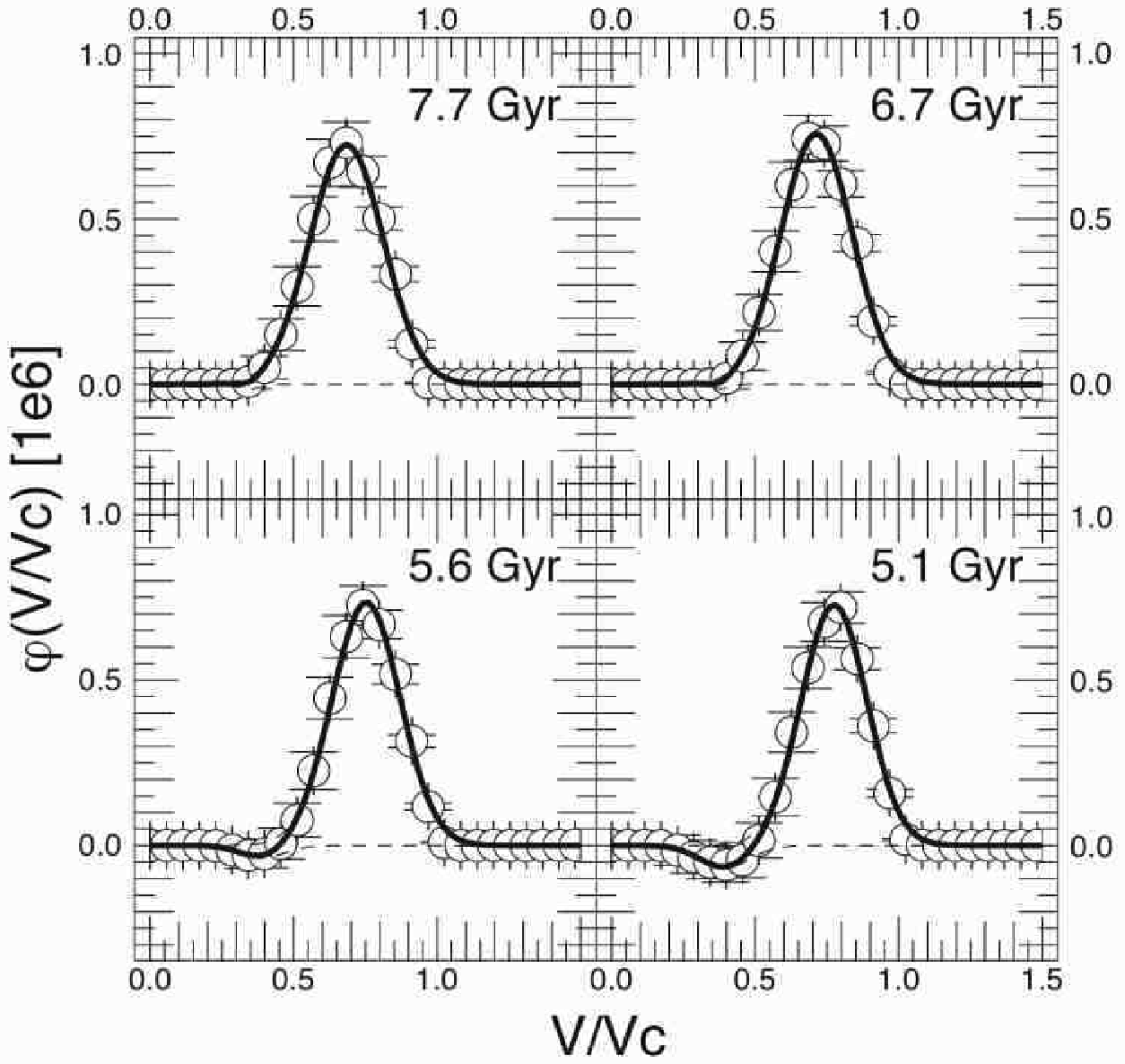}}
\resizebox{0.8\columnwidth}{0.8\columnwidth}{\includegraphics{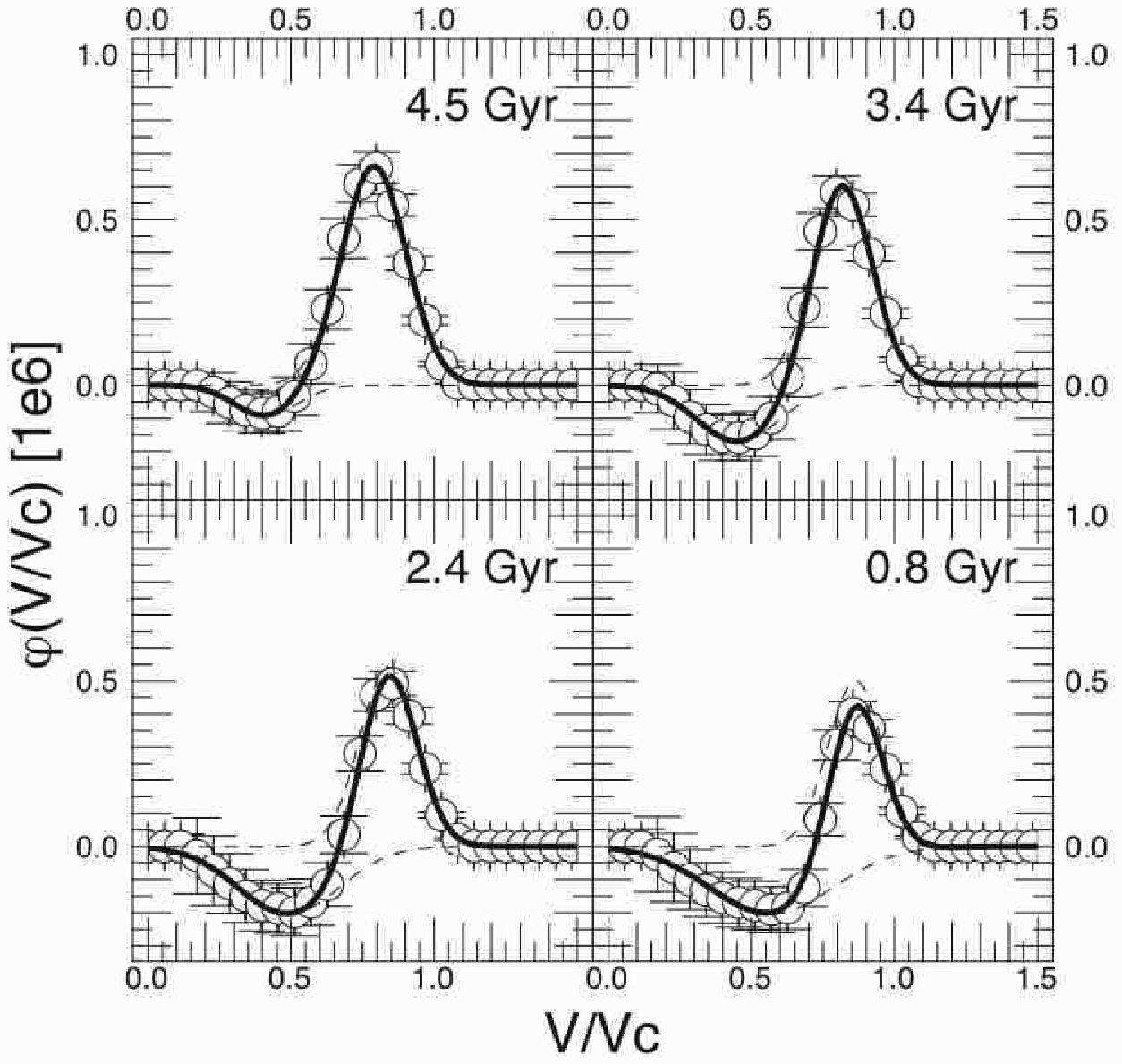}}
\caption{The time  evolution of the average  velocity distribution, $\langle
\varphi(v,t)  \rangle$, defined by  \eq{dfv}, for  $z<1$ (\textit{symbols}).
Ages are expressed as look-back  times (i.e. $t=0$ for $z=0$). Velocities are
given relative  to the  halo's circular velocity  at $z=0$.  Y-axis  unit is
$5\cdot10^9  \mathrm{M}_\odot$/kpc$^2$/Myr$/V_c$.    Error  bars  stand  for
$3-\sigma$ errors.  Here  $\varphi(v)$ is fitted by the  sum of two Gaussian
with opposite signs (\textit{solid  line}).  Each Gaussian's contribution is
also shown (\textit{dashed lines}).}
\label{f:distribErr} 
\end{figure}

Given  the  source coefficients  $c^{\M{m}}_{\alpha,\M{m}  '}$, the  average
velocity distribution  $\langle \varphi(v,t) \rangle$  (defined by \eq{dfv})
is  easily  computed  since  it  only  involves  $\langle  c^0_{\alpha,0}(t)
\rangle$.  The ensemble average  of $\langle c^0_{\alpha,0} \rangle$ and the
related    ensemble    dispersion    $\sigma(c^0_{\alpha,0})(t)\equiv\langle
(c^0_{\alpha,0}-\langle  c^0_{\alpha,0} \rangle)^2\rangle$  are  derived by
fitting the  $c^0_{\alpha,0}(t)$ distribution by a  Gaussian function.  From
these two quantities, it follows:
\begin{equation}
\langle  \varphi(v,t)\rangle=4\pi  v^2\sum_\alpha g_\alpha(v)\langle  c^{\bf
0}_{\alpha,{\bf 0}}\rangle,
\end{equation}
and
\begin{equation}
\sigma\left[\varphi(v,t)\right]=4\pi v^2\sqrt{\sum_\alpha  g_\alpha(v)^2
  \sigma(
c^{\bf 0}_{\alpha,{\bf 0}})^2},
\end{equation}
which  are respectively  the ensemble  average and  r.m.s.  of  the velocity
distribution.  The time evolution  of $\langle \varphi(v,t)\rangle$ is given
in   \Fig{distribErr}  and   \Fip{distribvimage}.   Errors   on  $\langle
\varphi(v,t)\rangle$ are computed as
\begin{equation}
\Delta\left[\langle \varphi(v,t)\rangle\right]=3\frac{\sigma[\varphi(v,t)]}{\sqrt{N_{\mathrm{halos}}}}.
\label{e:deferr}
\end{equation}
At `early  times' ($t>5$ Gyr), the  distribution is unimodal  with a maximum
around $0.7 V_c(z=0)$.  No outflows can  be detected at any velocity and the
infalling    dark   matter    dominates.    At    later    times,   $\langle
\varphi(v,t)\rangle$  drops below zero  for velocities  around $0.4 V_c(z=0)$.
Outflows  dominate  at `low'  velocities.   Meanwhile  the  amplitude of  the
previous peak  decreases and  shift to higher  velocities.  The  fraction of
infall relative to  the total amount of material  passing through the sphere
drops from $1.$ to $0.6$ between $t=8$ Gyr and $t=0.8$ Gyr.

\begin{figure} 
\centering           
%\resizebox{0.8\columnwidth}{0.8\columnwidth}{\includegraphics{distribv}}
\resizebox{0.8\columnwidth}{0.8\columnwidth}{\includegraphics{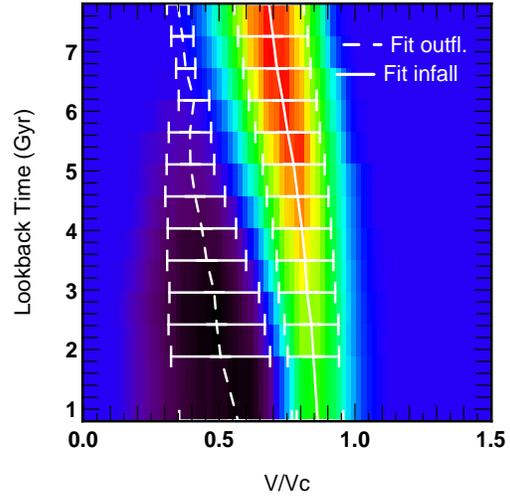}}
\caption{The  time evolution  of the  average velocity  distribution  in the
$t$-$V/V_c$  plane.    Red  symbols  stand   for  positive  values   of  the
distribution (i.e.  infall) while blue  symbols stand for negative  ones (i.e.
outflows).  Each of these components is fitted by a Gaussian function in the
$V/V_c$ space.  The time evolutions of the  mean of the  Gaussians are given
by the two lines  (\textit{solid} for infall, \textit{dashed} for outflows).
The Gaussians r.m.s. are also shown as bars.}
\label{f:distribvimage} 
\end{figure}

This behavior  is likely to  be due to  our measurement at  a \textit{fixed}
radius, $R_{200}(z=0)$.  At `early  times', this measurement radius is bigger
than the  actual virial radius of  halos.  Thus no  sign of `virialization'
(outflows  consecutive  to  accretion)   is  detected.   Later,  the  actual
$R_{200}$ gets closer  to the measurement radius.  Outflows  pass through the
measurement radius as a sign  of internal dynamical reorganization.  The fact
that  accretion  intrinsically decreases  with  time  would provide  another
explanation of this trend.  This  decrease can actually be traced in 
\Fig{distribErr},   \Fip{distribvimage}   and  \Fip{distribvcoeff}   .

The global  behavior of $\langle  \varphi(v,t)\rangle$ can be  modeled by
summing   two   Gaussians   representing   the  infalling   and   outflowing
components~:
\begin{eqnarray}
\langle \varphi(v,t)\rangle&=& \frac{q_{\imath,3}(t)}{q_{\imath,2}(t)\sqrt{2\pi}}
\exp\left({-\frac{(v-q_{\imath,1}(t))^2}{2q_{\imath,2}(t)^2}}\right)\nonumber\\&&+
\frac{q_{o,3}(t)}{q_{o,2}(t)\sqrt{2\pi}}\exp\left({-\frac{(v-q_{o,1}(t))^2}{2q_{o,2}(t)^2}}\right).\label{e:deuxgauss}
\end{eqnarray}
Subscripts  $\imath$   and  $o$  stand   for  infall  and  outflow.    Note  that
$q_{\imath,3}(t)\ge 0$ and $q_{o,3}(t)\le 0$.  The coefficients time evolution is
given  in   \Fig{distribvcoeff}, where  $t$  should  be  expressed   in  Gyr  and $\varphi(v,t)$  unit   is  $5\cdot10^9  \mathrm{M}_\odot$/kpc$^2$/Myr$/V_c$.
Examples  of  fits  are  shown  as solid  lines  in  \Fig{distribErr}.   Note that  all the six coefficients  evolve roughly linearly  with time (see
\Fig{distribvcoeff}). Their linear  fitting parameters are given in
table   \ref{t:coeffgauss}.     Using   \eq{deuxgauss}   and    the   linear
parameterization     of     the     Gaussians     coefficients,     $\langle
\varphi(v,t)\rangle$ is reproduced accurately.
The only restriction concerns  the negative Gaussian's amplitude ($q_{o,3}$)
which should not  be greater than 0.  Since this  condition is not naturally
satisfied by a linear fit, it should be set by hand. 

The
evolution   of  the relative   positions of  two   Gaussians     is  given   in
\Fig{distribvimage}.
 For $t>5$Gyr, it is consistent with the `no
outflow' hypothesis, the  amplitude of the negative Gaussian  being close to
zero at  this epoch.   Both Gaussians  mean values seem  to drift  to higher
velocities  as a function  of time.   Even though  the relative  velocity of
accreted  material is  determined by  the initial  conditions  (namely large
scale  clustering), the  velocity of  an infalling  satellite  should partly
reflect the properties of the accreting body.  A dense massive halo will not
accrete like a  fluffy light one. In other words,  the velocity of infalling
material should reflect the actual  circular velocity of the accreting body.
As a  consequence it is expected  that accretion velocity  drifts with time
towards $V_c(z=0)$.

Furthermore, the mean values of both Gaussians evolves roughly linearly over
the  whole   time  range  with   comparable  rate  of  change   (see  \Fig{distribvcoeff}  and  the following  discussion).   As a  consequence,
their   relative   positions    remains   roughly   constant   (see   
\Fig{distribvimage}).   This indicates  that these  two components  may be
physically related, outflows being the consequence of a past accretion.  \citet{Mamon} mention the existence of a \textit{backsplash} population, rebounding through the virial radius and this population is known to have a different velocity (e.g. \citet{Gill}).  The outflows detected via our description of the source is consistent with this backsplash component. The
difference in velocity may be explained if outflows are representative of an
earlier  accretion with  a velocity  typical  of earlier  times.  Also  past
accreted  material  is influenced  by  the  halo's  internal dynamics.   Its
velocity  distribution would  be  `reprocessed' (e.g. via dynamical friction, tidal stripping or phase mixing) to  lower  velocities as  the material exits through the measurement radius. 

\begin{figure} 
\centering
\resizebox{0.8\columnwidth}{0.8\columnwidth}{\includegraphics{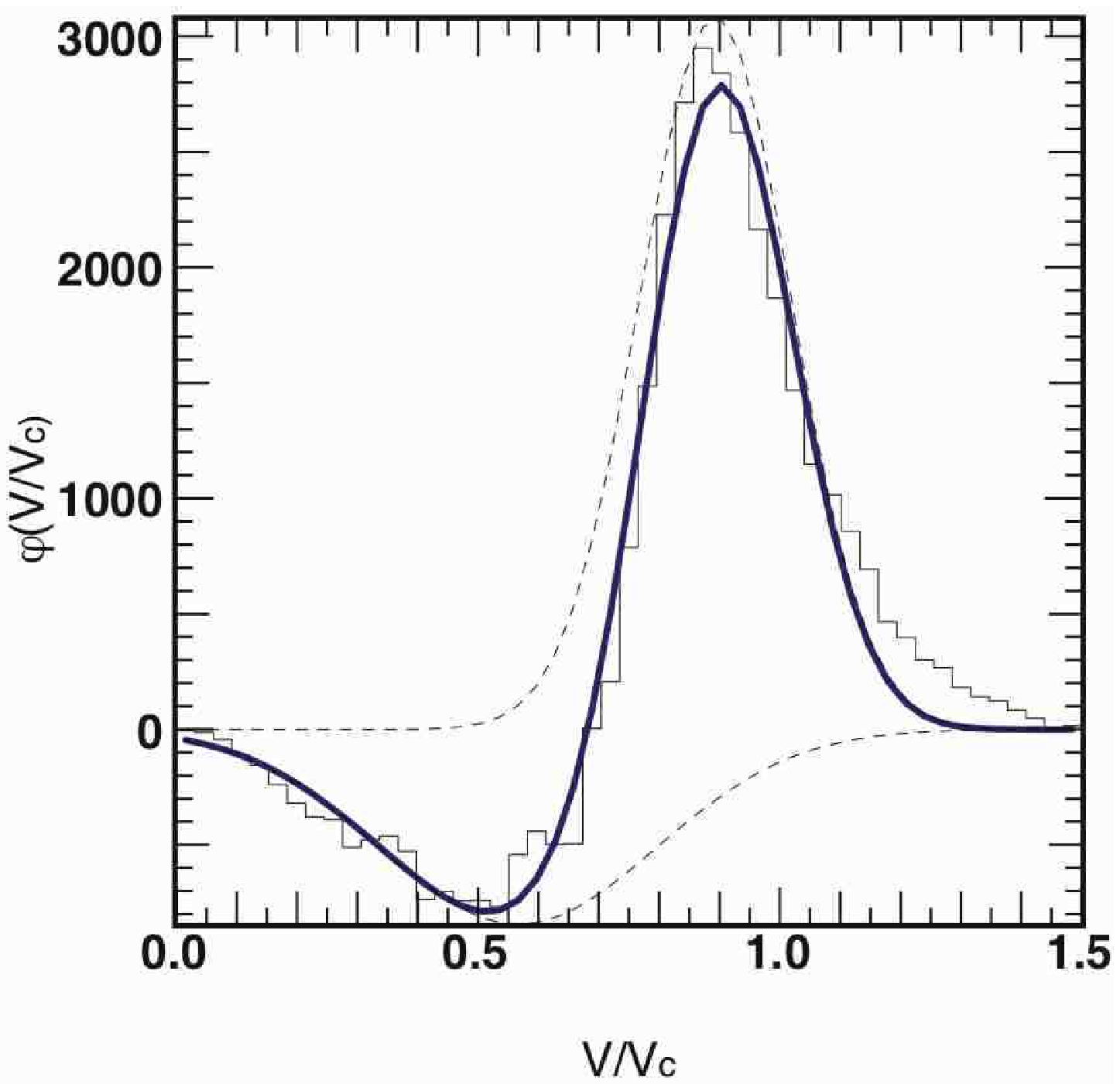}}
\resizebox{0.8\columnwidth}{0.8\columnwidth}{\includegraphics{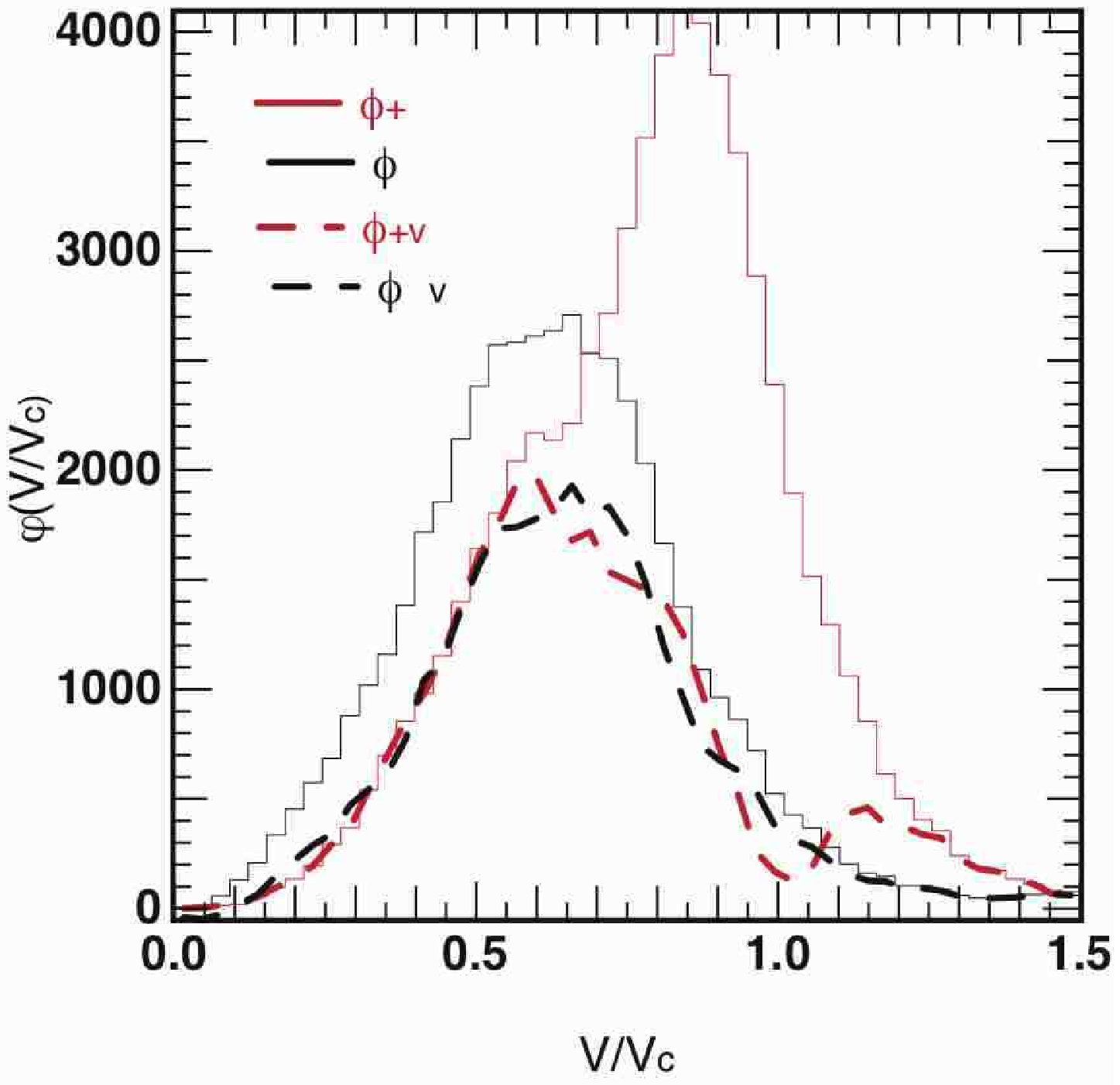}}
\caption{\textit{Top: } the net velocities distribution (histogram) measured from 300 haloes at t=3.4 Gyr. This distribution is representative of the distribution computed from coefficients and averaged over 15 000 haloes. Velocities are given in circular velocity unit while y-axis units are arbitrary. The two components are fitted by two gaussians (dashed lines). \textit{Bottom:} the separate velocities distribution of accretion (red histogram) and outflows (black histogram). The dashed curves represent the difference between these two distributions and their respective fits shown above. It results in two residual distributions, centered on the same velocity and displaying nearly the same shape. These two residual distributions describe the material which already experienced one passage through the virial sphere.}
\label{f:netabsflux} 
\end{figure}

However,  recall that the distribution shown here is a `net' distribution. In other words, it is quite plausible that an outflowing component may be completely canceled by an infalling component which has an exact opposite distribution. This effect is illustrated by \Fig{netabsflux} where the velocity distribution of infall and outflows are being shown separately. This distribution has been computed from 300 haloes at t=2.3 Gyr. This distribution is quite representative of the average ones, except a few high velocities events which skew the distribution of the infalling component and which are induced by outliers. If the gaussians fits are removed from these two separate distributions, two almost identical distributions appear for the two components, centered on $V/V_c\sim 0.6$.
These two identical distributions are related to the virialised component of infall, which already interacted with the inner region of the halo. The overall shape of these two distributions may provide insights on the typical dynamical state in the haloes' inner regions.  

\subsubsection{Impact parameters and incidence angles}

\begin{figure} 
\centering
\resizebox{0.8\columnwidth}{0.8\columnwidth}{\includegraphics{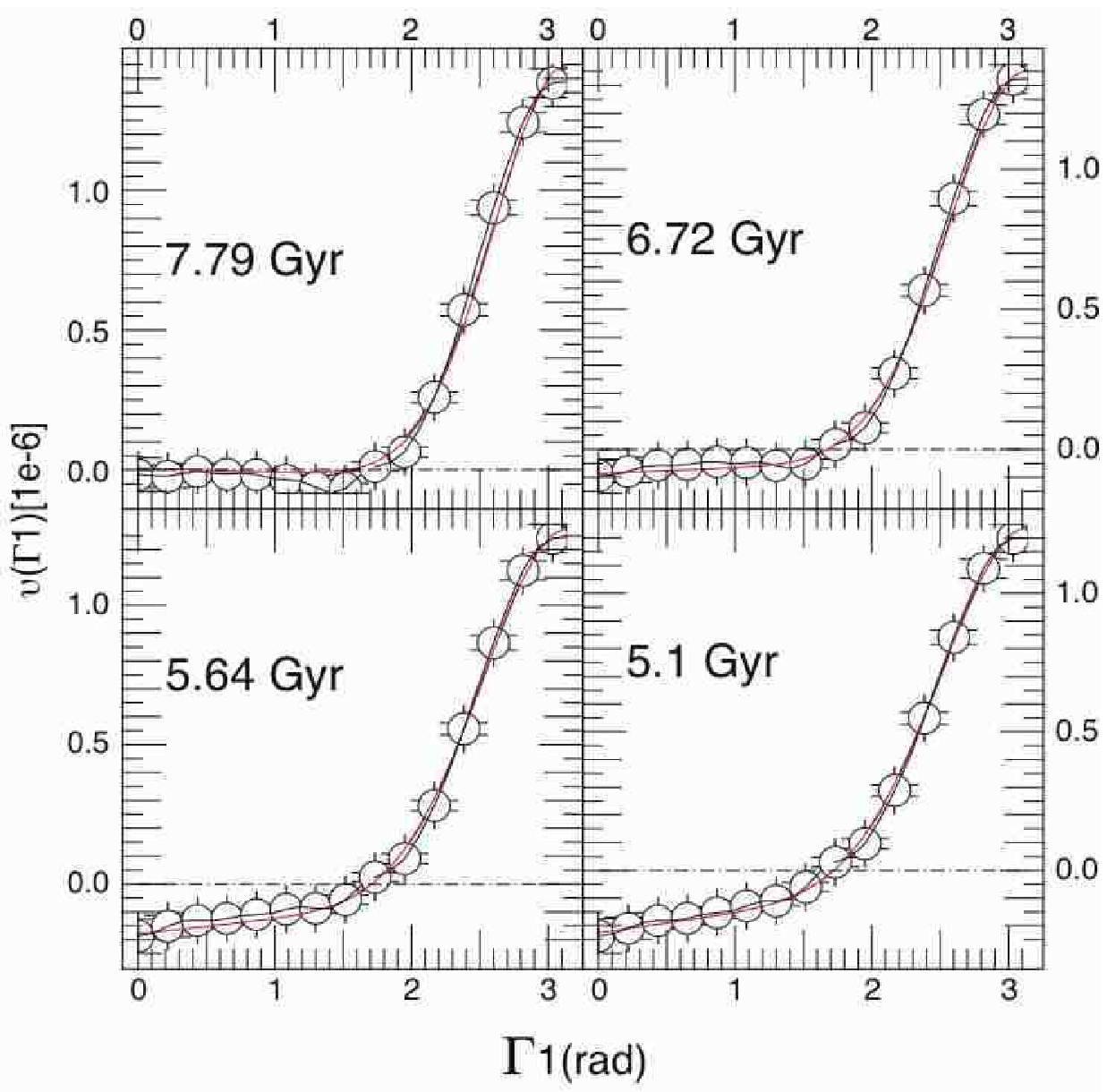}}
\resizebox{0.8\columnwidth}{0.8\columnwidth}{\includegraphics{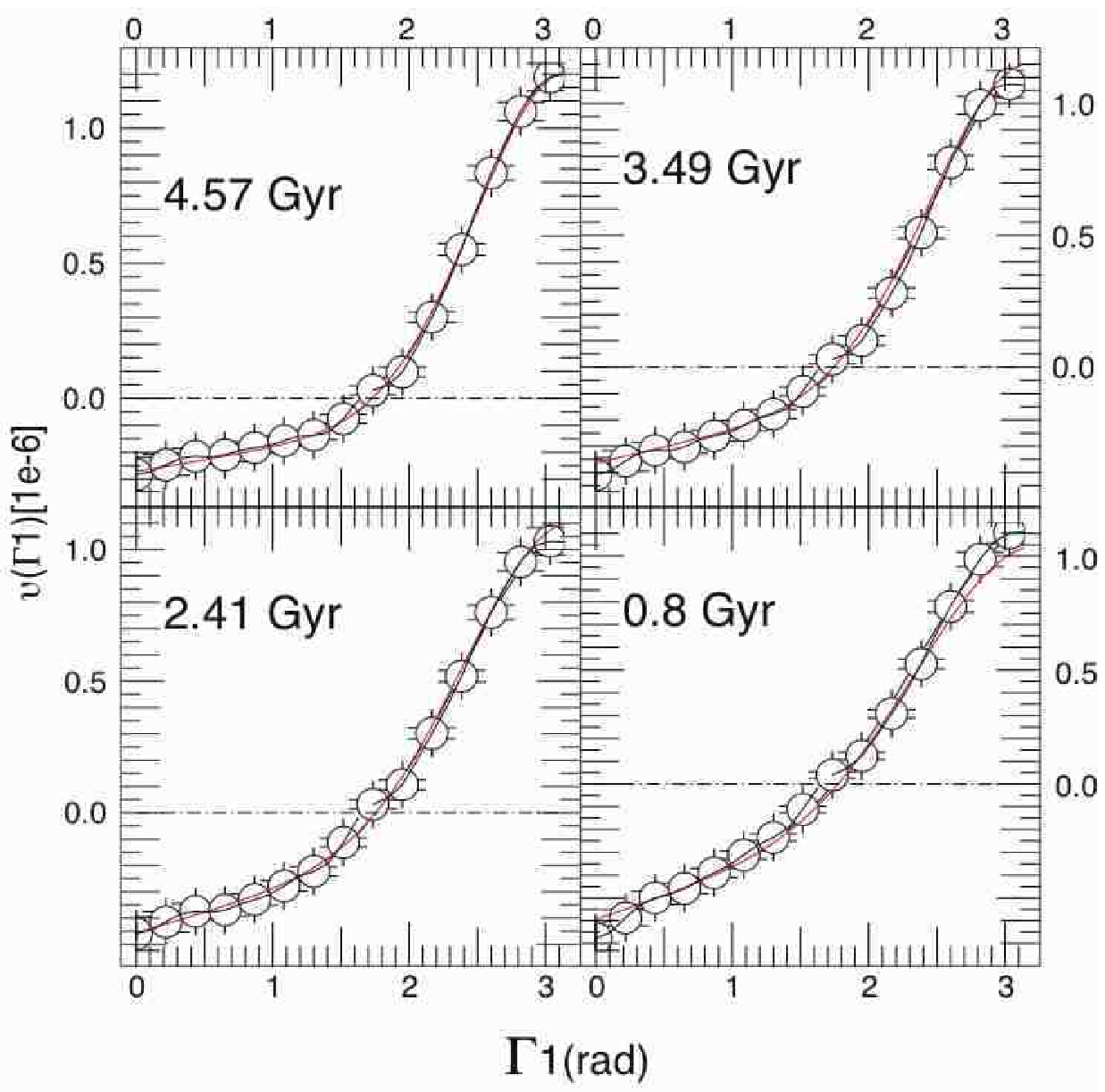}}
\caption{The  time evolution (symbols)  of the  distribution of  the average
incidence angle $\Gamma_1$ $\langle \vartheta(\Gamma_1) \rangle$, defined by
\eq{defq}. Ages  are expressed as  look-back time. Bars stand  for $3\sigma$
errors. Y-axis  unit is $5\cdot10^9  \mathrm{M}_\odot$/kpc$^2$/Myr. Outflows
are   counted   negatively,  leading   to   negative   values  of   $\langle
\vartheta(\Gamma_1)\rangle$ for  $\Gamma_1<\pi/2$.  The result  of the model
described in  \eq{modelgamma1} is also shown (red line).}
\label{f:evolangle} 
\end{figure}

The average distribution  of incidence angle, $\langle \vartheta(\Gamma_1,t)
\rangle$ has been computed following  the same procedure described above for
$\langle \varphi(v,t) \rangle$. Defining $\tilde{c}_{\ell'}(t)$ as:
\begin{equation}
\tilde{c}_{\ell'}(t) =2\pi\sqrt{4\pi}\sum_{\alpha}c_{\alpha,0,\{\ell',0\}}(\mu_\alpha^2+\sigma^2),
\end{equation}
yields
\begin{equation}
\langle \vartheta(\Gamma_1,t)\rangle=\sum_{\ell'}  Y_{\ell',0}(\bG) \langle \tilde{c}_{\ell'}(t)\rangle,
\end{equation}
and
\begin{equation}
\sigma(\vartheta(\Gamma_1,t))=\sqrt{\sum_{\ell'}           Y_{\ell',0}(\bG)^2
\sigma(\tilde{c}_{\ell'}(t))^2},
\end{equation}
where           $\langle            \vartheta(\Gamma_1,t)\rangle$          and
$\sigma( \vartheta(\Gamma_1,t)$     are    derived     by     fitting    the
$\tilde{c}_{\ell'}(t)$  distribution  by  a  Gaussian function.   Errors  on
$\langle   \vartheta(\Gamma_1,t)\rangle$ are computed  similarly to  errors on
$\langle  \varphi(v,t)\rangle$  (see \eq{deferr}).   The  time evolution  of
$\langle  \vartheta(\Gamma_1,t)\rangle$ is  shown in \Fig{evolangle}.
Since the  impact parameter  and the incidence  angle are simply  related by
$b/R_{200}=\sin  (\Gamma_1)$, $\langle  \vartheta(b,t)\rangle$ is  also easily
computed.

The  infall   ($\Gamma_1  >   \pi/2$  or  the   upper  branch   in  $\langle
\vartheta(b,t) \rangle$  diagrams) is clearly mostly  radial.  The infalling
part of the  distribution peaks for $\Gamma_1 \sim \pi$  instead of having a
uniform behavior and this trend can  be observed for all redshifts below 1.
The distribution slightly widens with but remains skewed toward large values
of $\Gamma_1$.  In the  $\langle \vartheta(b,t) \rangle$ representation, the
higher branch becomes flatter with time.  The outflows ( $\Gamma_1 < \pi/2$
or  lower branch in  $\langle \vartheta(b,t)  \rangle$ diagrams)  are mainly
undetectable at early  times,   as mentioned earlier.    As  time
increases,  the  outflow contribution  becomes  stronger  and radial  orbits
($\Gamma_1 \sim  0$) also appear to  be dominant.  However  the behavior of
the   `outflowing'  part   of   the  $\langle    \vartheta(\Gamma_1,t)\rangle$
distribution is almost  linear and does not peak.   Tangential orbits cannot
be neglected for this component.

The evolution of $\langle \vartheta(\Gamma_1,t)\rangle$ can be fitted by the
following  parametrisation:
\begin{equation}
\langle                    \vartheta(\Gamma_1,t)\rangle                    =
\frac{p_0}{\sqrt{2\pi}p_1(t)} \exp\left({-\frac{(\Gamma_1-\pi)^2}{2p_1(t)^2}}\right)+p_2(t)\Gamma_1+p_3(t),
\label{e:modelgamma1}
\end{equation}
where $p_0=2\cdot 10^{-6}$ in our  units. The `infalling' part is modeled as
a  Gaussian  while the  `outflowing'  part  is  fitted linearly.   The  time
evolution  of   the  three  parameters  $p_{k}(t),  k=1,2,3$ can be fitted by  a  linear evolution and  the related linear parameters
are  given in table  \ref{t:gamfit}.  The  evolution of  $p_{1}(t)$ confirms
that    the    `infalling'    part    of    the    distribution,    $\langle
\vartheta(\Gamma_1,t)\rangle$  widens with time.   
%No obvious  reason would
%explain why infalling  material, having its origin far  from the halo, would
%have a more tangential trajectory as time goes by.
  
This result implies that material experiences a circularisation as it interacts  with the halo.  Consequently,  orbits  are  more tangential  as
particles  \textit{exits}  and  \textit{re-enter}  the halo's  sphere.
Such an effect has already been measured by e.g. \citet{Gill}. Dynamical friction would provide a natural explanation for this evolution of the orbits, but this argument is refuted by e.g. \citet{Colpi} or \citet{Hashi}. \citet{Gill} mention the secular evolution of halos to explain this circularisation: the time evolution of the potential well induced by the halo would affect the orbits of infalling material and satellites. Other processes such as tidal stripping or satellite-satellite interactions may also modify the orbital parameters of dark matter fluxes. Clearly, the interactions between the infall and the halo drive this circularisation but the detailed process still has to be understood.
%It may  be explained by a  secondary infall made of  material which enters
%the sphere for the second (or  more) time. For example, high velocity objects 
%may need  more than one orbit to  be completely accreted by  the halo.  This
%material  would experience  a circularisation  of its  orbit induced  by the
%halo's  dynamical  friction.   

This
dynamical circularisation  could also explain  why the `outflowing'  part of
the $\langle \vartheta(\Gamma_1)\rangle$ (or $\langle \vartheta(b)\rangle$)
is  flatter than  the infalling  one: by definition this component interacted with the halo in the past, unlike most of the infall.   Finally, the  $\Gamma_1$ or  $b$
representation  explicitly separates  infall and  outflows. It  implies that
\textit{virialized} particles  which pass through  the sphere do  not `cancel'
each  other and  do  contributes  to the  distributions.   Such a  `relaxed'
material  is likely to  have a  non zero  tangential motion,  flattening the
distributions as its contribution becomes important. Since the actual size of the halo gets closer to the measurement radius as time advances, this component
contributes more with time and  its flattening effect on the incidence angle
(or impact parameter) distributions should increase as well.

\begin{figure} 
\centering           
\resizebox{0.8\columnwidth}{0.8\columnwidth}{\includegraphics{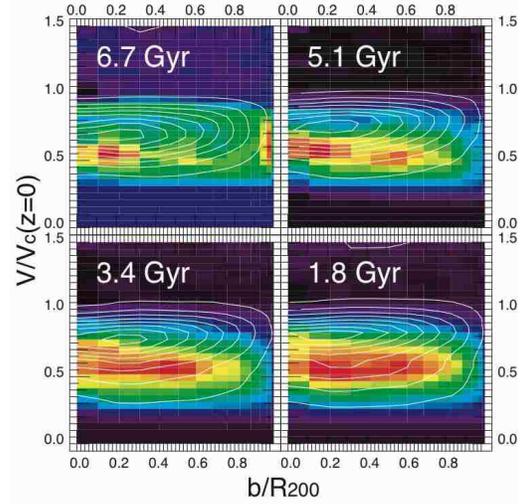}}
%\resizebox{0.8\columnwidth}{0.8\columnwidth}{\includegraphics{correlbvstat_out}}
\caption{ \textit{Top:} the  distribution, $\langle \wp(b,v)  \rangle$, of
particles in the $(b,v)$ subspace at lookback time $t=1.8,3.4,5.1$ and $6.7$
Gyr;  the infall  (contour plot)  and outflow  (density plot) are  represented
separately.   Beyond the  bimodal feature,  no residual  correlation appears.}
\label{f:bv} 
\end{figure}

\Fig{bv} presents the correlation between the velocity amplitude $v$ and the impact parameter $b$, at four different instants and for both the infall and the outflow component. Considering these two components separately, no correlation can be found: the incidence does not depend on the amplitude on first approximation. The only noticeable result comes from the fact that accreted material has systematically a higher velocity than outflows, which confirms the results obtained from the distribution of velocities only. Again, this effect
is related to the separate origin of these two fluxes, accretion, being
dominated by newly accreted material, and outflows, which were processed by the inner dynamics of haloes.

\begin{figure} 
\centering           
\resizebox{0.8\columnwidth}{0.8\columnwidth}{\includegraphics{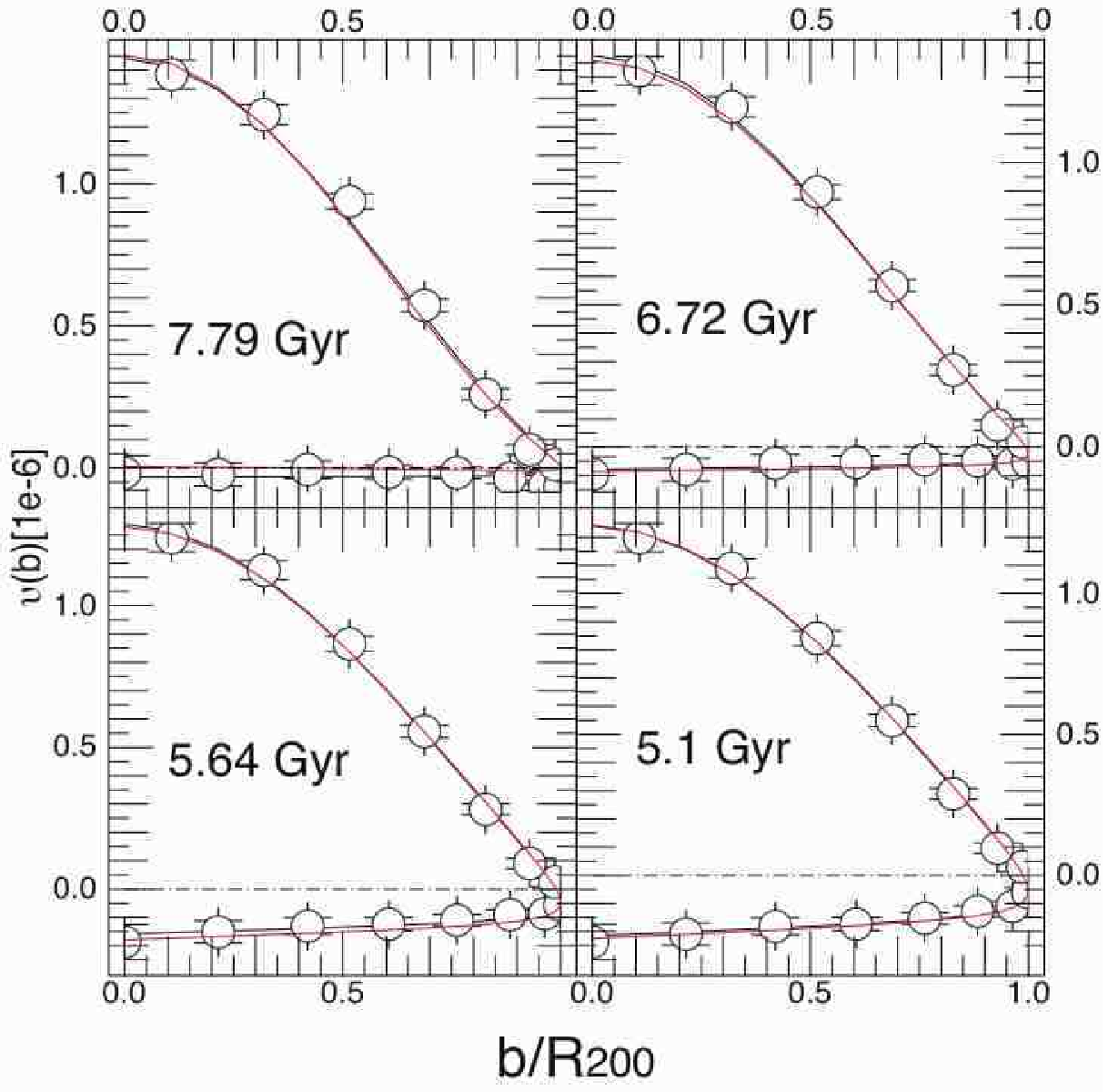}}
\resizebox{0.8\columnwidth}{0.8\columnwidth}{\includegraphics{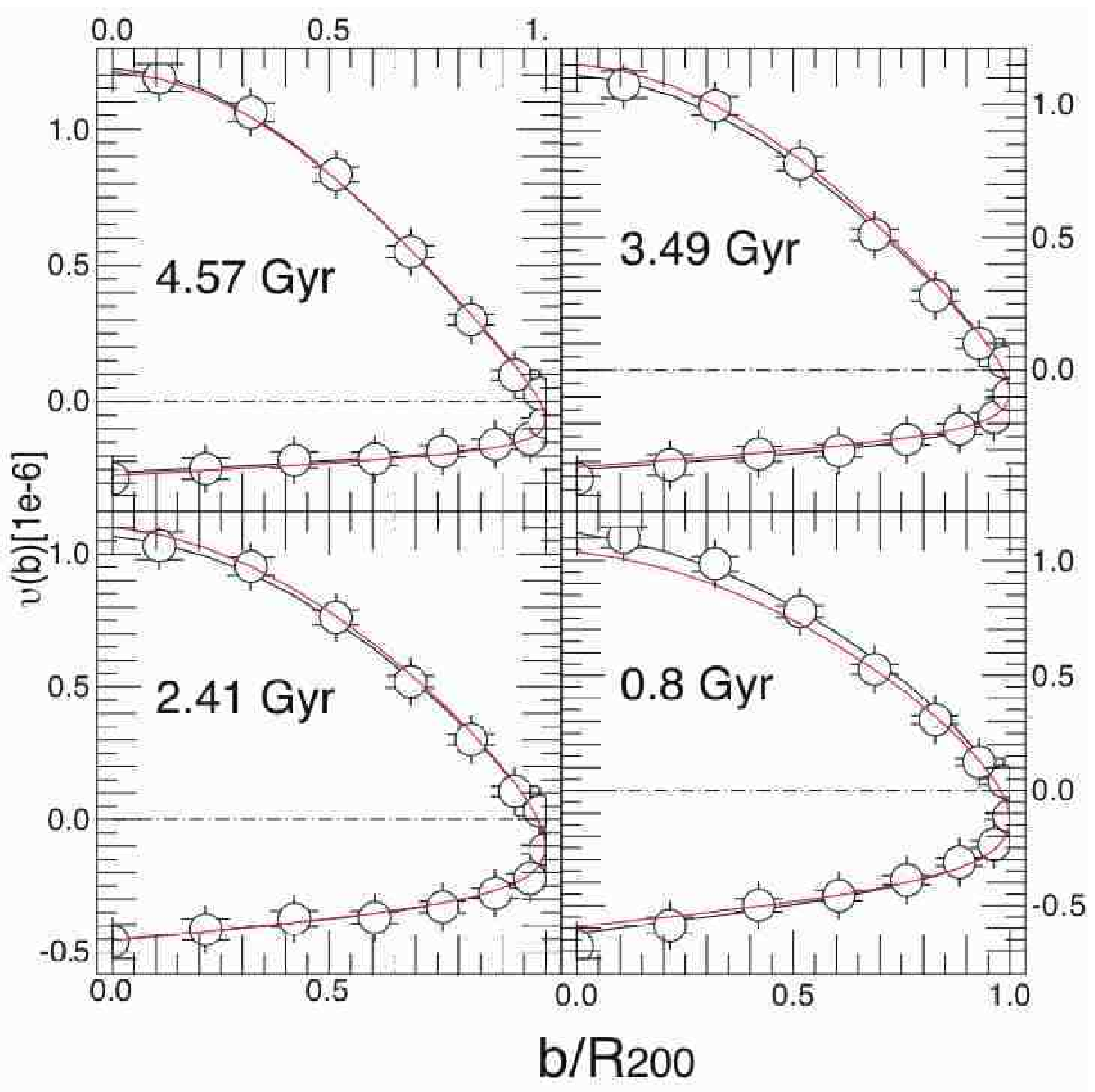}}
\caption{The  time   evolution  (symbols)   of  the  impact   parameter  $b$
distribution  $\langle \vartheta(b,t) \rangle$,  defined by  \eq{defq}. Ages
are expressed  as look-back  time. Bars stand  for $3\sigma$  errors. Y-axis
unit  is   $5\cdot10^9  \mathrm{M}_\odot$/kpc$^2$/Myr.   The   lower  (resp.
higher)  branch is  the  $\langle \vartheta(b,t)  \rangle$ distribution  for
outflows   (resp.   infall).   The   result  of   the  model   described  in
\eq{modelgamma1} is also shown (red line).}
\label{f:evolb} 
\end{figure}

%%%%%%%%%%%%%%%%%%%%%%%%%%%%%%%%%%%%%%%%%%%%%%%%%%
\section{Two point statistics}
%%%%%%%%%%%%%%%%%%%%%%%%%%%%%%%%%%%%%%%%%%%%%%%%%%
\label{s:stat2d}
%%%%%%%%%%%%%%%%%%%%%%%%%%%%%%%%%%%%%%%%%%%%%%%%%%

Let us now focus on  second order statistics, through the correlations on the virial sphere. The two point correlations are assessed through the angular power spectrum and the angulo-temporal correlation function for both the external potential, $\psi^{e}$ and the first moment of the source term, i.e. the flux density of mass, $\varpi_\rho$.

\subsection{External potential}
\label{s:statpot}
\subsubsection{Angular power spectrum}
\begin{figure} 
\centering           
\resizebox{0.8\columnwidth}{0.8\columnwidth}{\includegraphics{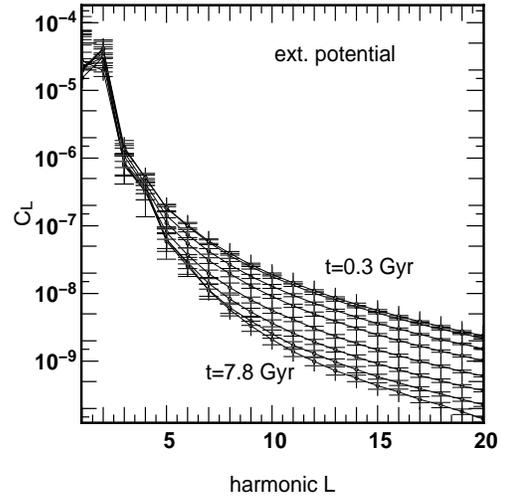}}
\caption{The angular  power spectrum  of the external  gravitational potential
$\psi^e(\bO,t)$.   Symbols  represent  the   \textit{mode}  of   the  $C_\ell^{\psi^e}$
distribution for  each harmonic $\ell$  and each time step. Times  are lookback
times.  Bars stand  for $3-\sigma$ errors on the mode  value. The large scales
contribution remains  constant with time while the  small scales contribution
smoothly increases with time. The bump
for  the  $\langle  C_2\rangle$   component  indicates  a  strong  quadrupolar
configuration for $\psi^e(\bO,t)$.}
\label{f:clpot} 
\end{figure}

The potential's angular power spectrum $C_\ell^{\psi^e}$ is computed for each
halo from the $\tilde b_{\ell,m}$ coefficients (\citet{aubert}):
%\begin{equation}
%C_\ell^{\psi^e}=\frac{1}{4\pi}\frac{1}{2\ell+1}\sum_m |\tilde b_{\ell m}|^2, 
%\end{equation}
%where~:
\begin{equation}
\tilde         b_{\ell,m}\equiv        \sqrt{4\pi}(\frac{b_{\ell,m}}{\langle
b_{00}\rangle}-\delta_{\ell 0}\frac{b_{0,0}}{\langle b_{00}\rangle}),
\label{e:defbtilde}
\end{equation}
related to the potential contrast~:
\begin{equation}
\delta_{[\psi^e]}(\bO)\equiv\frac{\psi^e(\bO) -\overline{\psi^e}}{\langle\overline{\psi^e}\rangle}=\sum_{\ell,m} \tilde b_{\ell,m} Y_{\ell,m}(\bO).
\end{equation}
The probability distribution  of $C_\ell^{\psi^e}(t)$ was weighted as described in appendix \ref{s:harmconv}.
For   each  time step   and   for   each  harmonic   $\ell$,      the
$C_\ell^{\psi^e}$  was fitted by  a  log-normal   distribution (see Fig. \ref{f:hbl}).   Let us   define  $\llangle
C_\ell^{\psi^e}\rrangle  (t)$ as the  mode of  the fitting  distribution.  The
time evolution of  the external potential's power spectrum  is shown in Fig.
\ref{f:clpot}. 
%The  time evolutions  of specific harmonics  are also  given in
%\Fig{clpotevolt}.  
Globally the  power spectrum is dominated by large
scales and is quite insensitive to  time evolution. However, two regime may be
distinguished. For low order harmonics, $\llangle C_\ell^{\psi^e}\rrangle (t)$
remains  mostly   constant.   For   smaller  scales  ($\ell   >5$),  $\llangle
C_\ell^{\psi^e}\rrangle (t)$ increases along  time. As a consequence the power
spectrum's  amplitude does  not change  but  its shape  evolves while  smaller
scales become more important relative to larger scales. 

These two regimes  reflect the twofold nature of tidal  interactions of a halo
with its environment. Small angular variations of the potential relate to small
spatial  scales and  presumably track the  presence  of objects which are
getting closer or going through the virial sphere. Since small scales
contribution increases, it suggests that these objects tend to get smaller
with time. It would be consistent with the global decrease of the accretion
rate, as long as the merger rate does note increase strongly during this
epoch. However, the rise of small scales may also be related to an increasing
contribution of weak and poorly resolved accretion events. In such a case, the isolated particles' contribution to the potential should 
be measured. This possibility is investigated in section \ref{s:cldens}.

%Conversely, we may also conclude that `large' scales (and therefore heavier) potential blobs were more frequent in the past.

Meanwhile, large scale  potential's fluctuations  ($\ell \le 4$) may reflect the `cosmic tidal field' resulting from the distribution of matter around the halo on scales larger than the halo's radius. The amplitude of such a tidal field should remain fairly constant, as indeed measured. Furthermore, the peripheral distribution of matter  is not spherically distributed but is rather elongated along some direction: haloes tend to be triaxial with their ellipsoid  aligned  with  the  surrounding  distribution   of  satellites.  The
intersect of an elongated distribution  of matter with the virial sphere would
induce a quadrupolar  component, as detected in our  measurements. These two effects cannot be easily disentangled, since they actually are two sides of the same effect. Large scale distribution of matter is responsible of both the 'cosmic tidal field' and the halo triaxiality (via the distribution of satellites). In other words, it is not clear whether the large scale behavior of $\llangle C_\ell^{\psi^e}\rrangle (t)$ reflects the tidal field or the halo's reaction to this tidal field.

\subsubsection{Angulo-temporal correlation}
\begin{figure} 
\centering           
\resizebox{0.8\columnwidth}{0.8\columnwidth}{\includegraphics{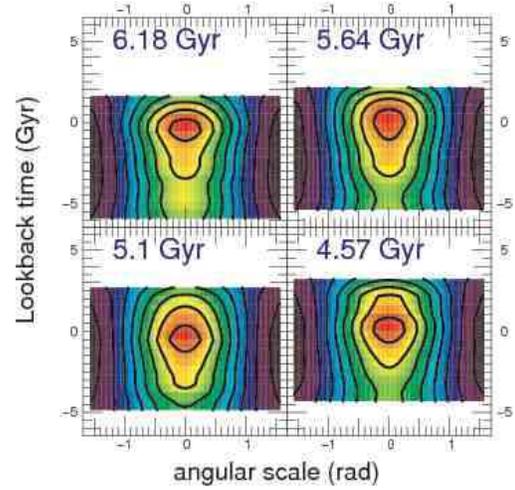}}
%\resizebox{0.8\columnwidth}{0.8\columnwidth}{\includegraphics{Correl_pot_2}}
\caption{The  angulo-temporal  correlation  function,  $w^e(\theta,\Delta
t)=\langle \delta_{[\psi^e]}(\bO,t)\delta_{[\psi^e]}(\bO+\Delta\bO,t+\Delta  t)\rangle$. Blue (resp.
red) colors stand for low (resp.  high) values of the correlation. Isocontours
are also shown. Large angular scale isocontours ($\theta \sim \pi/2$) have
large temporal extent, due to the quadrupole dominance over the potential seen
on the virial sphere.}
\label{f:correlpot} 
\end{figure}

\paragraph{Correlations and coherence time}
To investigate further these two regime of tidal interactions, let us compute the
angulo-temporal  correlation  function of  the  external potential contrast, defined as
\begin{equation}
\llangle
w^e(\theta,t,t+\Delta t)\rrangle=\llangle
\delta_{[\psi^e]}(\bO,t)\delta_{[\psi^e]}(\bO+\Delta\bO,t+\Delta t)\rrangle,
\end{equation}
which     is     related     to
$T_\ell^{\psi^e}(t,t+\Delta t)$
coefficients by:
\begin{equation}
w^e(\theta,t,t+\Delta t)=\sum_\ell
T_\ell^{\psi^e}(t,t+\Delta t)(2\ell+1)P_\ell(\cos(\theta)),
\label{e:corrdef}
\end{equation}
where
\begin{equation}
T_\ell^{\psi^e}(t,t+\Delta t)\equiv\frac{1}{4\pi}\frac{1}{2\ell+1}\sum_m\tilde
b_{\ell m}(t)\tilde b^*_{\ell m}(t+\Delta t).
\end{equation}
Here, $\theta$ stand for the angular distance between two points on the sphere located at $\bO$ and $\bO+\Delta\bO$. The   $T_\ell^{\psi^e}(t,t+\Delta t)$
coefficients were  computed for each  halo and each  pair of time step, for
each harmonic. The $T_\ell^{\psi^e}(t,t+\Delta t)$ distributions were fitted by a
log normal distribution and $\llangle T_\ell^{\psi^e}(t,t+\Delta t)\rrangle$
was deduced from it.  The corresponding $\llangle w^e(\theta,t,t+\Delta t)\rrangle$ are
shown in figure \Fig{correlpot}.  
%The  time   evolution  of   $\llangle
%w^e(0,t_1-t_2)\rrangle$ is also shown in \Fig{correlpotprofil}.

For large angular scales ($>45$ degrees), isocontours remain open during the whole
time range.   Large scales have a long  coherence time ($\sim 5$  Gyrs) and are
consistent  with  a  `cosmic  tidal  field' resulting  from  the  large  scale
distribution of  matter. The  latter is not  expected to  evolve significantly
with time  at our redhsifts and  the halo's triaxiality is also  a fairly constant
feature.  The innermost isocontours  are closed around the measurement time
$t_1$.   Small angular  scales  ($<45$ degrees)  have  shorter coherence  time
($\sim 1.5$ Gyr).   This is consistent with a contribution  to the potential due
to objects where  satellites pass by or  dive into the halo and  apply a tidal
field only for a short period.  

\begin{figure} 
\centering           
\resizebox{0.8\columnwidth}{0.8\columnwidth}{\includegraphics{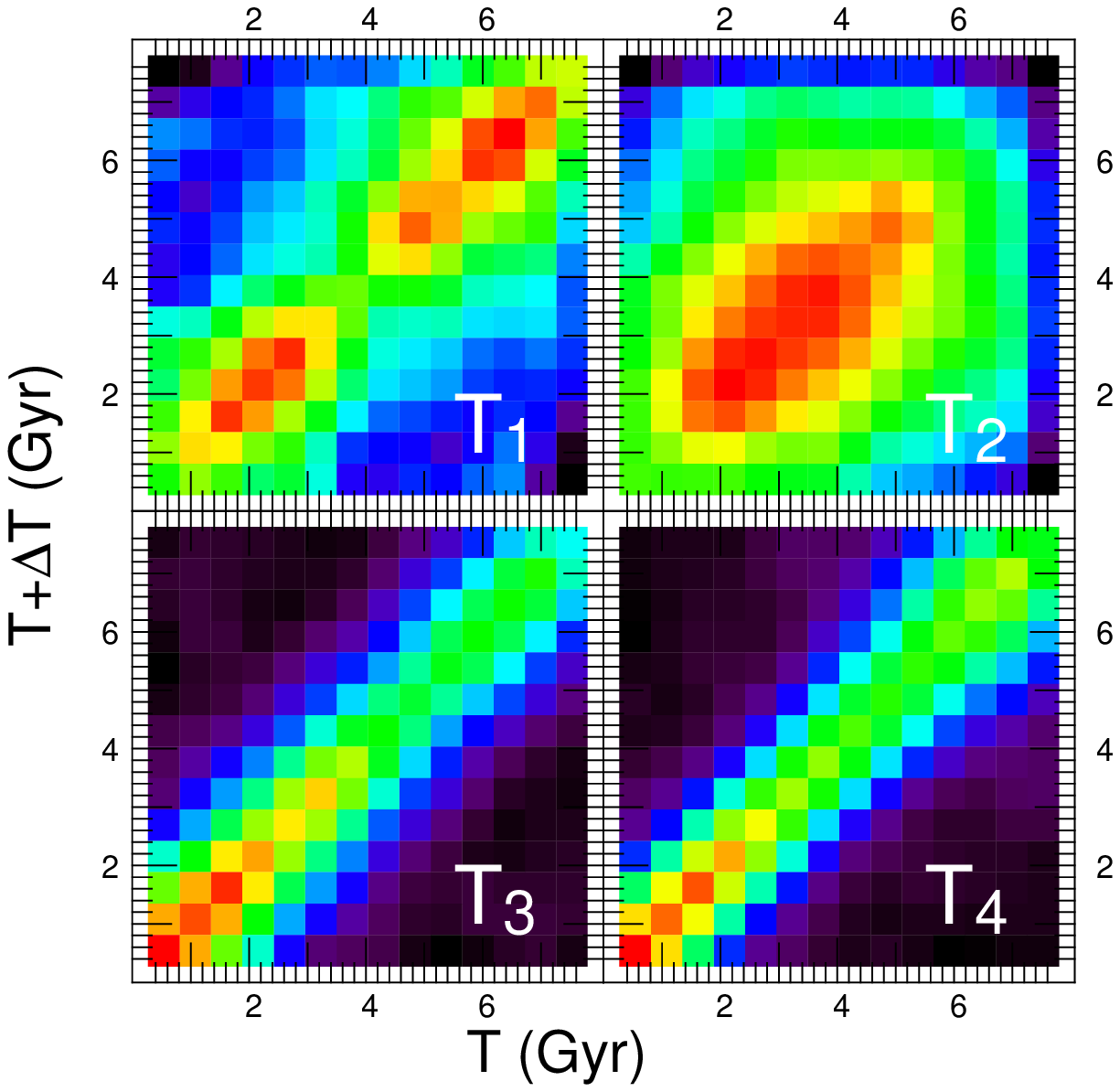}}
\resizebox{0.8\columnwidth}{0.8\columnwidth}{\includegraphics{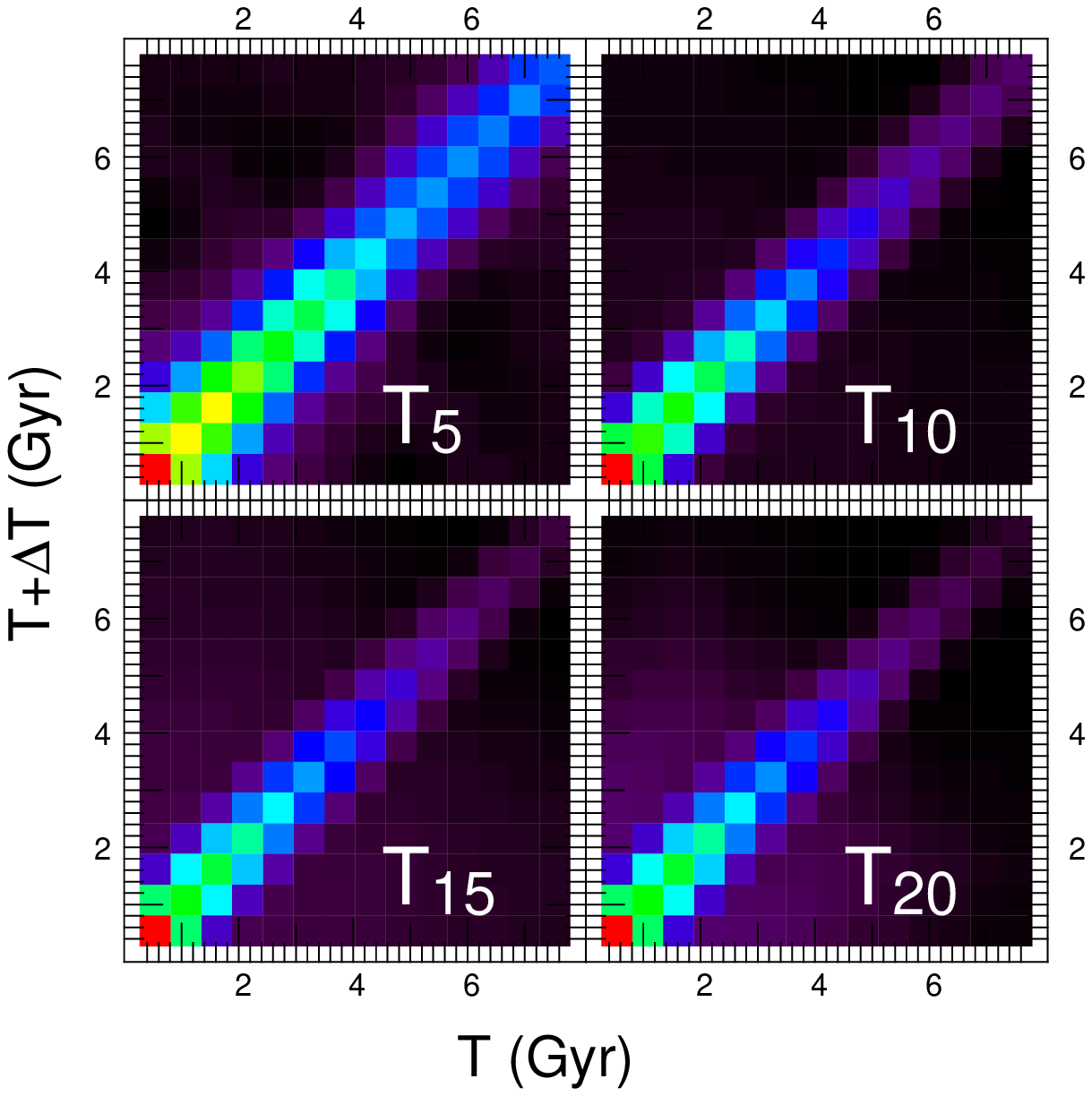}}
%\resizebox{0.8\columnwidth}{0.8\columnwidth}{\includegraphics{Correl_pot_3_theta0_centered}}
\caption{The   time   matrices    of   the   $T_\ell^{\psi^e}(t,t+\Delta   t)$
coefficients. Blue (resp.   red) colors stand for low  (resp.  high) values of
the coefficients.  The diagonal terms are equal to the angular power spectrum,
i.e.    $T_\ell^{\psi^e}(t,t)=C_\ell^{\psi^e}(t)$.   As   can  be   seen  from
\Fig{clpot}  fluctuations observed  for $T_1$  are within  the error  bars.  A
smooth  (resp. sharp)  decrease  of $T_\ell^{\psi^e}(t,t+\Delta  t)$ with  the
distance  to the diagonal  imply a  long (resp.  short) coherence  time. Here,
coherence time decreases with angular scale.}
\label{f:Matrixtell} 
\end{figure}

This difference between large and small scales
can   also    be   investigated   through    the   time   matrices    of   the
$T_\ell^{\psi^e}(t,t+\Delta  t)$   coefficients  (see  \Fig{Matrixtell}). The diagonal  terms describe  the time
evolution         of        the         angular         power        spectrum,
$T_\ell^{\psi^e}(t,t)=C_\ell^{\psi^e}(t).$    A    smooth   (resp.     peaked)
distribution  of values  around the  diagonal indicates  a long  (resp. short)
coherence time.   Clearly different scales have  different characteristic time
scales.   The  non-diagonal  elements  of  the  quadrupole  matrix  ($\ell=2$)
decrease slowly with the distance to the diagonal while the $T_{20}$ matrix is
almost  diagonal. Not surprinsingly, the smaller the  angular  scale, the  smaller the  coherence time: a small 3D object passing through the sphere is likely to have a small angular size on the sphere. 

\begin{figure} 
\centering           
\resizebox{7cm}{7cm}{\includegraphics{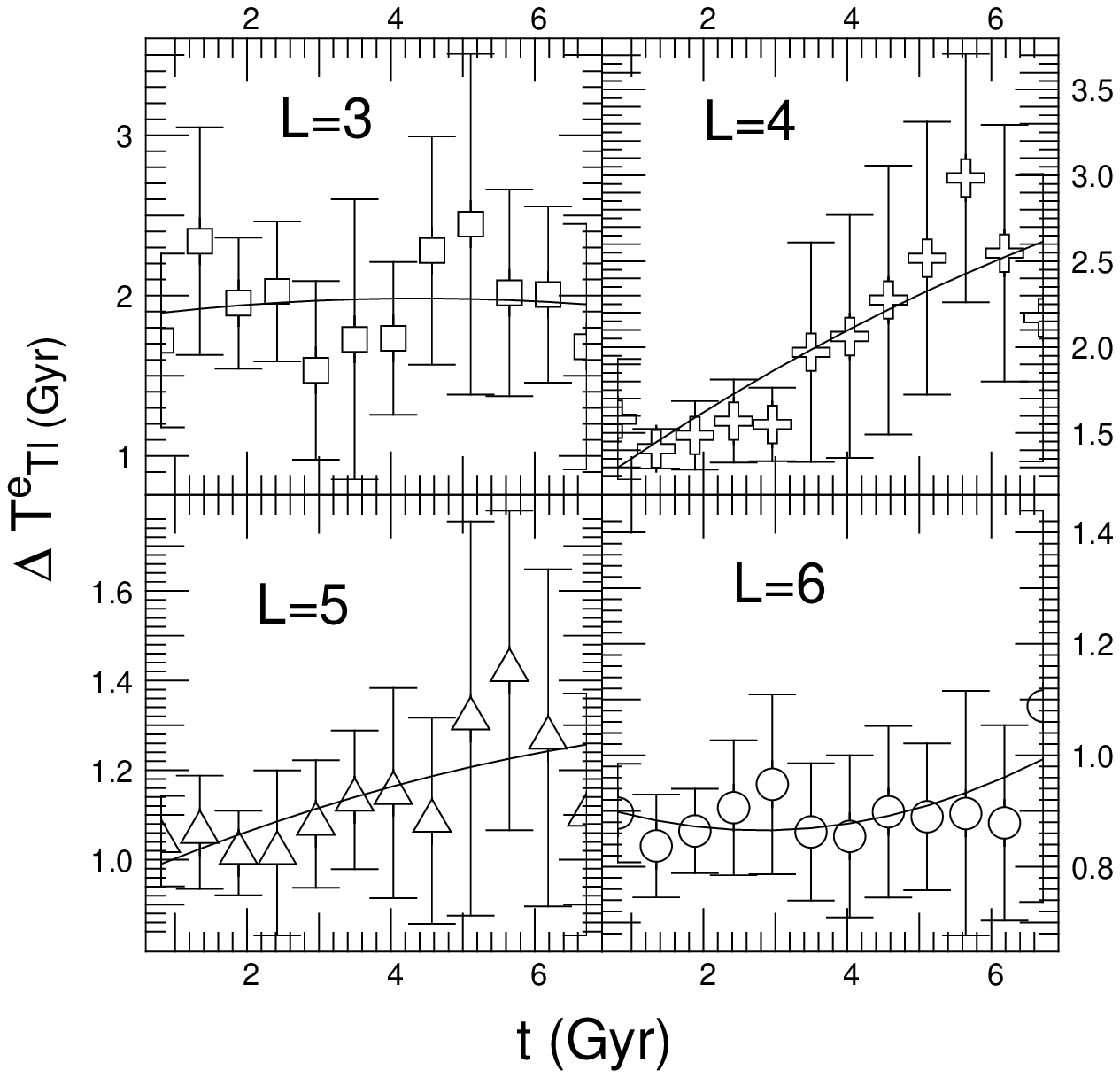}}
\resizebox{7cm}{7cm}{\includegraphics{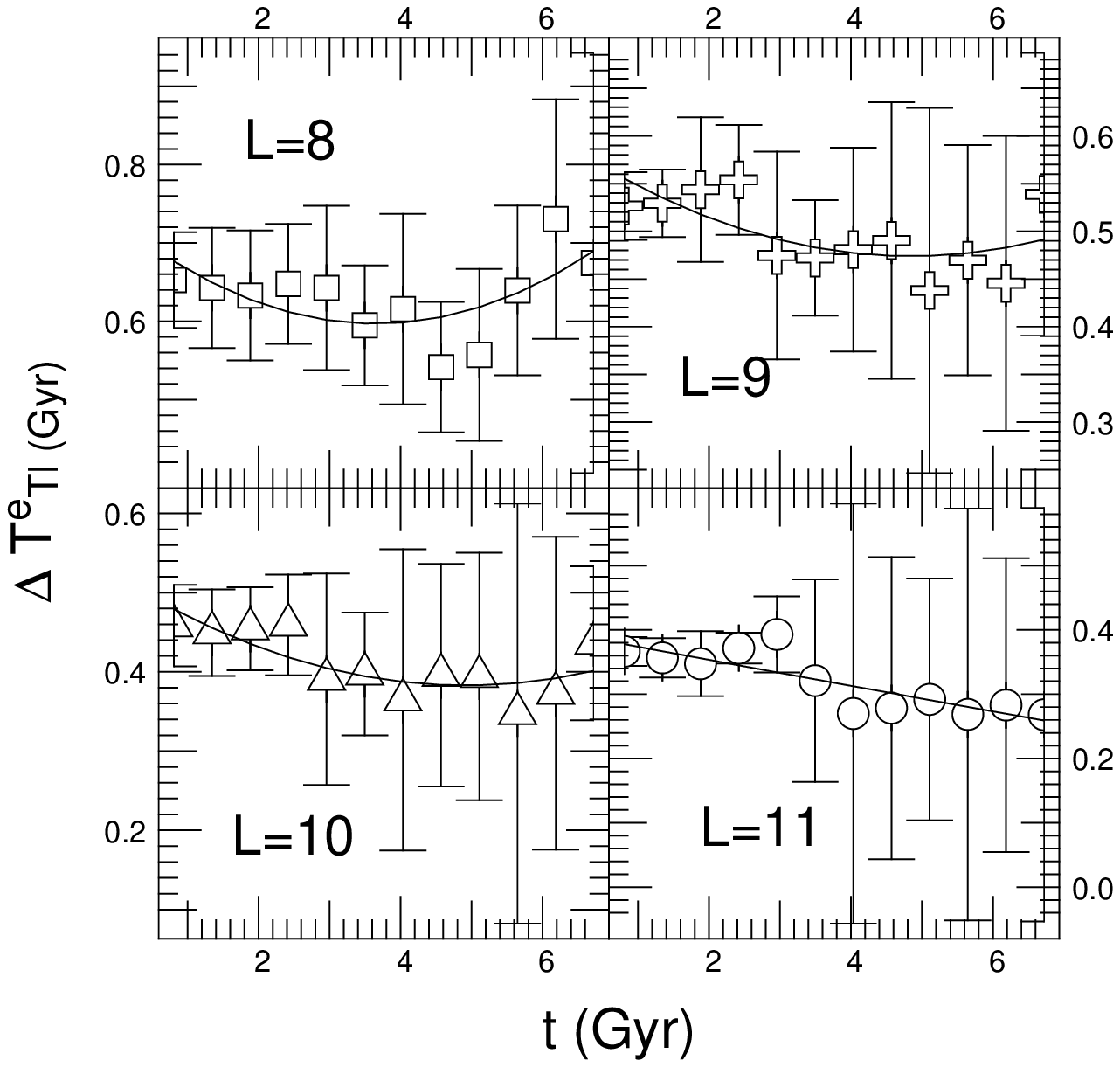}}
\resizebox{7cm}{7cm}{\includegraphics{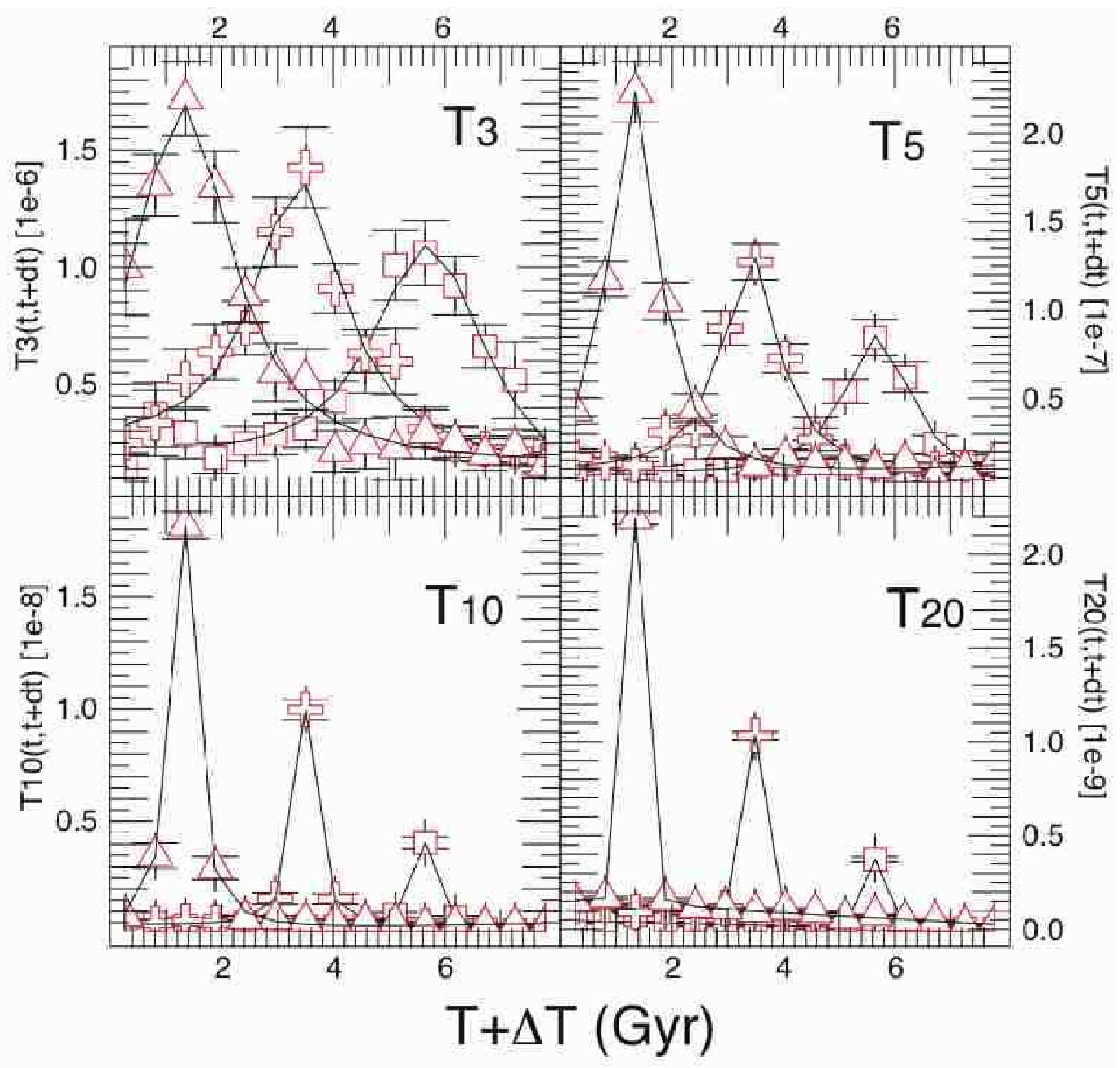}}
%\resizebox{0.8\columnwidth}{0.8\columnwidth}{\includegraphics{Correl_pot_3_theta0_centered}}
\caption{\textit{Top and Middle:} the  time   evolution  of   the  characteristic  time   scale  $\Delta
T_{T_\ell^e}$,  obtained  by   fitting  $T_\ell^{\psi^e}(t,t+\Delta  t)$  with
\eq{fitpottl}. Symbols  are the measurements while bars  stand for $3\sigma$
fitting  error bars.  The  second order  fit  of the  time  evolution of  each
$T_\ell^{\psi^e}$ is also  shown. Except
for the $\ell=4$  and  (marginally)  for the $\ell  =5$   modes,  no  time  evolution is
observed. The time resolution is 0.53 Gyr. \textit{Bottom:} examples of $T_\ell^{\psi^e}(t,t+\Delta  t)$ fitted by Lorentzian functions.}
\label{f:tcaractl} 
\end{figure}

For a  given  $\ell$ and  a  given $t$,  the correlations  coefficients
$T_\ell^{\psi^e}$ can  be fitted by  a Lorentzian
function defined by
\begin{equation}
T_\ell(t,t+\Delta t)=\frac{q_3^{\mathrm{Te}}(t)}{2/\pi}\frac{q_2^{\mathrm{Te}}(t)}{(\Delta
t-q_1^{\mathrm{Te}}(t))^2+(q_2^{\mathrm{Te}}(t)/2)^2}+q_4^{\mathrm{Te}}(t).
\label{e:fitpottl}
\end{equation}
where  the  characteristic  time   scale,  $\Delta  T_{T_\ell^e}$,  is  given  by
$q_2^{\mathrm{Te}}(t)$ and the  reference time $t$ is equal  to $q_1^{\mathrm{Te}}(t)$. Example of
fits are shown if \Fig{tcaractl}. 
%The time evolutions of $\bf q^{\mathrm{Te}}(t)$
%coefficients are fitted by second order polynomials which parameters are given
%in    appendix.    

For example,  the time evolution of $\Delta  T_{T_\ell^e}=q_2^{\mathrm{Te}}(t)$ is given
in  \Fig{tcaractl} for  different $\ell$  values.  Given  the error  bars, the
characteristic  time scales  are constant  over time  (except for  the $\ell=4$
mode). In the prospect  of the regeneration of the potential, the stationarity hypothesis
can then  be considered as valid  for most of the  angular scales.  Meanwhile,
the $\ell=4$ potential fluctuations display a decreasing $\Delta T_{T_\ell^e}$
with time.   The same effect exists  at a $1\sigma$ level  for $\ell=5$.  One
interpretation would  be that satellites achieve higher  velocities along
time: for a given typical size a faster satellite would spend less time to be
accreted  and the associated  potential would  be detected  on a  smaller time
scale. This picture is supported by the results shown in \ref{s:kinematic},
where  the   mean  velocity  of  infalling  material   increases  along  time.
Another possibility would be that $\ell=4$ fluctuations had a longer radial extent in the past. Since there is no reasons for potential fluctuations to have such a property, one could imagine \textit{successive} potential fluctuations which overlapped, leading to an apparent longer radial extent. This possibility is further investigated in the following paragraphs, by comparing the coherence time variation of the potential fluctuations to the evolution of the infall's typical velocity .

\begin{figure} 
\centering           
\resizebox{0.8\columnwidth}{0.8\columnwidth}{\includegraphics{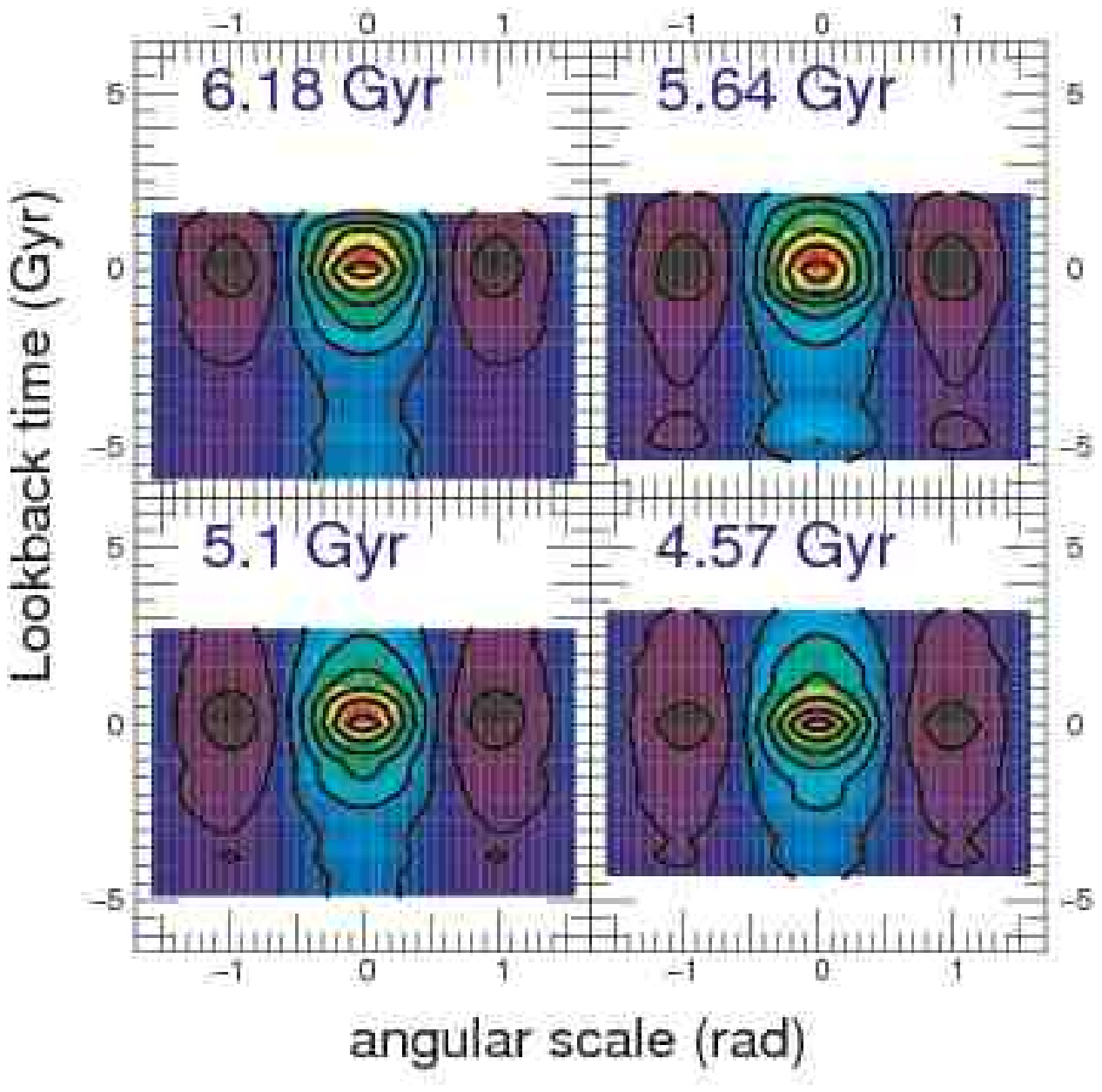}}
\resizebox{0.8\columnwidth}{0.8\columnwidth}{\includegraphics{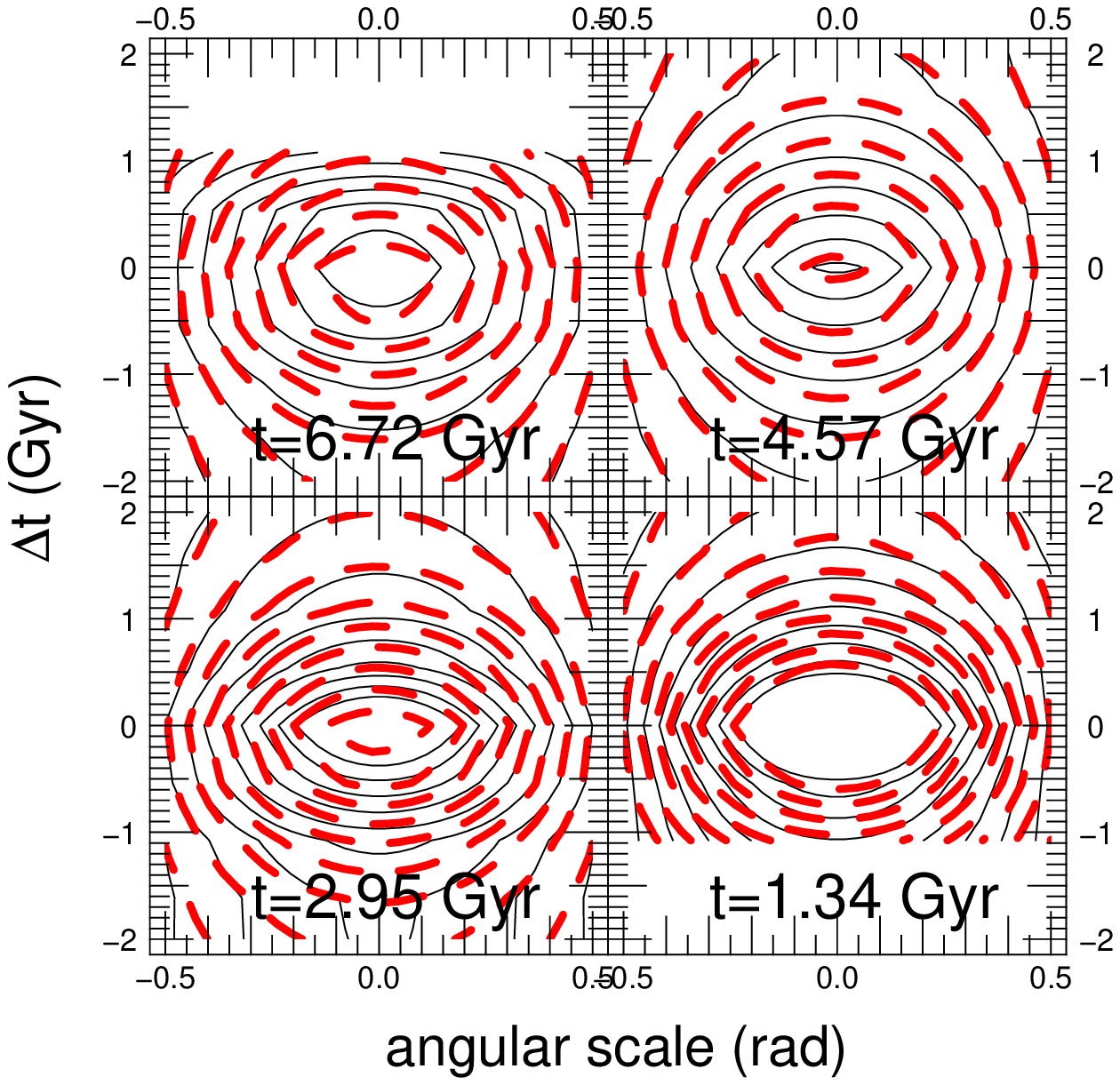}}
%\resizebox{0.8\columnwidth}{0.8\columnwidth}{\includegraphics{Correl_pot_nodipole_2}}
\caption{\textit{Top:} the  angulo-temporal  correlation  function, $w^e(\theta,\Delta
t)=\langle  \delta_{[\psi^e]}(\bO,t)\delta_{[\psi^e]}(\bO+\Delta\bO,t+\Delta t)\rangle$.  The dipole
($\ell  =1$)  and the  quadrupole  ($\ell=2$)  components  were removed.  Blue
(resp.   red)   colors   stand   for   low  (resp.   high)   values   of   the
correlation. Isocontours are also shown. The main axes of the `ellipses' centered on ($\Delta
\bO=0$, $\Delta t=0$)  give indications on the characteristic  time and angular
scales of $\psi^e(\bO,t)$. \textit{Bottom:} comparison between the
measured 2D  correlation function (\textit{solid lines}) and  the fit obtained
using  the \eq{sine} (\textit{dashed   lines}).}
\label{f:correlpotnodipole} 
\end{figure}

\begin{figure} 
\centering           
%\resizebox{0.8\columnwidth}{0.8\columnwidth}{\includegraphics{Correl_pot_theta0_centered_resol_caractime}}
\resizebox{0.8\columnwidth}{0.8\columnwidth}{\includegraphics{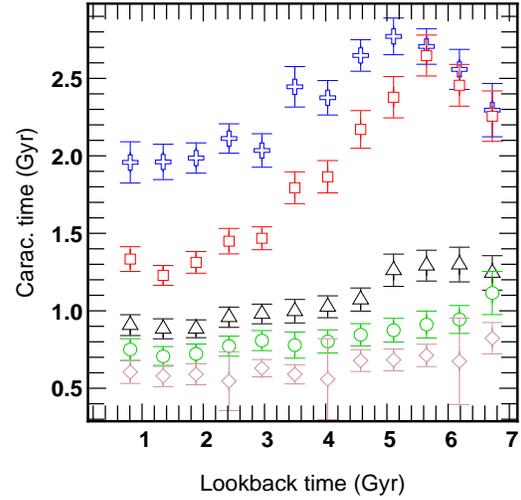}}
%\resizebox{0.8\columnwidth}{0.8\columnwidth}{\includegraphics{Ctimepotsuperp}}
\caption{{The cosmic evolution of the potential's coherence time. This characteristic
time  scale  is  obtained  by  fitting the 2D correlation  function  $w^e(\theta,\Delta t)$ with the function given in \eq{sine}.  The correlation function is
computed using  harmonics coefficients with $\ell  \ge 3$ (\textit{crosses}),
$\ell \ge 4$ (\textit{squares}), $\ell \ge 5$ (\textit{triangle}), $\ell \ge 6$
(\textit{circle})  and  $\ell  \ge  7$  (\textit{diamonds}).  Bars  stand  for
$3\sigma$ fitting errors. The time resolution is 0.53 Gyr.}} 
\label{f:correlpotnodipoleprofilT} 
\end{figure}

\paragraph{Correlations without the dipole and the quadrupole}
In  order to focus  on the  coherence time  of small  angular scales,    the   correlation   function  $\llangle   w^e(\theta,t,\Delta
t)\rrangle$  was also
computed without  the  dipole  ($\ell=1$) and  the  quadrupole  ($\ell=2$)
component  of  the potential  (see  also  \eq{corrdef}).  The  angulo-temporal
correlation  function   is  shown  in   \Fig{correlpotnodipole}.   Again,  the
isocontours of  the correlation  function are closed  around $\Delta  t=0$. This
shows that the  potential on the sphere have a  finite coherence time.  In contrast to 
coherence  times  measured  on  the $T_\ell^{\psi^e}$  coefficients,  all  the
angular scales are mixed and the typical time scales are those of structures
as they are `really' seen from a halocentric point-of view where 'potential blobs'
appear  and disappear on  the sphere.   To evaluate  the related  typical time
scale  $\Delta  T_{\psi^e}$,  $\llangle w^e(\Delta  \Omega,t,\Delta
t)\rrangle$ was  fitted  with a 2D function for different values of $t$.  The model used
is given by~:
\begin{eqnarray}
\llangle w^e(\theta,t,\Delta t)\rrangle&=&
q_6^{\mathrm{we}}(t)+\frac{q_5^{\mathrm{we}}(t)}{2\pi}\frac{q_4^{\mathrm{we}}(t)}{(\Delta t-q_3^{\mathrm{we}}(t))^2+(\frac{q_4^{\mathrm{we}}(t)}{2})^2}\nonumber\\&&\times\frac{\sin(2\pi(\Delta\Omega-q_1^{\mathrm{we}}(t))/q_2^{\mathrm{we}}(t))}{\Delta
\Omega},
\label{e:sine}
\end{eqnarray}
where the angular  dependence is fitted by a cardinal  sine function while the
time  dependence is  fitted  by a  Lorentzian  function.  Examples  of 2D  fits  are shown in \Fig{correlpotnodipole}.  
The  correlation function was also computed  using only  harmonics with  $\ell \ge  4,5,6,7$ and the   same  fitting  procedure is applied.   The  evolution   of  the  resulting
characteristic  time scales $\Delta  T_{w^e}=q_4^{\mathrm{we}}(t)$ is  shown in
\Fig{correlpotnodipoleprofilT}.   Bars stand for  $3\sigma$ fitting  errors. 

Note that $\Delta T_{w^e}$ tends  to decrease with time for every truncation
order. The $\ell \ge 3$ and  $\ell \ge 4$ correlation function displays a rise
of $\Delta T_{w^e}$ before it drops  to lower values.  Furthermore $\Delta T_{w^e}$ tend to decreases with $\ell_\R{min}$, suggesting that the $\ell_\R{min}$ contribution  dominates each $w^e$ reconstruction.  Correlation functions with $\ell_\R{min} \ge 5$ show marginal $\Delta T_{w^e}$ variation but  recall that
our  time resolution  is 0.53  Gyr, hence  any fluctuations  on  smaller scales
should be taken in caution. 
Still, the
0.81 Gyr variation observed for $\ell  \ge 3$ between $t=5.1$ Gyrs and $t=0.8$
Gyr is  significant and  so is the  variation observed  for $\ell \ge  4$ (1.3
Gyr).   

\paragraph{A longer coherence length}
As mentioned above,  the variation of the characteristic time  scale can be explained by the measured increase in mean velocity.  Conversely, the decrease of coherence
time may be the consequence of  smaller potential blobs as time passes: a `large'
(three  dimensional)  potential  takes  longer  to disappear  than  a  smaller
one. One crude approximation could be~:
\begin{equation}
\frac{L_1}{L_2}=\frac{V_1 \Delta T_1}{V_2 \Delta T_2}.
\EQN{simple}
\end{equation}
where $L$  is the radial size of  the potential blob, $V$  its radial velocity
and  $\Delta T$  its coherence  time. It is assumed  that the  $\ell  \ge 4$
truncation is representative of the  potential due to infalling objects, i.e.
$\Delta  T_1/\Delta  T_2\sim1.95$.   Let us  also consider  that  the  radial
velocity  variation is  equal to  the one  measured for  the mean  velocity of
infall  (  see section  \ref{s:kinematic}):  $V_1/V_2=0.77$.   Following \eq{simple}, it suggests  that
$L_1\sim1.5L_2$, i.e.  the  radial size decreases with time. The  same calculation
with $\ell \ge 3$ leads to $L_1\sim1.1L_2$~: the results remain
qualitatively the same. In other words,
the \textit{coherence length} was longer in the past and the velocity variation cannot explain the variation of coherence time. The only other way to explain a longer coherence length involves potential blobs falling successively through the sphere, coming from roughly the same direction. To induce a decreasing coherence time, these blobs would have to be either bigger before or either more numerous. Such a crude picture is coherent with the measured decrease of accretion with time and the anisotropic nature of accretion by haloes (see e.g. \citet{Knebe}, \citet{aubert}, \citet{zentner}).

%%%%%%%%%%%%%%%%%%%%%%%%%%%%%%%%%%%%%%%%%%%%%%%%%%
\subsection{Flux  density of mass~: $\varpi_\rho\equiv \rho v_r$}
%%%%%%%%%%%%%%%%%%%%%%%%%%%%%%%%%%%%%%%%%%%%%%%%%%
\label{s:cldens}
The mode $\llangle C_\ell^{\varpi_\rho}  \rrangle$ of the distribution of the $\varpi_\rho$ angular power spectrum  is computed using  \Eqs{c2a}{defcl}. In order to deal
with  adimensional quantities,  the reduced  harmonic  coefficients, $\tilde
a_{\ell,m}$, are defined as:
\begin{equation} 
\tilde         a_{\ell,m}\equiv        \sqrt{4\pi}(\frac{a_{\ell,m}}{\langle
a_{00}\rangle}-\delta_{\ell 0}\frac{a_{0,0}}{\langle a_{00}\rangle}).
\label{e:defatilde}
\end{equation}
The accretion contrast, $\delta_{[\varpi_\rho]}$, and the $\tilde a_{\ell,m}$
coefficients are linked by:
\begin{equation}
\delta_{[\varpi_\rho]}(\bO)\equiv\frac{\varpi_\rho(\bO) -\overline{\varpi_\rho}}{\langle\overline{\varpi_\rho}\rangle}=\sum_{\ell,m} \tilde a_{\ell,m} Y_{\ell,m}(\bO).
\end{equation}

\subsubsection{Angular power spectrum}
Given $ \langle a_{00}\rangle(t)$, the angular power spectrum $C_\ell^{\varpi_\rho}(t)$ is
computed for each halo. At each  time step and for each harmonic order $\ell$,
the $C_\ell^{\varpi_\rho}(t)$  distribution was fitted by a log-normal
distribution (see Fig. \ref{f:cl2}). The probability distribution  of $C_\ell^{\varpi_\rho}(t)$ is  weighted as described in appendix \ref{s:harmconv}.

The  evolution of $\llangle C_\ell^{\varpi_\rho}(t) \rrangle$  with time is
shown  in \Fig{clt}.   The shape  of $\llangle  C_\ell^{\varpi_\rho}(t)  \rrangle$ remains
mostly the same with time and is fitted by a simple function~:
\begin{equation}       
\llangle C_\ell^{\varpi_\rho} \rrangle (t) = q^{\varpi_\rho}_1(t)+\frac{q^{\varpi_\rho}_2(t)}{(\ell+q^{\varpi_\rho}_3(t))^2}.
\label{e:modelcl}
\end{equation}
The time evolutions  of $q^{\varpi_\rho}_1$, $q^{\varpi_\rho}_2$ and
$q^{\varpi_\rho}_3$  are shown in \Fig{clcoeffit} and 
can be fitted by decreasing exponential:
\begin{equation}
q^{\varpi_\rho}(t)=h+k\exp\left({-\frac{t}{u^2}}\right)\,.
\label{e:decexp}
\end{equation}
Only  the  dipole  ($\ell=1$)  harmonic  does  not  fit  with  the  previous
functional  form  and  is  systematically  lower than  contribution of the  other  harmonics.  If  particles  velocities   were  measured  in  an  absolute
referential,  the dipole  strength would  reflect the  halo's motion  in the
surrounding matter.   Also strong mergers may  cover a 180  degrees angle on
the  sphere  and would  contribute  to  the  dipole.  Since  velocities  are
measured  in the  halo's rest  frame and  strong mergers  are  excluded, the
dipole strength is substantially lowered, as is  measured.

   The  values  for  $h,k,u$  are  given  in  table
\ref{t:coeffexp}.  The   offset  $q^{\varpi_\rho}_1$  of   $\llangle  C_\ell^{\varpi_\rho}(t)  \rrangle$
increases  with  time.  From  \eq{defatilde},  one  can  see that  $\llangle
C_\ell^{\varpi_\rho}(t) \rrangle$ is proportional to the square of the accretion contrast.
If  the power spectrum experiences  a global  shift  toward higher
values  with time  it implies  that  the accretion  contrast increases  with
time.  Since the  average  velocity does  not  vary strongly  with time,  this
suggests that objects are getting denser with time. This effect
is similar to the global increase  of the 3D power spectrum $P(k)$ with time
due to density growth. Also, the $q^{\varpi_\rho}_2$ coefficient is found to evolve as $q^{\varpi_\rho}_1$. This illustrates the
fact  that  the amplitude of  $\llangle  C_\ell^{\varpi_\rho}(t)  \rrangle$   remains  mainly
constant.   The $q^{\varpi_\rho}_3$  coefficient should  be seen  as a  typical  scale and
varies slightly from $q^{\varpi_\rho}_3=6$ at $z=1$ to $q^{\varpi_\rho}_3=11$ at $z=0.1$.  The $\llangle
C_\ell^{\varpi_\rho}(t) \rrangle$ becomes marginally `flatter' as time passes, implying that
small  scales   contribute  more  to  the  spatial   distribution  of  $\varpi_\rho(\bO,t)$, consistently with the evolution of $\llangle  C_\ell(t)^{\psi^e}  \rrangle$.  
The flat power spectrum measured for $\varpi_\rho$ on small scales suggests that isolated particles contribute significantly and increasingly with time. In other words, the accretion becomes low enough to be poorly resolved in terms of particles. 

\subsubsection{Resolution in mass and particle number}
\label{s:reso}
\begin{figure} 
\centering           
\resizebox{0.8\columnwidth}{0.8\columnwidth}{\includegraphics{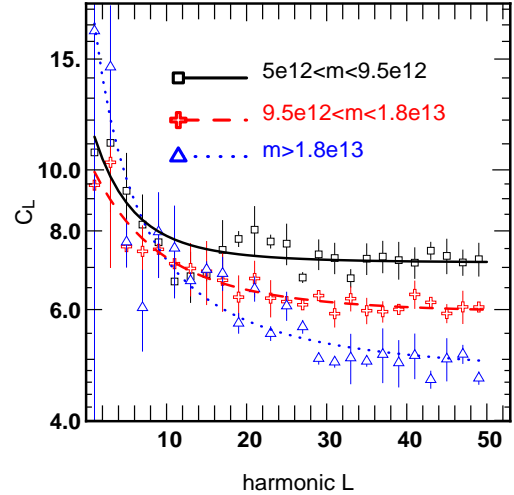}}
\caption{The average angular  power spectrum, $\llangle C_\ell^{\varpi_\rho}\rrangle (t)$,
at $t=0.8$  Gyr (symbols)  for three different  class of  masses.  $\llangle
C_\ell^{\varpi_\rho}\rrangle (t)$ is taken as the  mode of the log-normal function used to
fit the $C_\ell^{\varpi_\rho}$  distribution.  Bars stand for $3\sigma$  errors.  Mass are
expressed  in solar  masses. For  a given  $\ell$ the  corresponding angular
scale is $\pi/\ell$.   The three measurements are fitted  by \eq{modelcl} (solid
line). The power spectrum gets  flatter for small halos. Accretion by small
halos is dominated by small objects or even isolated particles.}
\label{f:flat} 
\end{figure}

\begin{figure} 
\centering           
\resizebox{0.8\columnwidth}{0.8\columnwidth}{\includegraphics{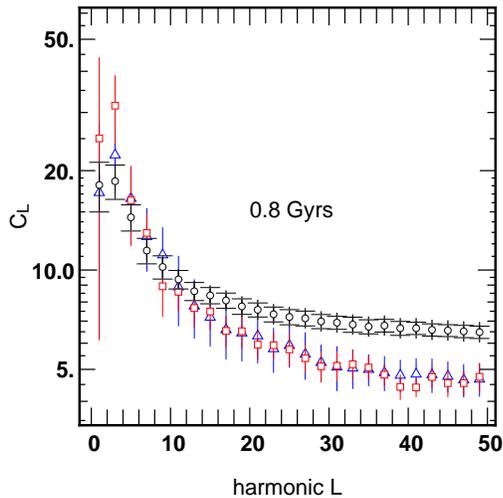}}
\caption{The average angular  power spectrum, $\llangle C_\ell^{\varpi_\rho}\rrangle (t)$,
at $t=0.8$  Gyr for three different  mass resolution in the simulations.  $\llangle
C_\ell^{\varpi_\rho}\rrangle (t)$ is taken as the  mode of the log-normal function used to
fit the $C_\ell^{\varpi_\rho}$  distribution. For  a given  $\ell$ the  corresponding angular
scale is $\pi/\ell$. Circles stand for the measurements performed of the main set of simulations  
(50 Mpc/h, $128^3$ particles) while error bars stand for $3-\sigma$ errors. Square and triangles stand for measurments performed on simulation with higher mass resolution, resp. 50 Mpc/h-$256^3$ particles (1532 haloes analysed) and 20 Mpc/h-$256^3$ particles (545 haloes analysed). In these two cases, error bars stand for 1$\sigma$ errors.
}
\label{f:conv} 
\end{figure}

In order to assess these environments/resolution effects,  $\llangle
C_\ell^{\varpi_\rho}(t)  \rrangle$ was computed for  three   different classes of  mass  (see  Fig. 
\ref{f:flat}) at  a look-back time of  800 Myrs.  For the  heaviest halos, the
power spectrum is peaked toward  low $\ell$ values.  The contribution of large
scales is quite important. For  smaller masses the power spectrum gets flatter
and all  scales almost contribute equally  for the lightest class  of mass.
Recall  that  the harmonic  decomposition  of  a  Dirac function  leads to
$C_{\ell}=\mathrm{constant}$, thus a  flat power  spectrum indicates  that isolated
particles  contribute significantly  to  the distribution  of  matter on  the
sphere.   The relative  behavior  of the  three  $\llangle  C_\ell^{\varpi_\rho}(t) \rrangle$
confirms that larger halos still  experience important mergers (i.e. on large
scales)  while  small  ones  are  in  quiet  environments  at  our  simulation
resolution.  The effect of the mass resolution  on the angular structure of accretion was  also investigated with two smaller sets of simulations : the first one involves 10 simulations with $256^3$ particles in 50 Mpc/h boxes and the second in 5 simulations with the $256^3$ particles in 20 Mpc/h boxes. The related 
 $\llangle C_\ell^{\varpi_\rho} \rrangle$ mesured at a lookback time of 800
Myrs are shown in Fig. \ref{f:conv}. Here 1532 and 545 haloes satisfying the
conditions described in $\ref{s:selection}$ were detected in these two
additional sets of simulations. For clarity, $1\sigma$ error bars are shown
for the two high resolution measurments while the 'larger statistics' power
spectrum is still represented with $3\sigma$ bars. For large scales ($\ell
<10$), the three power spectra are consistent, thus suggesting that
convergence was achieved there. On smaller scales, the two higher resolution
spectra differ significantly from the one measured using the other set of simulations (50 Mpc/$128^3$ particles)~: high $\ell$ hold significantly less power. This confirms that the lack of resolution tend to overestimates the importance of small scales and implies that 
 the study of $\varpi_\rho$ requires simulations at higher resolution in order to understand e.g. the detailed statistics of small infalling objects. Interestingly, the two higher resolution simulations have identical $\llangle C_\ell^{\varpi_\rho} \rrangle$, given the admittedly large error bars. This suggests that statistical convergence at scales $\ell <50$ does not requires extremely resolved simulations  and simulation boxes with a mass resolution only 8-10 times greater than those used in this paper should  suffice.

\begin{figure} 
\centering           
\resizebox{0.8\columnwidth}{0.8\columnwidth}{\includegraphics{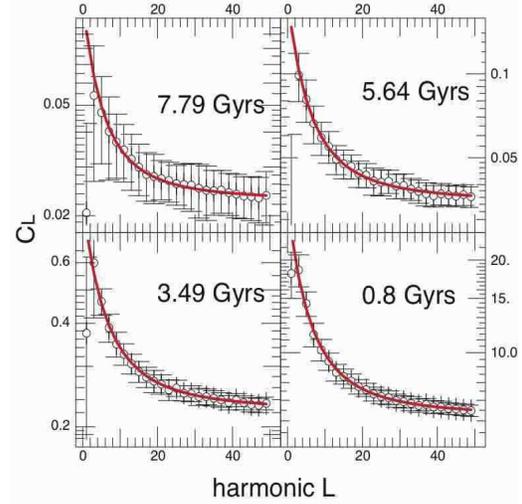}}
\caption{The average angular  power spectrum, $\llangle C_\ell^{\varpi_\rho}\rrangle (t)$,
at  $t=0.8, 3.5, 5.7,  7.8$ Gyr  (symbols). $\llangle  C_\ell^{\varpi_\rho}\rrangle(t)$ is
taken  as the  mode of  the  log-normal function  used to  fit the  $C_\ell$
distribution.   Bars stand  for $3\sigma$  errors.  For a  given $\ell$  the
corresponding angular scale  is $\pi/\ell$. $\llangle C_\ell^{\varpi_\rho}\rrangle(t)$ may
be fitted by a generic model given by \eq{modelcl} (solid line).}
\label{f:clt} 
\end{figure}

Finally, it clearly appears from \Fig{clt} and \Fig{clcoeffit} that the angulo-temporal correlation function related to $\llangle C_\ell^{\varpi_\rho}\rrangle (t)$ are dominated by the overall shift of the angular power spectrum towards higher values and consequently no coherence time should be detectable. It is  shown in appendix \ref{s:alter} how the evolution of $\llangle C_\ell^{\varpi_\rho}\rrangle (t)$ is related to mass biases and possible resolution effects. The previous measurements on the potential were not sensitive to these effects because of the smoother nature of the field. 

In appendix \ref{s:alter},  the measured secular evolution is avoided by using an alternative definition of  $\llangle C_\ell^{\varpi_\rho}\rrangle (t)$. It is  found that the angular power spectrum of $\varpi_\rho$ can be fitted by a simple power-law (see Fig. \ref{f:cltnonorm}) at every time :
\begin{equation}
\llangle C_\ell^{'\varpi_\rho}\rrangle (t)\sim\ell^{-1.15}.
\end{equation}
The corresponding angulo-temporal correlation function is given in Fig.  \ref{f:correldens}. As expected, there is a shorter coherence time for the $\varpi_\rho$  field than for the potential, because of the `sharper' nature of the former. Overall, these results strongly suggest that class of mass and resolution biases should be systematically investigated beyond what was shown here.
\section{A scenario for the accretion of a typical halo at $z\le1$ }\label{s:scenario}
Let us draw  a summary of the previous results, in order to get a synthetic picture of the flux properties at the virial radius.  A typical halo in our  sample has a mass of $10^{13} \R{M}_\odot$ and a radius $R_{200} \sim 500 \R{kpc}$ at z=0. It is embedded in a quasi-stationary gravitational potential, $\psi^e$. Such a potential is highly quadrupolar and is likely to be induced by the large scale distribution of matter around the halo. 
The halo accretes material between $z=1$ and $z=0$ at a rate which declines with time. At high redshift, only accretion is detected at $R_{200}$. It is mainly radial and occurs at a velocity close to $75\%$ times the circular velocity. As time advances, accretion of new material decreases, while outflows become significant.
 Outflows occur at lower velocities and on more circular orbits. A fraction of the outflowing component is due to a backsplash population made of material which already passed through the virial sphere. Another fraction of outflows corresponds to 'virialized' material of the halo which goes further than $R_{200}$ and is being 'cancelled' by its infalling counterpart.   
The clustering on the sphere of the gravitational potential drifts toward smaller scales, while the clustering of matter follows marginally the same trend at our level of resolution. It reflects the increasing contribution with time of weak/diffuse accretion, poorly sampled at our resolution.  In parallel, the coherence time of potential fluctuations is found to be decreasing with time by the halocentric observer. This decrease may be related with the accretion of satellites, where objects were numerous enough to `overlap' in the past,
 which implies that accretion occurs mainly in the same direction on the virial sphere.

This scenario seems consistent  with most of the past studies made on the subject using the full 3D information contained in simulations. The decline of accretion rate has been already measured by e.g.\citet{VDB2002} even though our measurements are in a slight quantitative disagreement. The `rebound' of matter through the Virial sphere has already been measured by e.g. \citet{Mamon} or \citet{Gill}. Furthermore,  the velocity bimodality is recovered together with the circularisation of orbits measured by  \citet{Gill} in high resolution simulations. Finally,  the variation of the potential's coherence time
is found to be related to the anisotropy of accretion, already demonstrated by e.g. \citet{Knebe}, \citet{aubert} or \citet{zentner}.

%%%%%%%%%%%%%%%%%%%%%%%%%%%%%%%%%%%%%%%%%%%%%%%%%%%%%%%%%%%%%%%
%%%%%%%%%%%%%%%%%%%%%%%%%%%%%%%%%%%%%%%%%%%%%%%%%%%%%%%%%%%%%%%
\section{Summary \& Discussion }
\label{s:conclusion}
%%\subsection{Summary}

This paper, 
 the second in the series of three,
  presents measurements of 
  the  detailed statistical properties of dark matter flows 
 on small  scales ($\le 500$ kpc) in  the near environment of  haloes using a
 large set  of  $\Lambda$CDM cosmological simulations.  
 The  purpose of this investigation was  twofold:  (i)
 characterize statistically (via one  and two points statistics) the detailed
 (angular and kinematic)  incoming fluxes of dark matter  entering the virial
 sphere of a biased (described in \Sec{selection})  sample of halos undergoing minor
 mergers for the broader interest of astronomers concerned with the environment
 of galaxies; 
% (and in particular measure  the statistical properties of the first
% three  moments,  $\varpi_\rho$, (this paper)  $\varpi_{\rho  \M{v}}$,  and  $\varpi_{\rho
% \sigma  \sigma}  $ (paper II)); 
    (ii) compute  the  first  two  moments of  the  linear
 coefficients,  $c_\M{n}$,  (resp.   $b_\M{n}$)  of the  source  term  (resp.
 external  potential)  entering \Eqs{linresp}{defcor} (paper~I).  
 
  We concentrated on  flows at the  haloes' virial
 radius while describing the infalling  matter via fluxes densities through a spherical
 shell. In parallel, we measured the statistical properties   of the tidal potential 
 reprojected back onto the boundary.
The  statistical one  and two point  expectations of the  inflow were
tabulated both kinematically and  angularly on the $R_{200}$ virtual sphere.
  All
 measurements were carried for 15000  haloes undergoing minor (as defined)
 mergers  between redshift  z=1  and  z=0.  The  two points  correlations are
 carried both angularly  and temporally for the flux  densities and the tidal field.
 We  also provided  a
 method  to re-generate  realization of  the field,  via  \Eq{sourceexpr} and
 \Ep{defPPi}).

%The current paper focuses on putting statistical constrains on the environments properties. 
%We provided a method to constrain the statistical properties of environments which relies on measuring the source term which feeds haloes and the surrounding tidal field.
% The method is halocentric, where the potential and the incoming fluxes of dark matter  are measured on the virial sphere. 
 
% We have shown how such a representation can be used as an entry to a perturbative theory of the dynamics of open haloes.  
We briefly demonstrated how a perturbative description of the dynamics of halos 
 can propagate the statistical properties of environments down to the statistical properties of the halo's response. 
 The description of the environment involved the projection of the potential and the source on a basis of functions. This basis allowed us to decouple the time evolution from the angular and velocity dependance of these two quantities. Hence, the accretion and the tidal potential were completely described by the  projection coefficients and their statistical properties, which depend on time only.  We also discussed how the  flux densities of matter, momentum and energy can be related to the source and its expansion coefficients. 
 We restricted ourselves to the one and two statistical description of the tidal field $\psi^e$ and the  flux density of mass $\varpi_\rho$ and postponed to paper III (\citet{aubert2}),  the full description of 
 the higher moments. Since these measurements will be used as an entry to a perturbative description of the inner dynamics  of haloes, only objects with quiet accretion history were selected as discussed in \Sec{stat}.
 Throughout this biased sample of haloes we made statistical measurements of the kinematic properties of accretion and derived results on the following quantities~:
 \begin{itemize}
\item the evolution  of the accretion rate at the virial sphere: the net accretion is found to decrease with time, probing both the increasing contribution of outflows and the decline of strong interactions.
\item the evolution of the net velocity distribution of the accretion~: infall exhibits a typical velocity of $0.75 V_c$. A backsplash component is detected at recent times with a significant outflowing component at a lower velocity ($\sim 0.6 V_c$).
\item the evolution of the impact parameters/incidence angle distribution of the infall. The infall is found to be mainly radial while outflows are on more circular orbits.
\item the angulo-temporal two-point correlation of the external potential on Virial sphere. The potential appear to be mainly dominated by a strong and constant quadrupole. The coherence time of smaller angular scales provide hints of an anisotropic accretion.
\item the angular power spectrum of accreted matter. The clustering is dominated by small angular scales, possibly 
 at the  resolution limit.
\item  the angulo-temporal correlation of the flux density of mass. The coherence time appears shorter than for potential fluctuations, as expected.
\end{itemize}
These results can be interpreted in terms of properties of accreted objects or
of smooth accretion and are coherent with previous studies
(e.g. \citet{1998MNRAS.300..146G}, \citet{Mamon}, \citet{Gill}, \citet{VDB2002}, \citet{Knebe},
\citet{aubert},\citet{2005MNRAS.358..551B}). These studies were mainly
focussed ont the properties of accreted satellites.
{Properties of accreted subhalos could also be directly derived from
these results, once a clear definition  allow us to distinguish structures
within the general flow of matter. Substructures are expected to form a
distinct ``phase'' of the accreted fluid: for instance the velocity dispersion
is expected to be quite different in compact objects than in the smooth
accreting component. This phase separation will be assessed in paper III  where a 
systematic comparison of the current approach 
with an analysis in terms of
pre-identified satellites will be carried, in the spirit of
\citet{aubert}. 
The contribution of outflows, the lack of a standard definition for
subhaloes, resolution issues and the fact that properties are measured at one
radius (which could be statistically propagated toward inner regions) are 
all issues which must be
assessed before a complete and rigorous comparison can be
performed. Yet, the current agreement between a  fluid description of the
environment and these above mentioned published results are clearly encouraging hits for the
reliability of the method presented here. 
%To this qualitative agreement,  it also
%adds some statistical significance and consequently allows quantitative
%constrains on the generic behaviour of dark matter accreted by haloes.
}

These kinematic signatures provide insights on the processes which occur in the inner regions of haloes. In particular, the kinematic discrepancies between the different components of the mass flux should be understood in terms of dynamical friction, tidal stripping or even satellite-satellite interactions within the halo.
The kinematical properties of accreted matter may be transposed to the kinematical properties of satellites observed around galaxies. Newly accreted material exhibits kinematic signatures (radial, high velocity trajectories) different from the ones measured for matter which already interacted with the halo (tangential, low-velocity orbits). Admittedly, it is not straightforward to apply directly these results to the  luminous component (see paper~I for a discussion of thresholding),  and to see how projection effects may affect the distributions.  Nevertheless,
    the corresponding observationnal measurements on satellites should provide information on the past history of these objects.

As discussed  in paper~I, these measurements can be used
    as an entry to the perturbative theory of the  response of the open halo. 
 Phenomena related to accretion can be consistently assessed via this framework: dynamical friction, tidal stripping and phase mixing. 
  With the statistical description of the tidal field presented in this paper only, 
 we may already implement the theory presented in paper~I in the regime of pure tidal
 excitation.
The complete knowledge of the source
 (which will be completed in paper III)  should 
 considerably extends the realm of application provided by this theory. 
 Specifically, we have shown in paper~I 
 that the internal  dynamics  of sub structures within
 galactic  haloes (distortion,  clumps as traced by Xray emissivity,  weak lensing, dark matter annihilation, tidal streams
..), and the implication for the disk (spiral structure, warp, etc...)   could be 
predicted within this framework.
Conversely
 the knowledge of the observed properties of a statistical sample of 
 galactic halos could be used  to
 (i) constrain observationnaly the statistical nature of the infall 
 (ii) predict the observed distribution and correlations of upcoming surveys, (iii) time reverse  the observed distribution of clumps, and finally
 (iv) weight the relative importance of the  intrinsic (via the unperturbed distribution function) 
 and external (tidal and/
or infall) influence of the environment in determining the fate of galaxies.

The current measurements reduce the degree of freedom that still exists in the setting of numerical experiments in a galactic context. For instance, given that the  structure of the external tidal field is found to be simple, it can easily be modelled as an external component in numerical simulations (or even in analytical studies). It would provide a simple but  statistically relevant contribution of the large scale structures to the dynamical states of haloes.The temporal coherence of the first $\ell>2$ angular harmonics of the tidal field should allow one to draw more accurate representations of external contribution to the field that would include the fluctuations due to smaller structures. The kinematics of accretion is not random as well and the distribution of velocities at $R_{200}$ follows a gaussian-shaped curve which caracteristics evolve with time and exhibit a certain distribution of  the impact parameter. These results put prescriptions that could be used to generate encounters between satellites and galactic disks that follow the ones measured in cosmological simulations at large radii. We also presented first constrains on the angulo-temporal correlation function of the accretion. Even though it is not completely clear yet how resolution will eventually affect these results, such functions contains some glimpse of informations regarding the angular distribution of encounters with external systems but also regarding the frequency of accretion events. This frequency can also be probed by the temporal coherence of the fluctuations in the tidal field. The apparent contradiction that exists between the observed number of discs and the predicted large number of mergers may be solved by a better knowledge of the frequency of the latter: it may be low enough to solve this contradiction. In this context, simulations of successive mergers between a galaxy and satellites should be consistent with large scale simulations an we provide first constrains on the rate of minor encounters at the halo's outer boundary.

As argued in \citet{pich}, we emphasize that an a priori discrimination between `objects' and diffuse matter may not constitute the best way to describe accretion: it is not clear that luminous matter is always attached to dark matter overdensities, there is no unambiguous definition of substructures and their state change with time under the influence of tidal shocks or dynamical friction. The generation of objects that follow the current results is admittedly the easiest way to proceed but should be followed by a more general description of matter in terms of 'fluid approach', where 'objects' only constitute a specific phase of such a fluid. 
 The statistical measurements on both $\psi^e$ and $\varpi_\rho$ allow the
 regeneration of synthetic environments. Knowing the average evolution and the
 angular power spectra of these quantities, the generation of spherical maps
 in the gaussian regime is straightforward. We describe such a regeneration
 procedure in appendix~\ref{s:gen}. Such maps would efficiently provide
 realistic  environments of halos, consistent (up to two-points statistics)
 with those measured in cosmological simulations and could be `embedded' into
 simulation of galaxies.   Again, virialised structures would be naturally
 included (since they have their own statistical signature on the virial
 sphere) without relying on any adhoc prescription on their nature.

{Extensions
 to non gaussian fields are also possible (following e.g. \citet{Contaldi}) but
 would rely on higher order correlations. It was assumed
 throughout this investigations
  that fields could be approximated as gaussian fields,
 fully described by their two-points statistics.
%  This assumption can be
% removed as long as three-points (or more) correlations are constrained and
% such studies will be performed in future works. 
Yet a simple visual
 inspection of $\varpi_\rho$ maps reveals that they are not strictly
 gaussian,  a finding confirmed by preliminary analysis of their bispectrum. Furthermore, paper I  demonstrated that a dynamical description
 which takes into account non-linear effects, such as dynamical friction,
 requires higher-order correlations. Therefore, extensions to non-gaussian
 fields are in order in the long run.
 }

  %However, such fluctuations are likely to be coupled to the accretion and should therefore be associated with a description of accretion.
%Regarding the accretion, we first put constrains on its kinematics. 

%Coupled to the measure of anisotropy described by \citet{aubert}, one could enhance the level of realism of such simulations by following these prescriptions. 

 It should  again be emphasized  that some aspects  of the present  work are
 exploratory  only, in  that  the resolution  achieved ($M_{\mathrm{halo}}  >
 5\cdot10^{12} M_\odot$) is somewhat high for $L_{\star}$ galaxies.  In fact,
 it will be interesting to confirm that the properties of infall do not 
 asymptote for lower
 mass   ($M_{\mathrm{halo}}  <5\cdot10^{12}   M_\odot$)  together   with  the
 intrinsic properties of  galaxies. In addition a systematic  study of biases
 induced by our estimators of  angular correlations should be conducted.  For
 a fixed halo mass, our lack  of resolution implies that we over estimate the
 clumsiness of the infall.

As demonstrated in \Sec{reso}, the limited resolution (both spatially and in mass) of our simulations
 appeared to be an issue for some of the results presented here 
 (e.g. the angular power spectrum of $\varpi_\rho$). Systematic use of 
 higher resolution simulations (in the spirit of section~6) will be
  required to fully assess these limitations.
   In particular, a fraction of the accretion detected as a weak/diffuse component may be associated to unresolved
   objects; the influence of small mass satellites should therefore be explored. 
With the prospect of deducing the  properties of galaxy
 from halos environments, lower mass haloes are more likely to host only one galaxy, making them more suitable for such a study.
Cosmological simulation of 
  small volumes also tend to prevent the formation of rare events which may be relevant for the representativity of the study: for example, some disks seem to indicate that they were formed in `very quiet' environments. The right balance between resolution and volume should be found.
 Aside from these biases induced by simulations, we also introduced selection criteria on both the mass or the  accretion history of haloes and the influence of these arbitrary choices on our statistical distribution should be assessed precisely. 
% It should also be stressed that  we did not take into account the anisotropy
 %of the  infall, as discussed  in \citet{aubert}, nor the  extra polarization
 %induced  by  the  presence  of  an embedded  disk,  which  will  undoubtedly
 %reinforce the corresponding polarization and anisotropy.
 
 Eventually  using hydrodynamical codes which include baryonic  effects in  simulations and  introducing
  the physics of gas  in our model,  we would  construct a  complete semi-analytic  tool to 
  study the detailed inner  dynamics of galaxies.

\subsubsection*{Acknowledgments}

 {\sl  We   are  grateful   to  S.  Colombi and J. Devriendt   for  useful comments
 and  helpful  suggestions.  We  would  like  to  thank D.~Munro  for  freely
 distributing   his  Yorick  programming   language  (available   at  {\em\tt
 ftp://ftp-icf.llnl.gov:/pub/Yorick}),   together  with   its   {\em\tt  MPI}
 interface,  which  we used  to  implement  our  algorithm in  parallel.  
DA thanks the Institute of Astronomy for their hospitality and funding from a Marie Curie
studentship.
 We
 acknowledge support from the Observatoire de Strasbourg
 computer facility and 
 the {\em HORIZON} project ({\em\tt http://www.projet-horizon.fr}). Finally,
 we would like to thank the anonymous referee for useful remarks on the manuscript. }

\bibliographystyle{mn2e}
\bibliography{aubertpichon06}
%%%%%%%%%%%%%%%%%%%%%%%%%%%%%%%%%%%%%%%%%%%%%%%%%%%%%%%%%%%%%%%
%%%%%%%%%%%%%%%%%%%%%%%%%%%%%%%%%%%%%%%%%%%%%%%%%%%%%%%%%%%%%%%
\appendix
%\onecolumn
\section{Harmonic convergence}
\label{s:harmconv}
\begin{figure} 
\centering           
\resizebox{7cm}{7cm}{\includegraphics{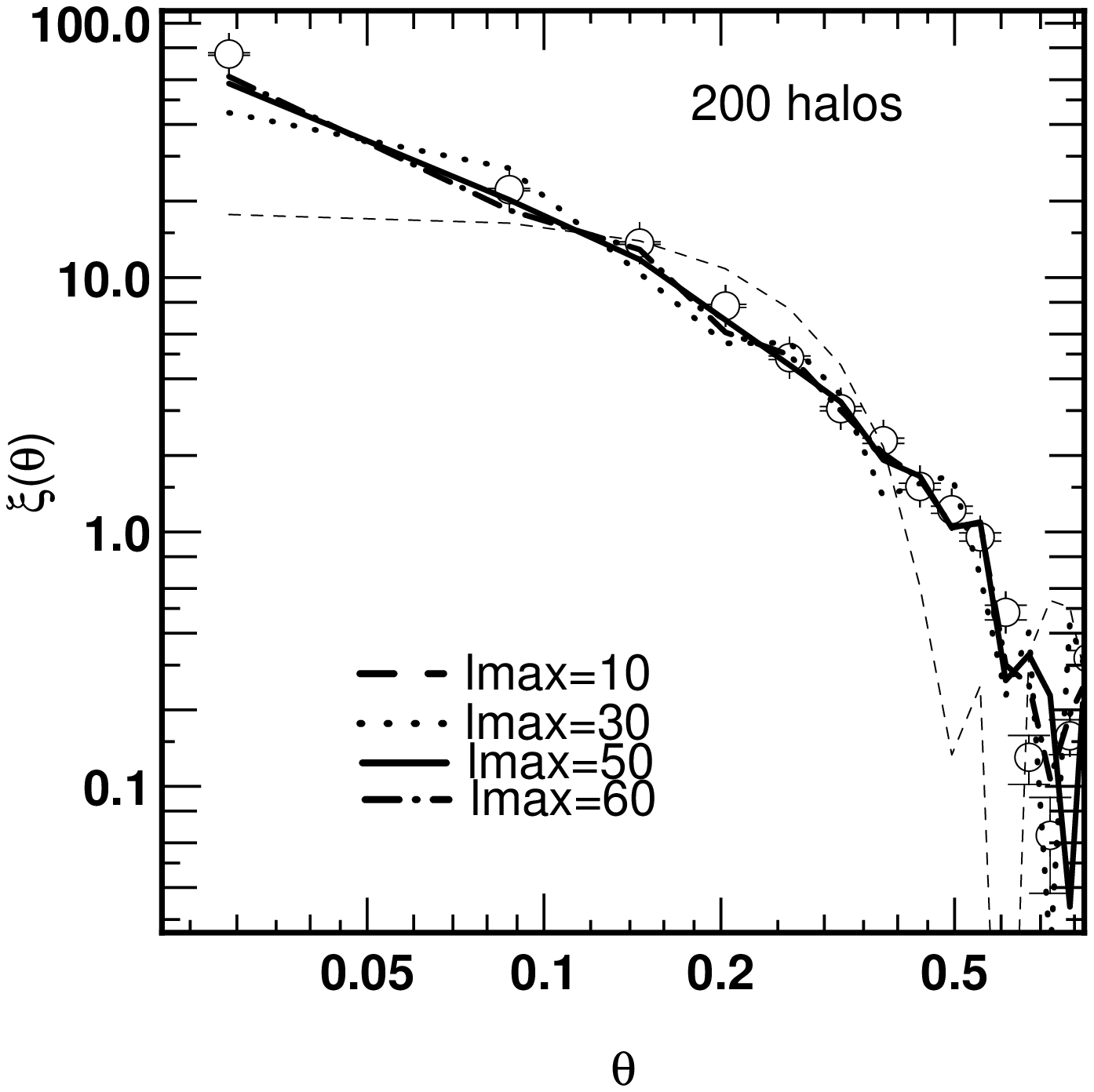}}
\caption{The  average  angular  two-point  correlation  function,  $\llangle
\xi(\theta)   \rrangle$  of   the  advected   mass  spherical   field  $\rho
v_r(\bO)$. The  correlation function is shown  as a function  of the angular
distance on  the sphere  $\theta$ given in  radians. Using the  positions of
accreted particles around 200 halos at z=1, the average correlation function
can be  computed (\textit{dots}).  Lines represent  the correlation function
deduced from the harmonic coefficients  of the $\rho v_r(\bO)$ fields around
the same  200 halos, with lmax= 10,  30, 50, 60. The  convergence is ensured
for lmax $\ge 50$.}
\label{f:convergence} 
\end{figure}
As explained  in section \ref{s:descsource},  the angular dependence  of the
source function  is expanded  over a basis  of spherical  harmonics $Y_{\ell
m}(\bO)$.   In   order  to  set   the  maximal  order  of   this  expansion,
$\ell_{\mathrm{max}}$, we compared the two-point correlation function of the
spherical  field   $\rho  v_r(\bO)$  to  the  one   inferred  from  harmonic
coefficients,  $a_{\ell  m}$, and  power  spectrum  $C_\ell$ (see  equations
(\eq{c2a}) and  (\eq{defcl})).  Within  a set of  randomly distributed
particles, let $d_{\mathrm{poisson}}(\theta)$  be the probability of finding
two particles  with an  angular separation $\theta$.  If $d(\theta)$  is the
same  probilty for a  given distribution  of particles  then its  two points
correlation function $\xi(\theta)$ is defined as:
\begin{equation}
\xi(\theta)\equiv\frac{d(\theta)}{d_{\mathrm{poisson}}(\theta)}-1.
\label{e:defxi}
\end{equation}
The correlation function $\xi(\theta)$ and the $a_{\ell m}$ coefficients are
related by (e.g. \citet{peacok})~:
\begin{equation}
\xi(\theta)=\sum_{\ell=0}^{\ell_{\mathrm{max}}}    C_\ell(2\ell+1)P_\ell(\cos
\theta),
\end{equation}
where $\theta$ is an angular distance on the sphere and $P_\ell(x)$ a
Legendre function. The average angular correlation function $\llangle
\xi(\theta) \rrangle_{\rm pair}$ is defined as:
\begin{equation}
\llangle  \xi(\theta)  \rrangle_{\rm  pair}  \equiv  \frac{1}{\sum_p  n_p^2}
\sum_p n_p^2 \xi_p(\theta),
\end{equation}
where $\xi_p(\theta)$ is the two-point correlation function of the $p$-th halo
computed using $n_p$  particles passing through the virial  sphere. From a set
of  200  halos  extracted  from  a simulation,  we  computed  $\llangle
\xi(\theta) \rrangle$  using different values  for $\ell_{\mathrm{max}}$
(see  Fig.  \ref{f:convergence}). From  the  same  set  of halos,  we  also
computed  the  average  two  point  correlation  function  directly  from  the
particles positions using \eq{defxi}.

From   Fig.  \ref{f:convergence},  it   clearly  appears   that  $\llangle
\xi(\theta) \rrangle_{\rm pair}$  has not converged for $\ell_{\mathrm{max}}
\le 30$.  For $\ell_{\mathrm{max}} \ge  50$ the actual two point correlation
function is  well reproduced. Since  no real difference can  be distinguished
between $\ell_{\mathrm{max}}=50$  and $\ell_{\mathrm{max}}=60$, we  chose to
limit the harmonic expansion of the source term to $\ell\le 50$.
 Note {that the truncation in $\ell_{\rm max}$ defines an effective resolution
beyond which the distribution is effectively coarsed grained.
}

%\section{The distribution of impact parameter, $\vartheta(b,t)$}
%
%\begin{figure} 
%\centering           
%\resizebox{7cm}{7cm}{\includegraphics{evoltimpact1_b}}
%\resizebox{7cm}{7cm}{\includegraphics{evoltimpact2_b}}
%\caption{}
%\label{f:evolb} 
%\end{figure}

\section{Angulo-temporal correlation}
Let us consider a spherical field $X(\bO,t)$ which can be expanded over the spherical harmonic basis:
\begin{equation}
  X(\bO,t)=\sum_{\ell m} x_{\ell m}(t) Y_{\ell m}(\bO).  
\end{equation}
The correlation $w^X$ between two successive realizations of $X$ is defined as:
\begin{equation}
  w^X(\bO,\bO',t,t')\equiv \langle X(\bO,t)X(\bO',t') \rangle\,,
\label{e:w1}
\end{equation}
where $\langle \cdot \rangle$ stands for the statistical average. If $X(\bO,t)$ is \textit{isotropic}, the correlation should not depend on $\bO$ or $\bO'$ but only on the distance $\theta$ between the two points. It implies that $w^X$ can be expanded on a basis of Legendre Polynomials $P_L(y)$~:
\begin{equation}
     w^X(\bO,\bO',t,t')= w^X(\theta,t,t') =\sum_L (2L+1) T_L P_L(\cos( \theta))\,.
\end{equation}
How are $T_L$ and $x_{\ell m}$ related ? Rewriting \eq{w1} as:
\begin{equation}
  w^X(\theta,t,t')=\sum_{\ell m} \sum_{\ell' m'}  \langle x_{\ell m}(t)x^*_{\ell' m'}(t')\rangle Y_{\ell m}(\bO) Y^*_{\ell' m'}(\bO'),
\end{equation}
one can write:
\begin{equation}
  \int \md \bO \md \bO' Y^*_{\ell_1 m_1}(\bO) Y_{\ell_2 m_2}(\bO') w^X=\langle  x_{\ell_1 m_1}(t)x^*_{\ell_2 m_2}(t')\rangle.
\end{equation}
Meanwhile, assuming isotropy, one can also write
\begin{eqnarray}
 && \int \md \bO \md \bO' Y^*_{\ell_1 m_1}(\bO) Y_{\ell_2 m_2}(\bO') w^X=\nonumber \\
&&\sum_L (2L+1) T_L \int \md \bO \md \bO' Y^*_{\ell_1 m_1}(\bO) Y_{\ell_2 m_2}(\bO') P_L(\cos( \theta))\nonumber\\
&=& \sum_{LM} (4\pi) T_L \int \md \bO \md \bO' Y^*_{\ell_1 m_1}(\bO) Y_{\ell_2 m_2}(\bO') Y_{L M}(\bO) Y^*_{L M}(\bO') \label{e:addition}\nonumber\\
&=&\sum_{LM} (4\pi) T_L \delta_{L\ell_1}\delta_{L\ell_2} \delta_{M m_1}\delta_{M m_2}\,,
\end{eqnarray}
where $P_L$ is expressed in terms of spherical harmonics using the spherical harmonics addition theorem. In the end, we get:
\begin{equation}
  T_\ell=\frac{1}{4\pi}  \langle  x_{\ell m}(t)x^*_{\ell m}(t')\rangle.
\end{equation}
For a given realization of $X(\bO,t)$, $T_\ell$ can be estimated by:
\begin{equation}
  T_\ell=\frac{1}{4\pi}\frac{1}{2\ell+1}  \sum_m x_{\ell m}(t)x^*_{\ell m}(t')\,.
\end{equation}

\section{Distributions}

\begin{figure} 
\centering           
\resizebox{7cm}{7cm}{\includegraphics{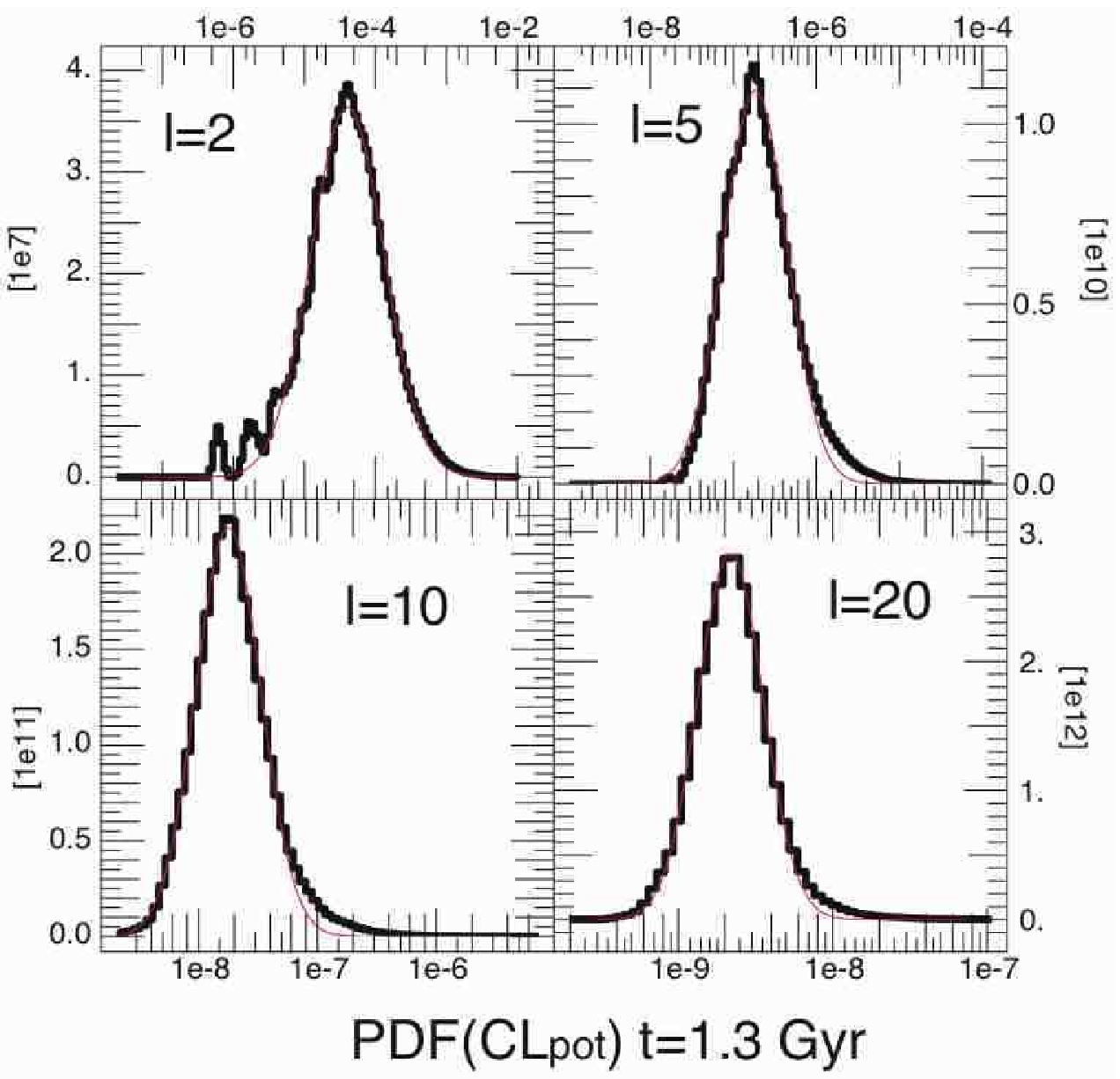}}
%\resizebox{7cm}{7cm}{\includegraphics{pdfcl_1}}
\caption{The probability distribution  of $C_\ell^{\psi^e}$ the  for  $\ell = 2, 5,
10, 20  $ at  the look-back time  $t=1.9$ Gyr. Note  that the X-axis  is sampled
logarithmically; the corresponding log normal fit of the mode is also shown.
}
\label{f:hbl} 
\end{figure}

\begin{figure} 
\centering           
\resizebox{7cm}{7cm}{\includegraphics{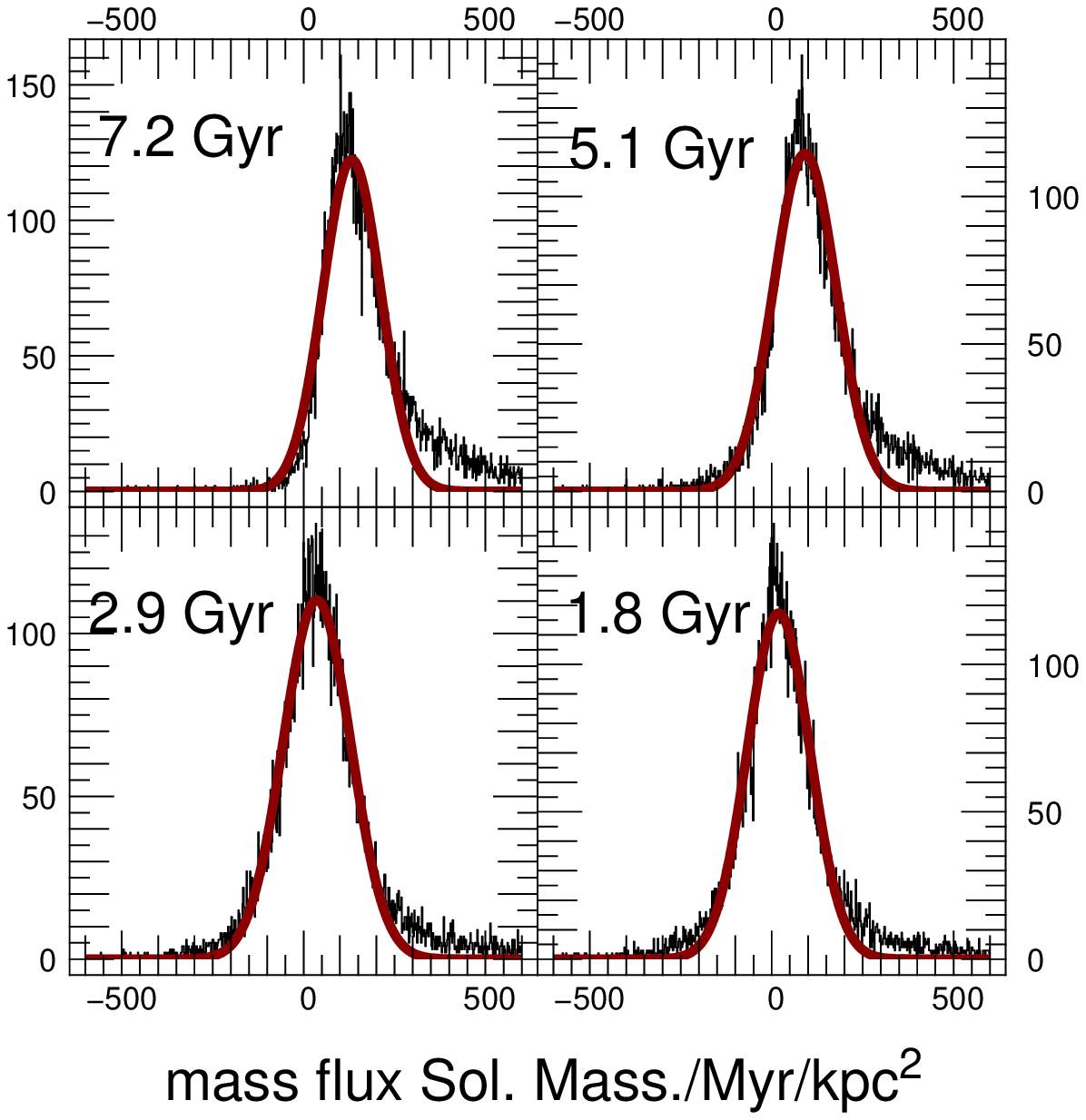}}
\caption{The probability distribution of the mean flux $\Phi^{M}(t)$ for four different times. The normal fit is also shown.
}
\label{f:a001} 
\end{figure}

\begin{figure} 
\centering           
\resizebox{7cm}{7cm}{\includegraphics{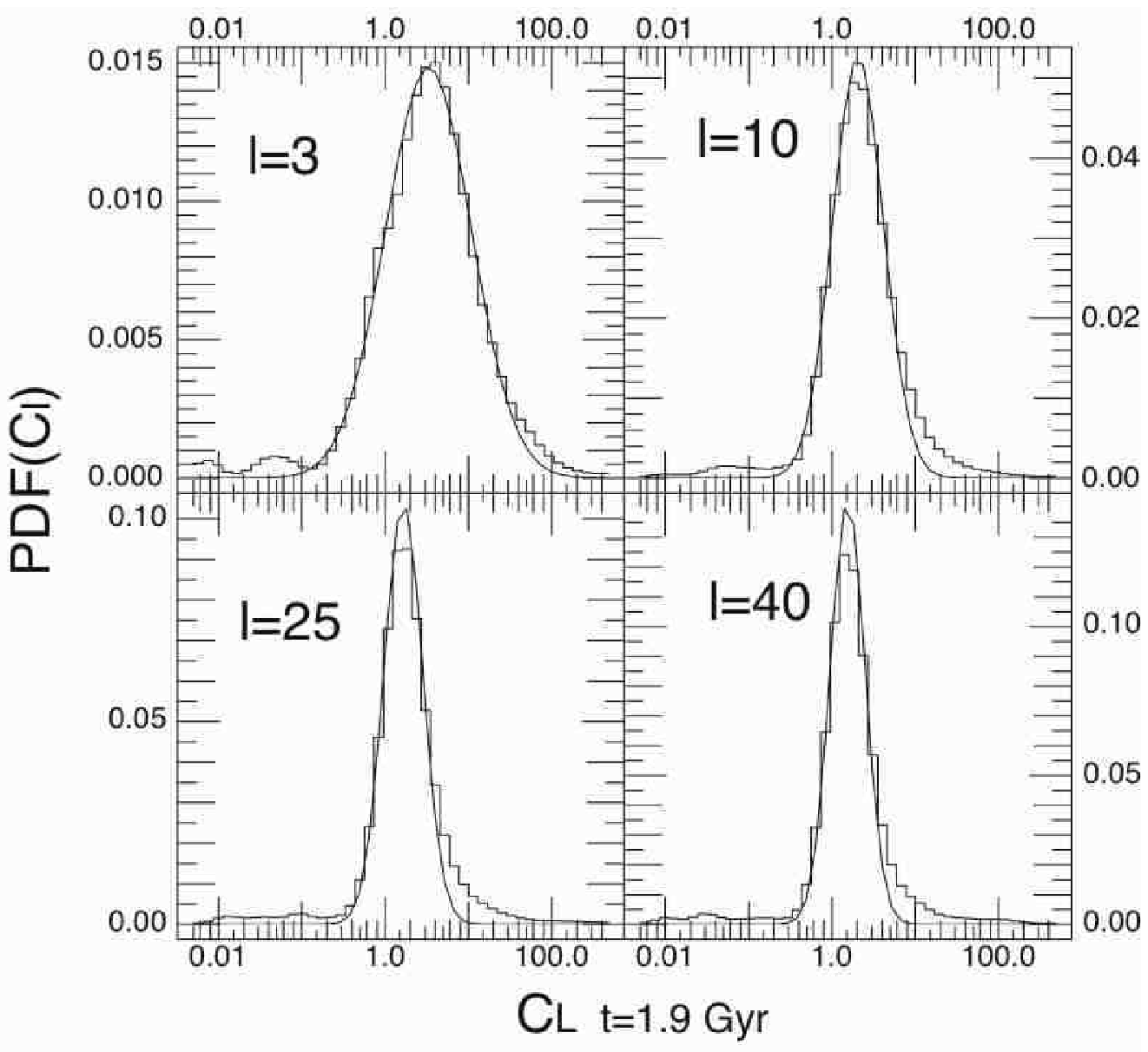}}
\caption{The probability distribution  of the $ C_\ell $ for  $\ell = 3, 10,
25, 40  $ at  the look-back time  $t=1.9$ Gyr. Note  that the X-axis  is sampled
logarithmically; the corresponding log normal fit of the mode is also shown.
%Since  the   $C_\ell$  are  quadratic   functions  of  the   $c_\M{n}$,  the
%distribution is quite skewed towards  larger values of $C_\ell$ as expected.
%Our fit,  $\llangle C_\ell \rrangle$,  provides a significantly  less biased
%estimator of the most likely value of $C_\ell$.
}
\label{f:cl2} 
\end{figure}

\begin{figure} 
\centering           
\resizebox{7cm}{7cm}{\includegraphics{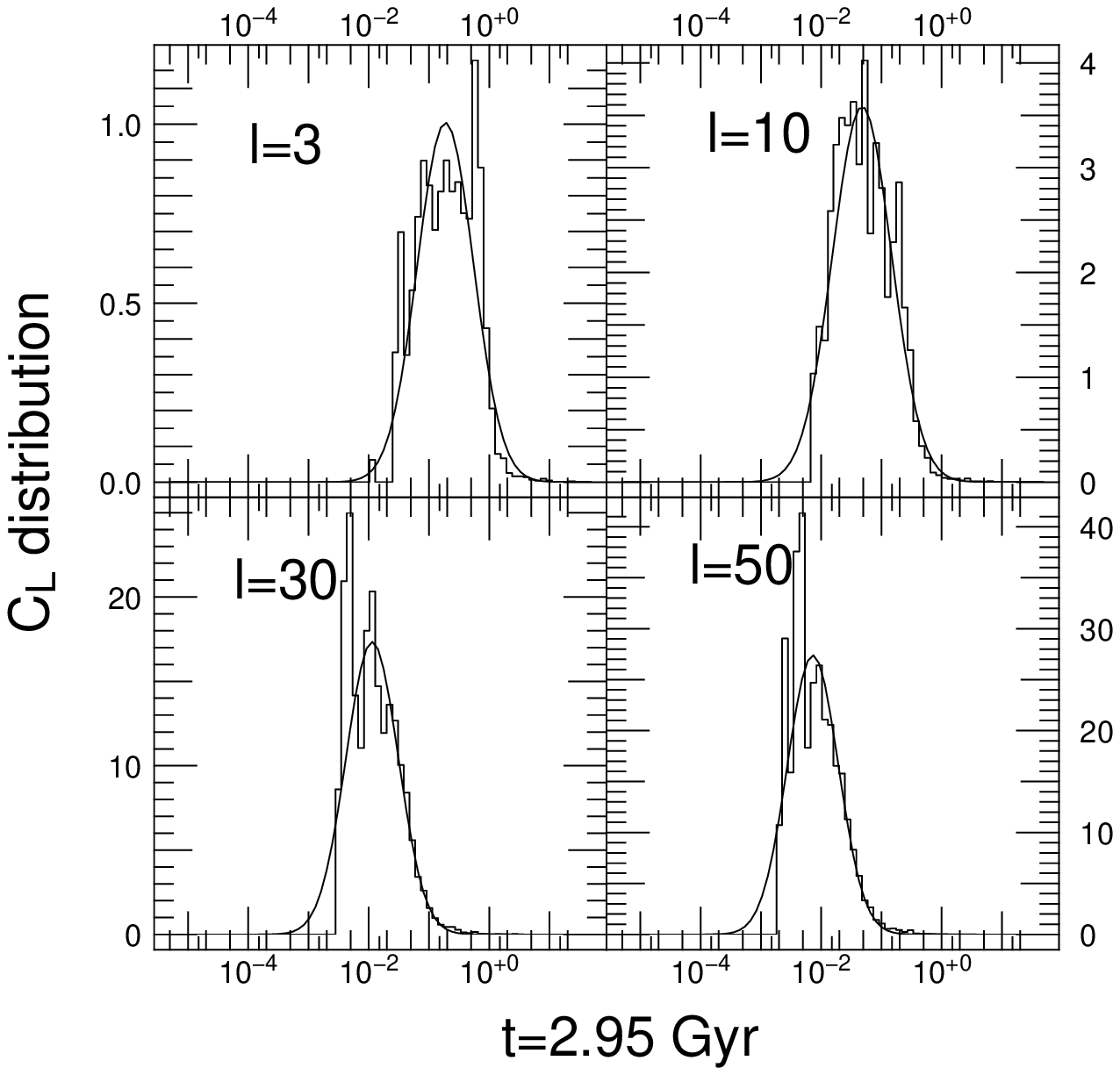}}
%\resizebox{7cm}{7cm}{\includegraphics{pdfcl_1}}
\caption{The probability distribution  of $C_\ell^{\varpi_\rho'}$ the  for  $\ell = 3, 10,
25, 40  $ at  the look-back time  $t=1.9$ Gyr. Note  that the X-axis  is sampled
logarithmically; the corresponding log normal fit of the mode is also shown.
%Since  the  $C_\ell^{\varpi_\rho'}$ are  quadratic   functions  of  the   $c_\M{n}$,  the
%distribution is quite skewed towards  larger values of $C_\ell$ as expected.
%Our fit,  $\llangle C_\ell \rrangle$,  provides a significantly  less biased
%estimator of the most likely value of $C_\ell$.  \Xtophe{ca te va ?}
}
\label{f:cl1} 
\end{figure}

\begin{figure} 
\centering           
\resizebox{7cm}{7cm}{\includegraphics{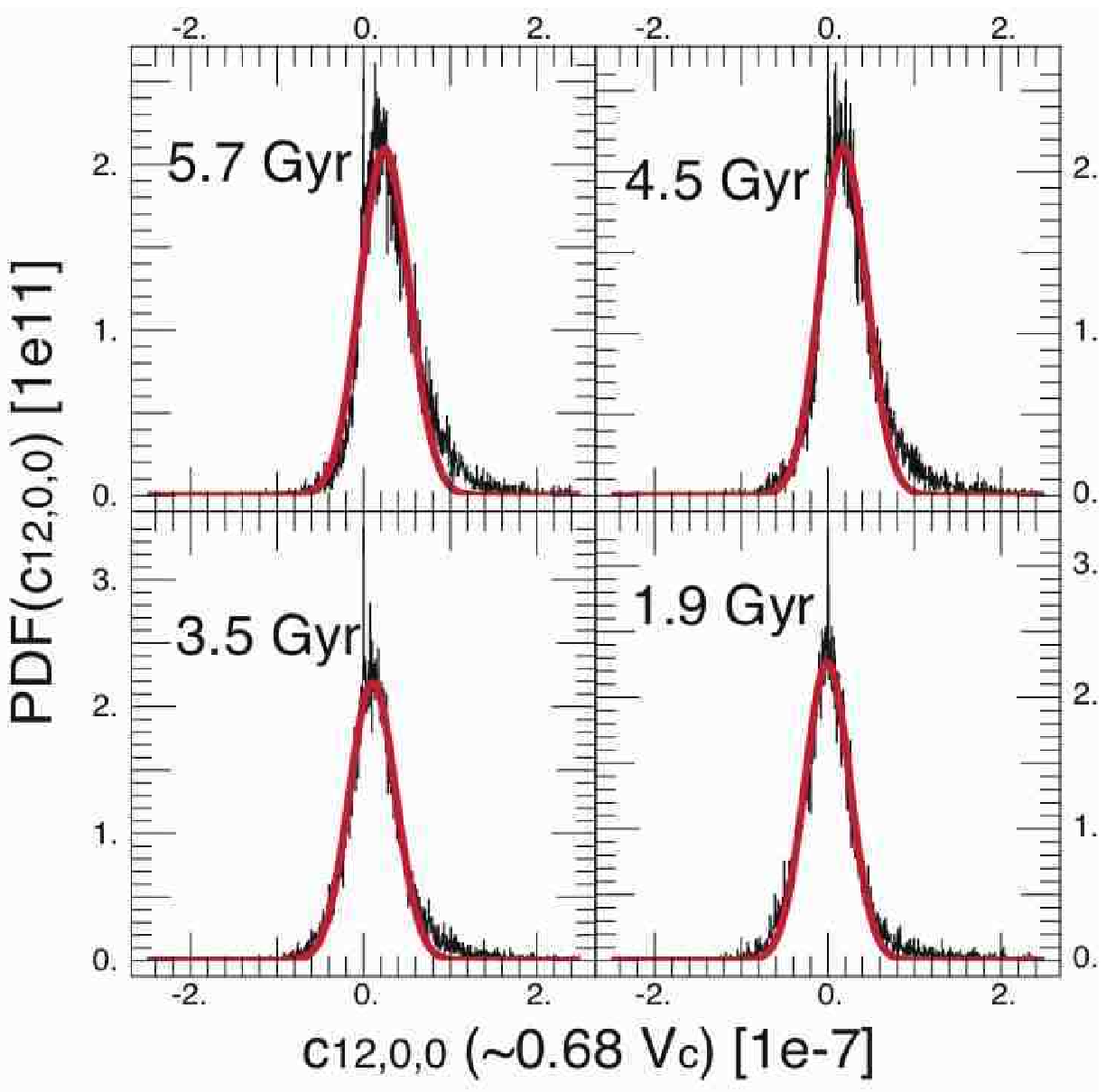}}
\caption{The  probability  distribution of  the  $c_{12,0,0}^{0,0}  $ for  the
look-back times $t=1.9,  3.5, 4.5$ and $5.7$ Gyrs.  Note  that the X-axis is
sampled linearly; the corresponding fit of the mode is also shown. }
\label{f:pdfv} 
\end{figure}

In this appendix, we present the various distributions fitted either by normal or log-normal PDF. For each quantity, the mode (or most probable value) of the distribution has been obtained from these fits. The gaussian distribution is defined by:
\begin{equation}
  \R{N}(x)= \frac{A}{\sigma\sqrt{2\pi}}\exp\left({-\frac{(x-\mu)^2}{2\sigma^2}}\right),
\end{equation}
while the mode is equivalent to the mean $\mu$.
The log-normal distribution is given by 
\begin{equation}
 \R{LN}(x)= \frac{A}{\sigma\sqrt{2\pi}}\exp\left({-\frac{(\log(x/\mu))^2}{2\sigma^2}}\right),
\end{equation}
while the mode is given by $\mu\exp(-\sigma^2)$.
The different fits mentioned in the main text are described in the following figures:
\begin{itemize}
\item \Fig{b001} shows the distributions of the harmonic coefficient $b_{00}(t)$ which is proportional to the potential averaged on the sphere. It is expressed in units of $GM/R$ where M is expressed in $10^{10} M_\odot$, R in kpc$h^{-1}$ and G=43007 in internal units.
\item \Fig{hbl} shows the distributions of the external potential's power spectrum for four different harmonic $\ell=2,5,10,20$ at t=1.3 Gyr. The distribution has been fitted by a log normal function.
\item \Fig{a001} shows the distributions of the mean flux $\Phi^{M}(t)$. The mean flux is proportional to the harmonic coefficient $a_{00}$. The distribution is fitted by a normal distribution. The normal model agrees well with the measured distribution at recent times but fail to reproduce the outliers tail at high redshift. Consequently the mode position is underestimated at these times.
\item \Fig{pdfv} shows the distributions of one of the coefficient involved in the computation of the velocity distribution $\phi(v)$. Four different times are being represented. The coefficient distribution has been fitted by a gaussian distribution.
\item \Fig{cl2} shows the distributions of the power spectrum values $C_\ell^{\varpi_\rho}$ for four different harmonic order $\ell$. The fits were made at $t=1.9$ Gyr. This distribution has been fitted by a lognormal distribution. 
\item \Fig{cl1} shows the distributions of the power spectrum values $C_\ell^{\varpi_\rho'}$ for four different harmonic order $\ell$. The fits were made at $t=2.95$ Gyr. This distribution has been fitted by a lognormal distribution. 
\end{itemize}

\begin{figure} 
\centering           
\resizebox{7cm}{7cm}{\includegraphics{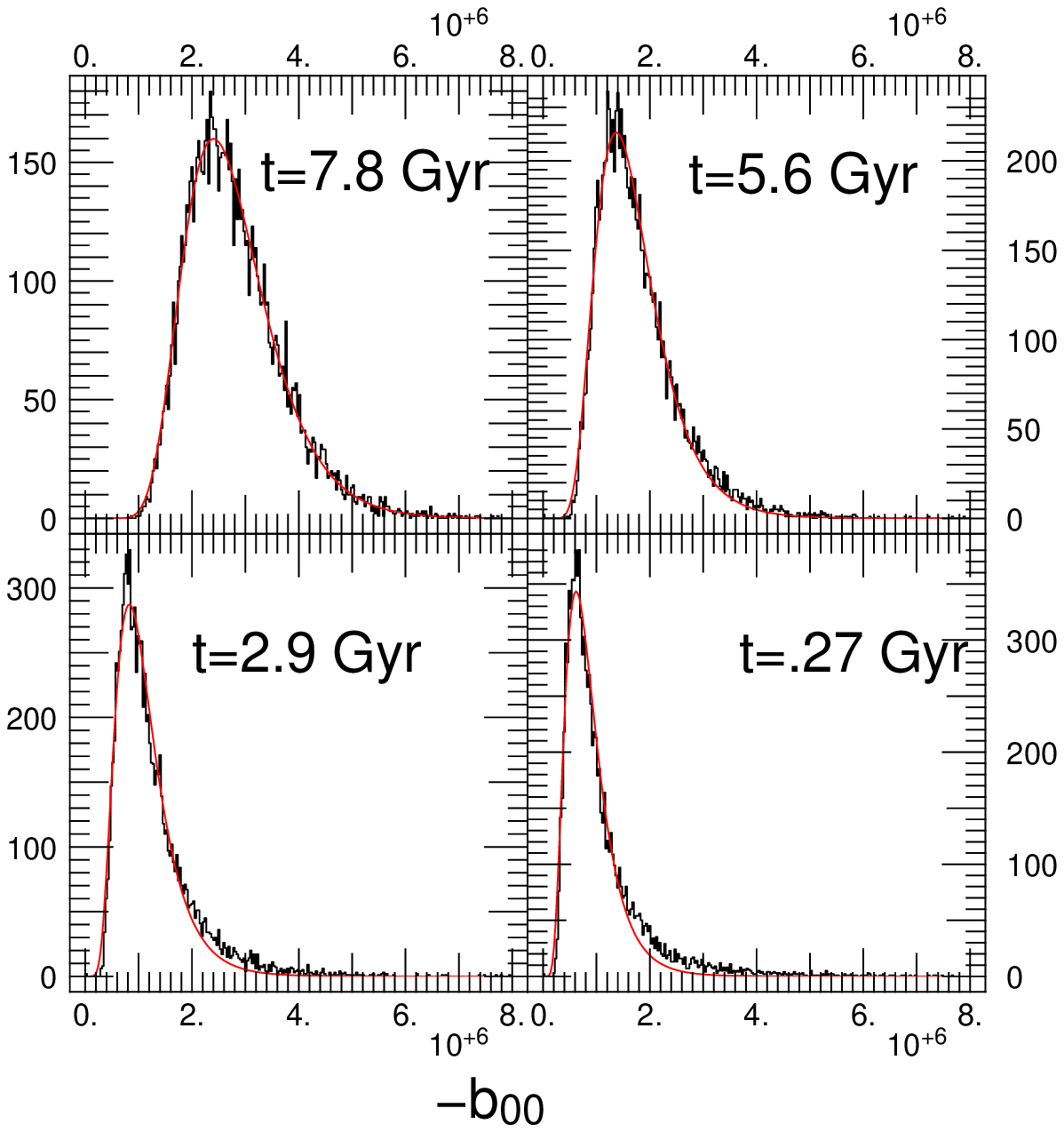}}
\caption{The probability distribution of $b_{00}$ at four different times. This coefficient is proportional to the external potential averaged on the sphere. The log-normal fit is also shown.}
\label{f:b001} 
\end{figure}

\section{Fits and Tables}
\begin{figure} 
\centering           
\resizebox{0.8\columnwidth}{0.8\columnwidth}{\includegraphics{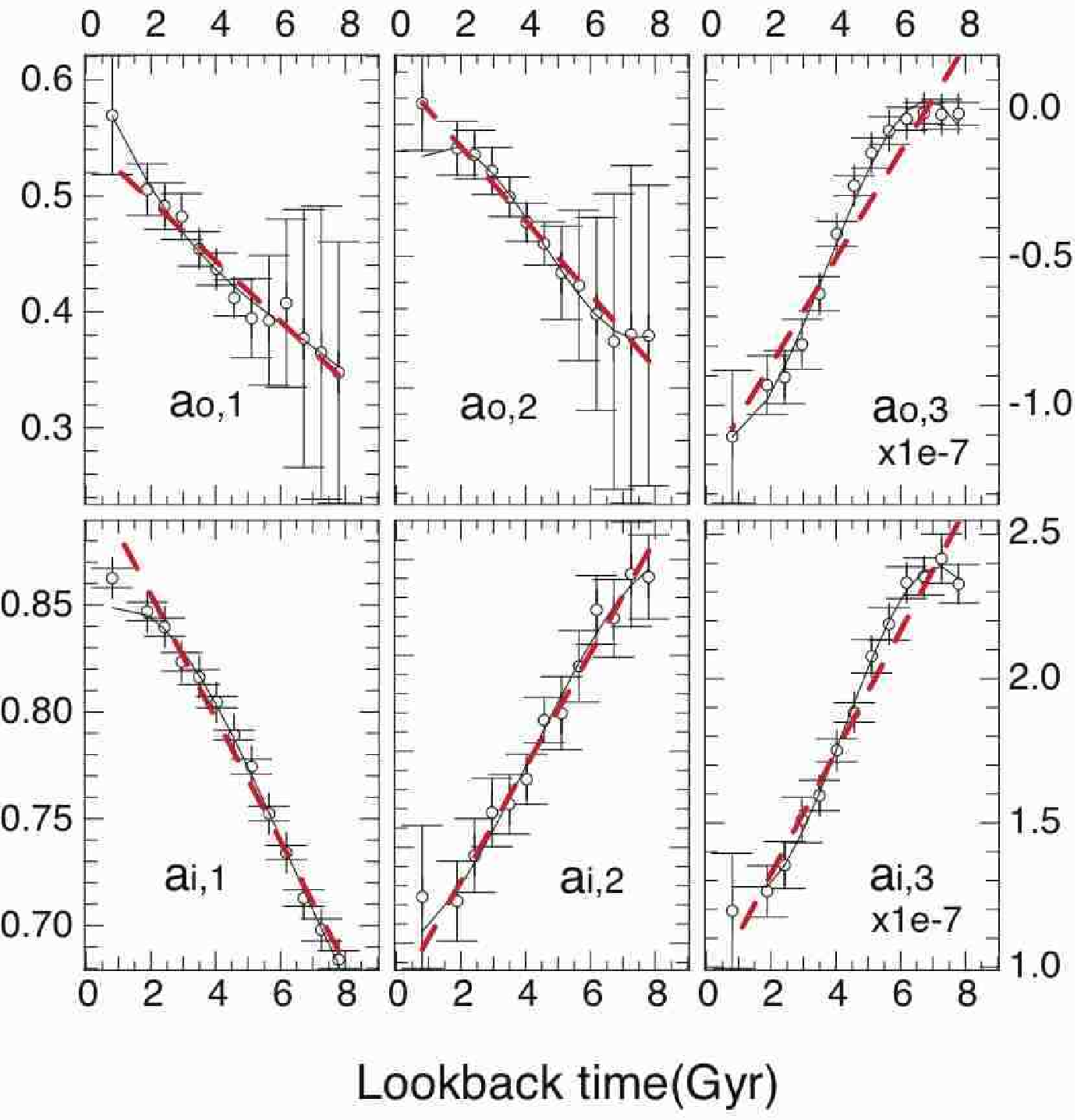}}
\caption{   time   evolution   of   the  fitted   coefficients   (given   in
  table~\ref{t:coeffgauss}) of the PDF $\langle\varphi(v,t) \rangle$ for the
  parameterization  suggested  in  \eq{deuxgauss} (\textit{symbols}).   Bars
  stand  for  the  $3\sigma$   errors  on  the  coefficients  determination.
  $q_{\imath,k}, k=1,2,3$ are respectively the  mean, the r.m.s and the amplitude
  of the  positive Gaussian  (i.e. infalling component).   $q_{o,k}, k=1,2,3$
  are  the corresponding values  for the  negative Gaussian  (i.e. outflowing
  component).   The  trend  is   accurately  fitted  by  a  linear  relation
  (\textit{red dashed line}).}
\label{f:distribvcoeff} 
\end{figure}

\begin{figure} 
  \centering           
  \resizebox{0.8\columnwidth}{0.8\columnwidth}{\includegraphics{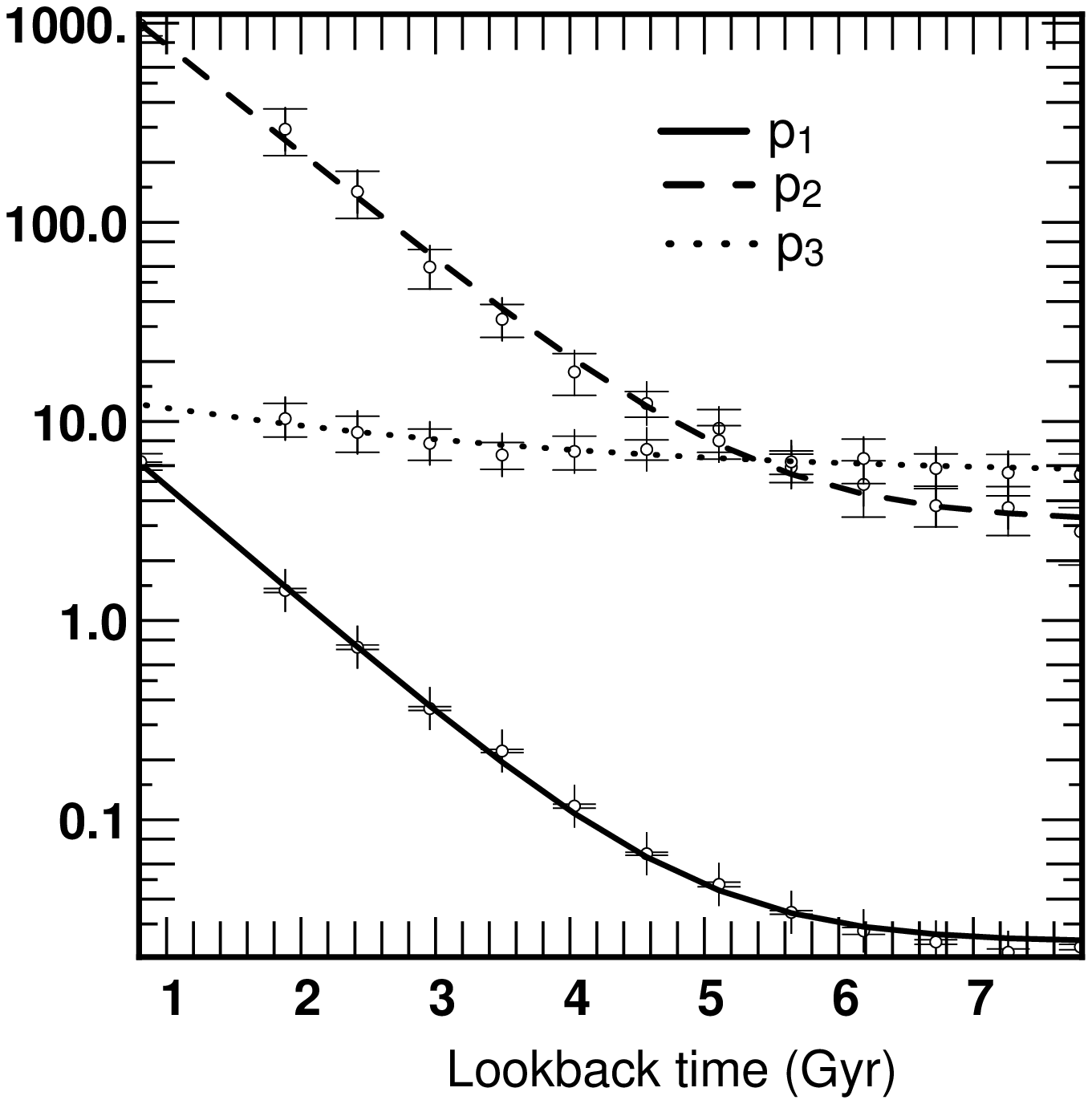}}
  \caption{Time  evolution of the coefficients  for the  $\llangle C_\ell^{\varpi_\rho}\rrangle (t)$
  model (see \eq{modelcl})(\textit{symbols}). The three time evolutions may be
  described   accurately   by   decreasing  exponentials   (see   \eq{decexp})
  (\textit{solid lines}).}
  \label{f:clcoeffit} 
 \end{figure}

In the main text, some statistics are fitted by simple laws and the time evolution of these statistics can be described by the time evolution of the fitting parameters. In this appendix, we present the time evolution of these parameters:
\begin{itemize} 
\item{\Fig{distribvcoeff} and Table \ref{t:coeffgauss} presents the time evolution of the velocity distribution of accretion. It can be fitted by two gaussians drifting toward higher velocities with time.}
\item{Table \ref{t:gamfit} summarize the time evolution of the distribution of incidence angles. This distribution can be fitted by a gaussian (for the inflows) and a linear relation (for the outflows).  The fitting parameters present a linear evolution with time.}
\item{\Fig{clcoeffit} and Table \ref{t:coeffexp} summarize the time evolution of the angular power spectrum of $\varpi_\rho$. The power spectrum can be fitted by \eq{modelcl} where fitting parameters decreases exponentially with time.}
\end{itemize}

\begin{table}
\centering
\begin{tabular}{@{}lcc@{}}
\hline
& $m$ & $n$\\
\hline
$q_{\imath,1}$& -0.028 & 0.912 \\
$q_{\imath,2}$& 0.0059 & 0.084 \\
$q_{\imath,3}$& 2.1 $10^{-8}$ & 9.03 $10^{-8}$\\
$q_{o,1}$& -0.026 & 0.54 \\
$q_{o,2}$& -0.028 & 0.23 \\
$q_{o,3}^{(*)}$& 1.79 $10^{-8}$ & -1.22 $10^{-7} $\\
\hline
\end{tabular}
\caption{Fitting parameters for the time evolution of Gaussian coefficients,
$q(t)=m\times  t{\mathrm{(Gyr)}}+n$.    These  coefficients,  together  with
\eq{deuxgauss}  allow   us  to  compute  the  time   evolution  of  $\langle
\varphi(v,t) \rangle$. (*) By definition, $q_{o,3} \le 0$. }
\label{t:coeffgauss}
\end{table}

\begin{table}
\centering
\begin{tabular}{@{}lcc@{}}
\hline
& $m$ & $n$\\
\hline
$p_1$& -0.038 & 0.83 \\
$p_2$& -3.12 $10^{-8}$ & 2.29 $10^{-7}$ \\
$p_3$& 8.59 $10^{-8}$ & -6.64 $10^{-7}$\\
\hline
\end{tabular}
\caption{Fitting  parameters  for   the  time  evolution  $p_{k}(t),k=1,2,3$
parameters.     $p(t)=m\times     t{\mathrm{(Gyr)}}+n$.     Together    with
\eq{modelgamma1}, these coefficients allow  to predict the time evolution of
$\langle \vartheta(\Gamma_1,t) \rangle$.}
\label{t:gamfit}
\end{table}

\begin{table}
\centering
\begin{tabular}{@{}lccc@{}}
\hline
& $h$ & $k$ &$u$\\
\hline
$q_{1}^{\varpi_\rho}$&0.0242 & 17.86&0.867 \\
$q_{2}^{\varpi_\rho}$& 3.153 & 2699. &0.8933 \\
$q_{3}^{\varpi_\rho}$& 5.4105 & 9.44 & 1.55 \\
\hline
\end{tabular}
\caption{Fitting  parameters  for  the  time evolution  of  $\llangle  C_\ell
\rrangle^{\varpi_\rho}$    model's   coefficients    (see  (\eq{modelcl})).   $q_{3}^{\varpi_\rho}(t)=h+k\mathe^{-\frac{t}{u^2}}$  with lookback  time
$t$  expressed in  Gyr.}
\label{t:coeffexp}
\end{table}

\section{Alternative contrast and angulo-temporal correlation}
%%%%%%%%%%%%%%%%%%%%%%%%%%%%%%%
\label{s:alter}
\begin{figure} 
\centering           
\resizebox{0.8\columnwidth}{0.8\columnwidth}{\includegraphics{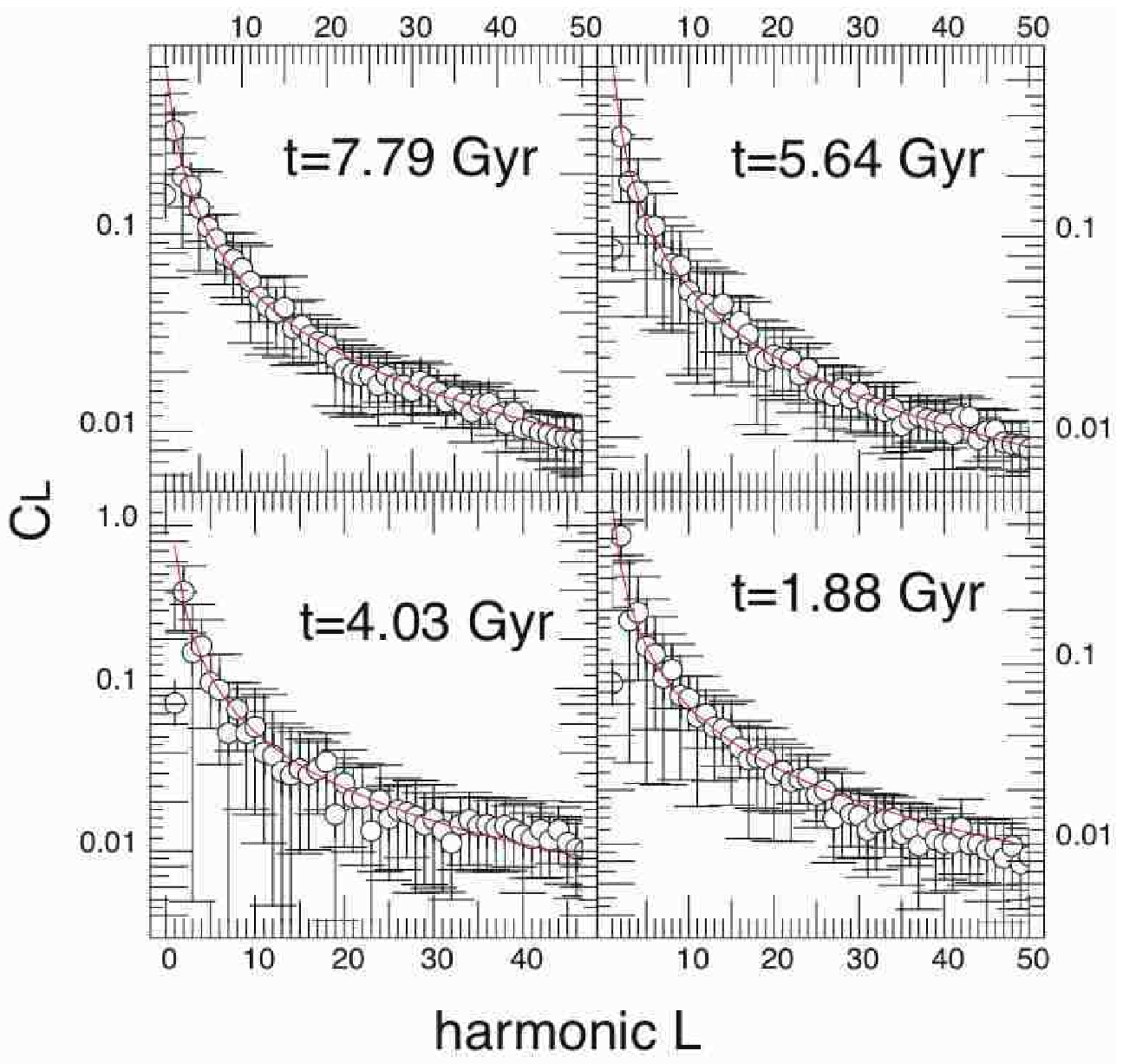}}
\caption{
The angular  power spectrum of the external potential, $\llangle C^{'\varpi_\rho}_\ell\rrangle (t)$,
at  four different lookback times (symbols). Harmonics coefficients were normalized using \eq{newdef}, halo by halo. $\llangle C^{'\varpi_\rho}_\ell\rrangle (t)$ is taken  as the  mode of  the  log-normal function  used to  fit the  $C^{'\varpi_\rho}_\ell$ distribution.   Bars stand  for $3\sigma$  errors.  For a  given $\ell$  the corresponding angular scale  is $\pi/\ell$. The power spectra maybe fitted by a generic model given by \eq{univ} (solid line). 
Unlike $\llangle C^{'\varpi_\rho}_\ell\rrangle (t)$, $\llangle C^{'\varpi_\rho}_\ell\rrangle (t)$ remains the same with time because of a different normalization.
}
\label{f:cltnonorm} 
\end{figure}
It appears clearly from the time evolution of $\llangle C^{\varpi_\rho}_\ell\rrangle(t)$ 
presented in the main text, 
that no angulo-temporal correlation can be computed, since it would be completely dominated by the secular evolution of the power spectrum. For this reason, an alternative definition of the flux density contrast has been used:
\begin{equation}
\delta'_{[\varpi_\rho]}(\bO)\equiv\frac{\varpi_\rho(\bO) -\overline{\varpi_\rho}}{\overline{\varpi_\rho}}=\sum_{\ell,m} \tilde a'_{\ell,m} Y_{\ell,m}(\bO),
\end{equation}
or in terms of harmonic coefficients:
\begin{equation}
\tilde         a'_{\ell,m}\equiv        \sqrt{4\pi}(\frac{a_{\ell,m}}{
a_{00}}-\delta_{\ell 0}).
\label{e:newdef}
\end{equation}
Using this new definition, the fluctuations of $\varpi_\rho$ on the sphere are measured relative to the mass flux of each individual halo. The main drawback of such a definition is that it couples the $a_{00}$ and the $a_{\ell m}$ coefficients from the beginning. For example, the resimulation of such fields would require the knowledge of conditional probabilities, i.e. what is the distribution of $a_{\ell m}$ for a given value of $a_{00}$, while the previous definition (given by \eq{defatilde}) only acted as specific choice of (non constant) units. Furthermore this new definition is more sensitive to outliers with low $a_{00}$ values.

\begin{figure} 
\centering           
\resizebox{0.8\columnwidth}{0.8\columnwidth}{\includegraphics{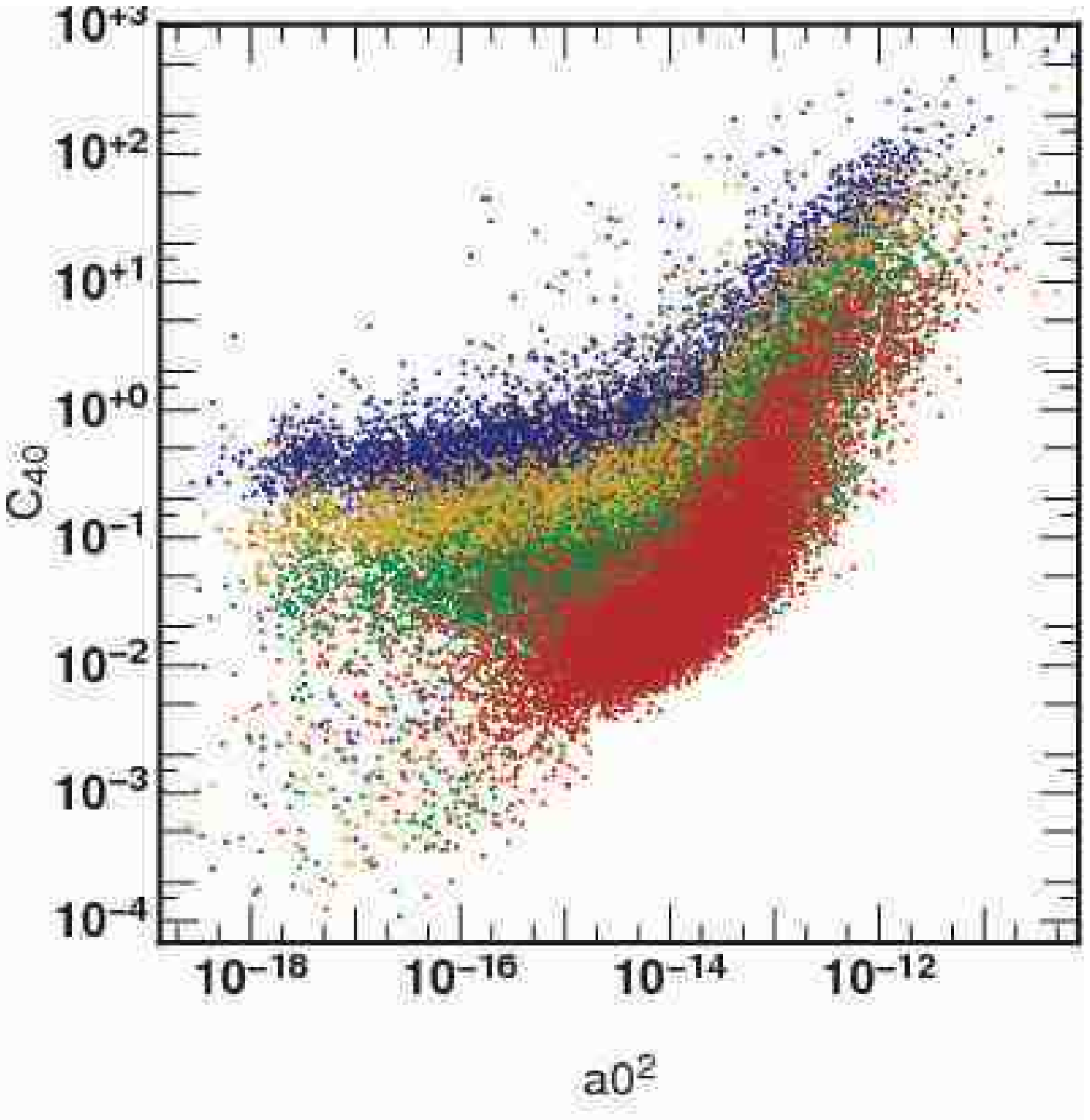}}
\resizebox{0.8\columnwidth}{0.8\columnwidth}{\includegraphics{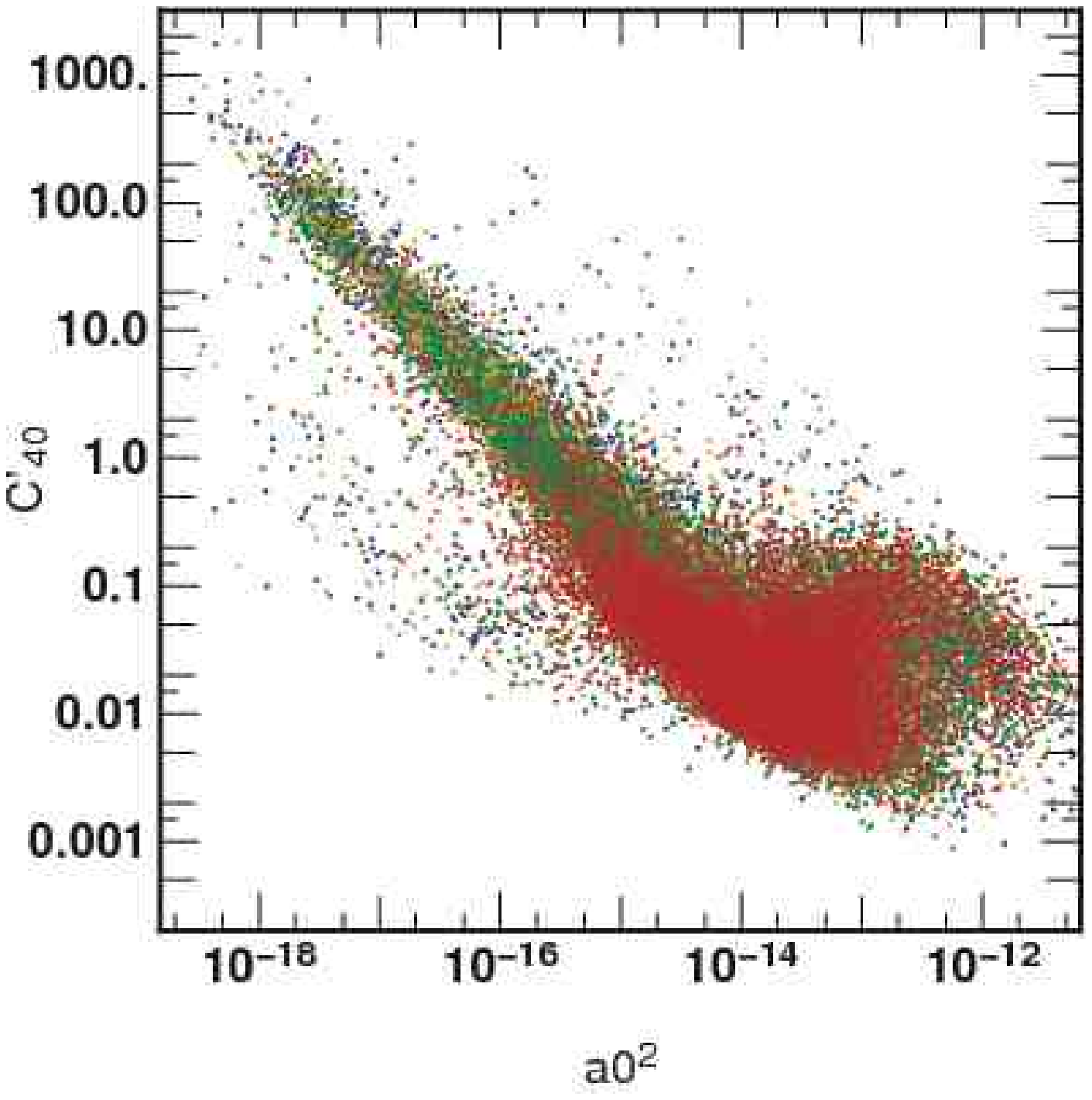}}
\caption{Scatter plots of the power spectra $C^{\varpi_\rho}_{40}$ (\textit{Top}) and $C'^{\varpi_\rho}_{40}$ (\textit{Bottom}) as a function of $a_{00}^2$. The four colors stand for different lookback times: t= 7.8 Gyrs (red), 5.6 Gyrs (green), 4. Gyrs (yellow) and 2.8 Gyrs (blue). The monopole $a_{00}$ scales like the integrated flux of matter. The quantities $C^{\varpi_\rho}_{40}$ and $C'^{\varpi_\rho}_{40}$ differ by the normalization applied to the harmonic coefficients $a_{\ell m}$. See the main text for more details.  }
\label{f:C40} 
\end{figure}

Still, the angular power spectrum $\llangle C^{'\varpi_\rho}_\ell\rrangle(t)$ of $\delta'_{[\varpi_\rho]}(\bO)$ is much more regular than the one obtained from the previous definition. Its overall amplitude remains constant over the last 8 Gyrs, while its shape seems to be less dominated by small scale contributions. This alternative power spectrum is well fitted by a single power law~:
\begin{equation}
    \llangle C^{'\varpi_\rho}_\ell\rrangle(t) = 0.75 \ell^{-1.15},
\label{e:univ}
\end{equation}
 for the whole time range covered by the current measurements. This constant shape suggest that harmonic coefficients scale like $a_{00}$, i.e. the mass flux. Such a scaling is not obvious, since a strongly clustered $\varpi_\rho$ field may coexist with a nil net flux (i.e. $a_{00}\sim 0$).  It also implies that the evolution measured on the previous definition of the power spectrum, $\llangle C^{\varpi_\rho}_\ell\rrangle(t)$,  is more related to the evolution of the average flux (traced by $a_{00}$) than to the modification of the fluctuations amplitude (traced by the others $a_{\ell m}$). Still, the evolution of $\langle a_{00} \rangle$ spans over one magnitude while the evolution of $\llangle C^{\varpi_\rho}_\ell\rrangle(t)$ spans over several order of magnitudes: this strongly suggests that two different populations of haloes contribute to the two type of power spectrum. In \Fig{C40}, the scatter plot of $C_{40}$ and $C'_{40}$ as a function of $a_{00}$ shows that the haloes which experience strong accretion dominate the peak of the $C'_{40}$ distribution while haloes with low accretion dominate the peak of the  $C_{40}$.  Furthermore $C_{40}$ does not scale anymore like $a_{00}$ as it drops below some level, providing hints of resolution and isolated particles effects. To conclude, $\delta'_{[\varpi_\rho]}(\bO)$ appears as better way to rescale the fluctuations' amplitude since it provides a more regular behavior of the power spectrum, but it is a more complex quantity to manipulate. Meanwhile,  $\delta_{[\varpi_\rho]}(\bO)$ is probably the correct way to to proceed but is clearly more sensitive to resolution effects, which should be assessed with bigger simulations in the future.

\begin{figure} 
\centering           
\resizebox{0.8\columnwidth}{0.8\columnwidth}{\includegraphics{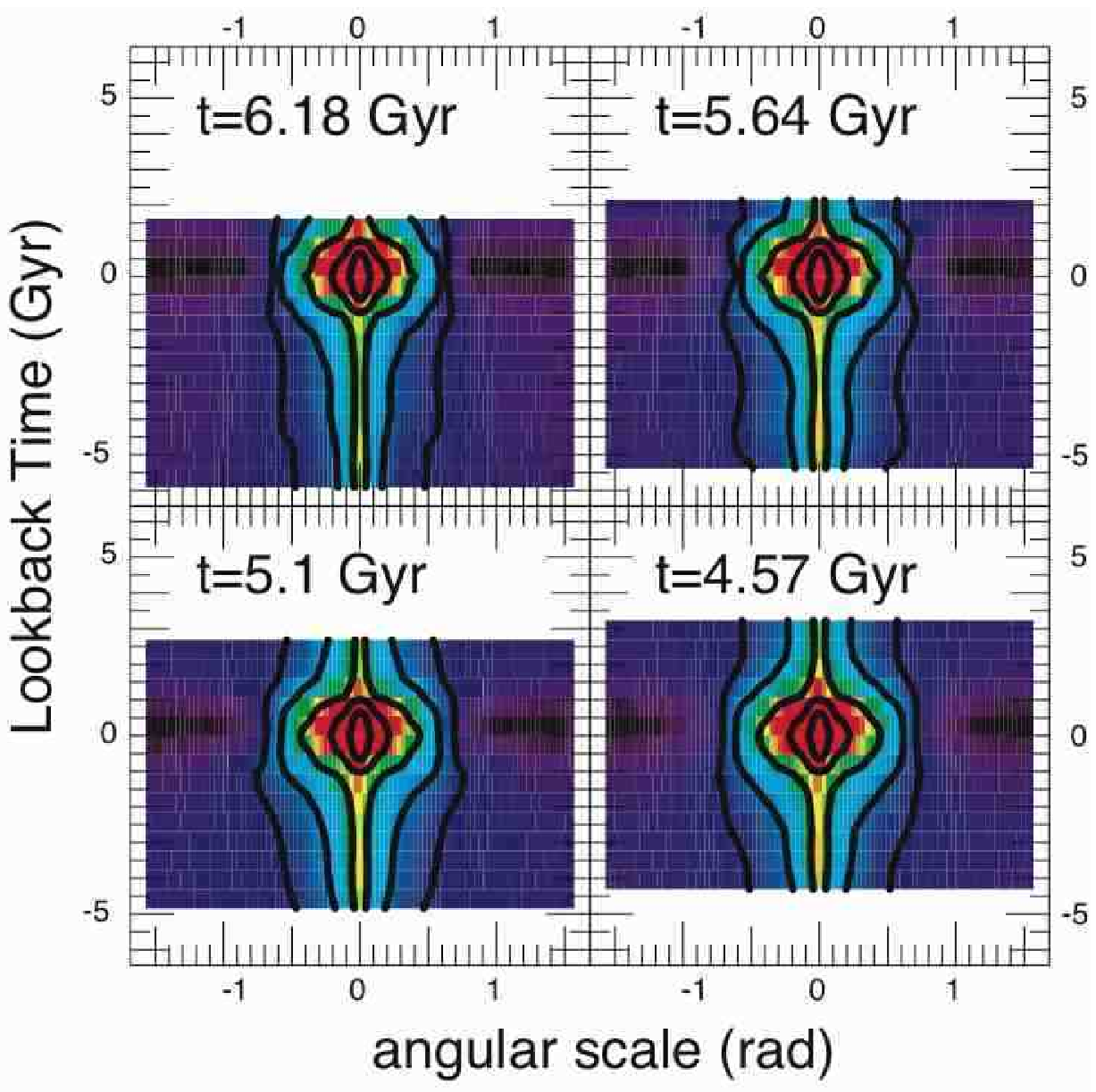}}
%\resizebox{0.8\columnwidth}{0.8\columnwidth}{\includegraphics{CorrelrhoVr_2}}
\caption{The  angulo-temporal  correlation  function,  $w^{'\varpi}(\theta,\Delta
t)=\langle \delta_{[\varpi]}(\bO,t)\delta_{[\varpi]}(\bO+\Delta\bO,t+\Delta  t)\rangle$. Blue (resp.
red) colors stand for low (resp.  high) values of the correlation. Isocontours
are also shown. Large angular scale isocontours ($\Delta \bO \sim \pi/2$) have
large temporal extent, due to the quadrupole dominance over the potential seen
on the virial sphere.}
\label{f:correldens} 
\end{figure}

Since the behavior of $\delta'_{[\varpi_\rho]}(\bO)$ is more regular than the previous contrast definition , the angulo-temporal correlation function of the flux density of mass has been computed from this new definition. Since  this definition is different from that used for the potential, we restrict ourselves to a qualitative description. The correlations are given in  \Fig{correldens}. Clearly, the correlation is more peaked around $\Delta t =0$ and more generally $w^{\varpi_\rho}$ is sharper than $w^{\psi}$. Note that no multipole $\ell$ has been removed during the computation of $w^{\varpi_\rho}$, implying that the quadrupole effect measured in the potential correlation is simply not detected for this field. This strongly suggest a large-scale `cosmic' origin for the quadrupolar tidal field rather than an artifact of the spherical intersection of an ellipse. Furthermore, the correlation time is smaller than the one measured for the potential, even compared to the correlation time of the potential without the $\ell=2$ component. This is coherent with the fact that density blobs should be `sharper' than potential blobs as they pass through the sphere.

%%%%%%%%%%%%%%%%%%%%%%%%%%%%%%%%%%%%%%%%%%%%%%%%%%%%
\section{Re-generating galactic time lines }
\label{s:gen}
%%%%%%%%%%%%%%%%%%%%%%%%%%%%%%%%%%%%%%%%%%%%%%%%%%%%

Given the  measurements given in  \Sec{stat1d} and \Sec{stat2d}, let  us
describe here how to regenerate realizations  of the history of the environment
of haloes, first  for the tidal field only, and then  for the full accretion
history.

%%%%%%%%%%%%%%%%%%%%%%%%%%%%%%%%%%%%%%%%%%%%%%%%%%%%
\subsection{Re-generating tidal fields }
\label{s:regen-tidal}
%%%%%%%%%%%%%%%%%%%%%%%%%%%%%%%%%%%%%%%%%%%%%%%%%%%%
%%%%%%%%%%%%%%%%%%%%%%%%%%%%%%%%%%%%%%%%%%%%%%%%%%%%

Let us first focus on the  generation of the potential tidal field generated
by fly by's, hence neglecting the influence of the infall through the virial
sphere.\footnote{note that this assumption is  not coherent with the way the
measurements  are carried,  since  it implies  that  the infalling  material
somehow disappear after crossing $R_{200}$.  }

First we consider two time variables: $T$ (the `slow time') which describes the
secular evolution of the field and $t_f$ (the 'fast time') which describes the
temporal evolution around a given value of $T$, hence describing high
frequencies variations. We assume that correlations exist only on small
time scales, while variations on the 'slow time' scale describe secular
drifts. Therefore, the field's regeneration should include
both correlations on short time scales and long term evolution.

Let us call  $\MG{{\hat\Psi}}={{\hat\psi}}_\M{m,\bo}(T)$ the temporal (with
respect to the fast time) and angular Fourier  transform of the potential, at fixed slow
time,  $T$.  The probability  distribution  of  $\MG{{\hat\Psi}}$, $  p(\MG{
{\hat\Psi}}) $ is given by
\begin{equation}
p(\MG{ {\hat\Psi}}               )=\frac{
               \exp\left[
    -\frac{1}{2}\left(\MG{{\hat\Psi}}-\langle\MG{{\hat\Psi}}            \rangle
    \right)^\T{}\cdot
    \M{C}_{\hat\Psi}^{-1}\cdot\left(\MG{{\hat\Psi}}-\langle\MG{{\hat\Psi}}
    \rangle     \right)      \right]}{\left(2\pi     \right)^{1/2}     {\rm
    det}^{{}^{1/2}}|\M{C_{\hat\Psi}}|}\,, \EQN{defPPsi}
\end{equation}
where the variance reads
\begin{equation}
\M{C_\Psi}(T)=    \left\langle        ({\hat\Psi}_{\M{m,\bo}}-\langle
{\hat\Psi}_{\M{m,\bo}}    \rangle)   .   ({\hat\Psi}_{\M{m,\bo}}-\langle
{\hat\Psi}_{\M{m,\bo}} \rangle) \right\rangle \EQN{defCPsi} \,,
\end{equation} 
and  the mean field obeys
\begin{equation}
\langle \MG{\hat\Psi} \rangle(T)=  \left\langle
  {\hat\Psi}_{\M{m,\bo}} 
\right\rangle \, . \EQN{defCPsi} 
\end{equation} 
Since the potential is isotropic, $\langle \MG{\hat\Psi} \rangle(T)$ is essentially zero (see also Fig. \ref{f:distaa}), while $\M{C_\Psi}$ stands for the angular power spectrum described in section \ref{s:statpot}.
Let  us  call $\{  {\hat\Psi}_{\M{m,\bo}}(T_i)  \}_{i\leq  N}$,  the set  of
sampled  $ \MG{\hat\Psi}$ a  fixed slow  time, $T_i$.   Relying on  a linear
interpolation  between  two such  realizations,  the corresponding  external
potential reads in real space:
\begin{equation}
{\psi}^e(t,\bO)= \sum_i \sum_\M{m}
\int \d\omega  \exp\left(\i \M{m}\cdot \bO+ \i \omega t \right)  
{\cal K}^\M{m}_i(t,\omega)
 \, . \EQN{defCPsiReal} 
\end{equation}
where the kernel, ${\cal K}^\M{m}_i(t,\omega)$, reads
\[
{\cal K}^\M{m}_i(t,\omega)=
\left[
{\hat\Psi}_{\M{m,\bo}}(T_{i+1}) \frac{t-T_i}{T_{i+1}-T_i}+
{\hat\Psi}_{\M{m,\bo}}(T_{i}) \frac{T_{i+1}-t}{T_{i+1}-T_i} \right]
  \times \nonumber 
\]
\begin{equation}
\qquad \Theta(T_{i+1}-t) \Theta(t-T_{i}) \, ,  \EQN{defKi}
\end{equation}
recalling that $\Theta(x)$  is the Heaviside function.

To sum up, the computation of ${\hat\Psi}_{\M{m,\bo}}$ following \ref{e:defPPsi} and
\ref{e:defCPsi} ensure that correlations on short periods are reproduced,
while the interpolation procedure allows to take in account the long period
evolution of the field. This  procedure can be  repeated for  an arbitrary  number of  virtual tidal
histories.  
%%%%%%%%%%%%%%%%%%%%%%%%%%%%%%%%%%%%%%%%%%%%%%%%%%%%
\subsection{Re-generating tidal fields and infall history }
\label{s:regen-all}
\begin{figure} 
\centering           
\resizebox{7cm}{7cm}{\includegraphics{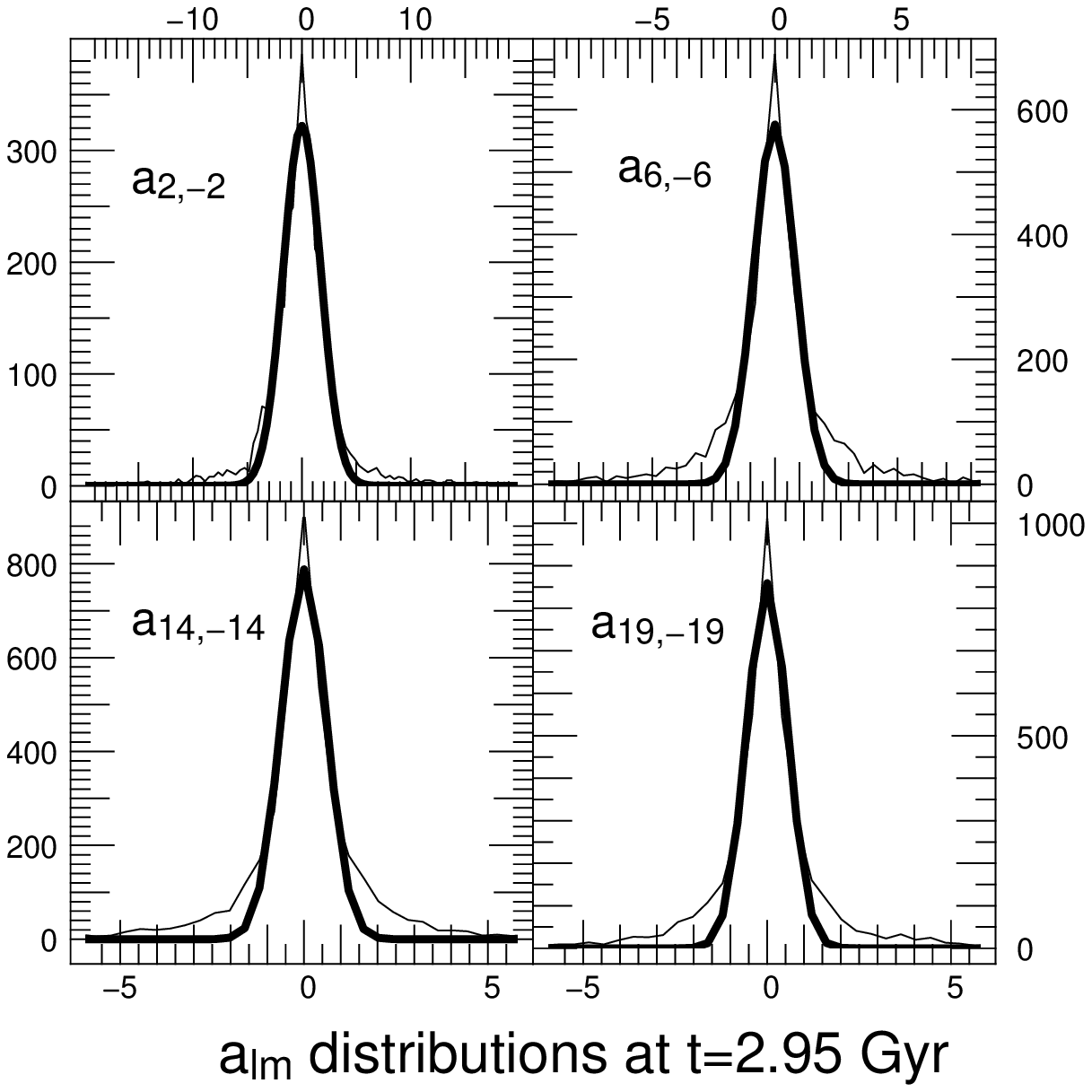}}
\resizebox{7cm}{7cm}{\includegraphics{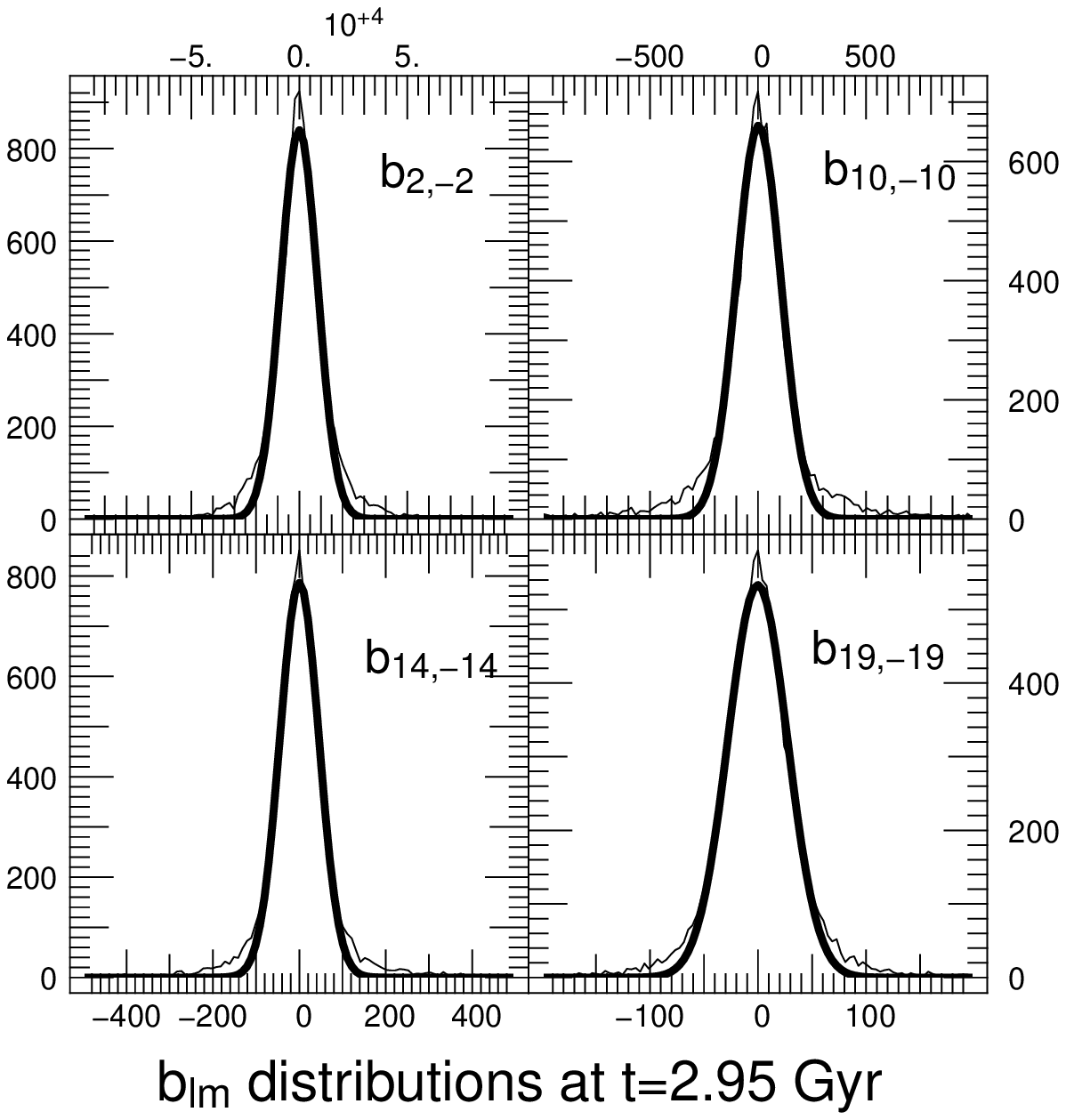}}
\caption{Probing the gaussianity of the harmonic expansions $a_{\ell m}$, describing the flux  density of mass (\textit{Top}) and $b_{\ell m}$, describing  the potential $\psi^e$ (\textit{bottom}). Units are arbitrary. Only the real part of the coefficients is shown here, but the imaginary parts have similar distributions. Clearly, the distributions are quasi-gaussians.}
\label{f:distaa} 
\end{figure}

%%%%%%%%%%%%%%%%%%%%%%%%%%%%%%%%%%%%%%%%%%%%%%%%%%%%
   Let us  assume briefly that the  fields are stationary both  in time and
angle, and  that their statistics  is Gaussian. As shown in Fig. \ref{f:distaa}, this assumption is essentially  valid for the expansions of the potential and the  flux density of mass. Let us call  the eleven
dimensional vector $\MG{\Pi}(t) \equiv \left(\varpi_\rho, \varpi_{\rho \bv},
\varpi_{\rho \sigma  \sigma} , \psi^e\right)$; and  $\hat \Pi_\M{m,\bo}$ the
temporal and angular Fourier transform of the fields.  The joint probability
of the field, $p(\MG{\hat \Pi}_\M{m,\bo})$, reads
\begin{equation}
p(\MG{                   {\hat\Pi}}_\M{m,\bo})=\frac{                    \exp\left[
    -\frac{1}{2}\left(\MG{{\hat\Pi}}-\langle\MG{{\hat\Pi}}            \rangle
    \right)^\T{}\cdot
    \M{C}_{\hat\Pi}^{-1}\cdot\left(\MG{{\hat\Pi}}-\langle\MG{{\hat\Pi}}
    \rangle       \right)      \right]}{\left(2\pi       \right)^{11/2}      {\rm
    det}^{{}^{1/2}}|\M{C_{\hat\Pi}}|}\,, \EQN{defPPi}
\end{equation}
where 
\begin{equation}
\M{C_\Pi}=\left[  \left\langle   ({\hat\Pi}^{i}_{\M{m,\bo}}-\langle  {\hat\Pi}^{i}_{\M{m,\bo}}  \rangle)
\cdot           ({\hat\Pi}^{j}_{\M{m,\bo}}-\langle {\hat\Pi}^{j}_{\M{m,\bo}}          \rangle)
\right\rangle\right]_{i,j\leq 11} \EQN{defCPi} \,,
\end{equation} 
and 
\begin{equation}
\langle \MG{\hat\Pi} \rangle= \left[  \left\langle
  {\hat\Pi}^{i}_{\M{m,\bo}} 
\right\rangle\right]_{i\leq 11} \,, \EQN{defCPi}. 
\end{equation} 
Since these fields are mostly isotropic, their expansions coefficients are nil on average. Hence, the quantity $\M{C_\Pi}$ stands for the angular power spectrum of the eleven fields. For respectively  the potential and the  flux density of mass, its \textit{temporal} evolution is described in section \ref{s:statpot} and \ref{s:cldens} . These measured power spectra are sufficient to generate environments restricted to the flux  density of mass and the potential. We emphasize that these two fields would not be coherent if no-cross correlations is specified. These cross-correlations and the power spectra for higher moments of the source are postponed to  paper III.    

Assuming the full knowledge of these 11 fields and their cross-correlation, it is therefore  straightforward to  generate for
each $(\M{m},\bo)$ a eleven dimensional vector which satisfies \eq{defPPi},
and repeat  the draw for all  modes (both $\bo$ and  $\bm$). Inverse Fourier
transform  yields   $\M{\Pi}(t)$.   Once  $\M{\Pi}(t)$  is   known,  we  can
regenerate the whole  five dimensional phase-space source as  a function of
time via  \eq{sourceexpr}.  This  process can also be  repeated for  an arbitrary number of  virtual halo histories.  The assumption of stationarity  in time
can   be   lifted  following   the   same   route   as  that   sketched   in
\Sec{regen-tidal} (see \Eq{defKi}).

%%%%%%%%%%%%%%%%%%%%%%%%%%%%%%%%%%%%%%%%%%%%%%%%%%%%%%%%%%%%%%%%%%%%%%
\section{From expansion coefficients to flux densities}
\label{s:appenflux}
%%%%%%%%%%%%%%%%%%%%%%%%%%%%%%%%%%%%%%%%%%%%%%%%%%%%%%%%%%%%%%%%%%%%%%

%%%%%%%%%%%%%%%%%%%%%%%%%%%%%%%%%%%%%%%%%%%%%%%%%%%%%%%%%%%%
\subsection{From expansion coefficients to advected  momentum}
\label{s:appen_v}
The phase space distribution of advected momentum is given by:
\begin{eqnarray}
\varpi_{\rho\bv}(\bO,v,\bG,t)&\equiv& s^e(\bO,v,\bG,t)\bv.\\
&=&\!\!\!\!\sum_{\alpha,\bm,\bm'} c_{\alpha \bm '}^{\bm}(t) g_\alpha(v)Y_{\bm}(\bO)Y_{\bm'}(\bG)\bv,
\end{eqnarray}
where the velocity vector may be written as a function of spherical harmonics:
\begin{eqnarray}
\bv&=&-v\sqrt{\frac{2\pi}{3}}(-Y_{1-1}^*(\bG)+Y_{11}^*(\bG))\be_\theta\nonumber\\
&&-\mathi v\sqrt{\frac{2\pi}{3}}(Y_{1-1}^*(\bG)+Y_{11}^*(\bG))\be_\phi\nonumber\\
&&v\sqrt{\frac{\pi}{3}} Y^*_{10}(\bG)\be_r.\label{e:vharm}
\end{eqnarray}
Then, one can write:
\begin{eqnarray}
\varpi_{\rho\bv}(\bO,t)&\equiv&\int\md v \md\bG v^2 s^e(\bO,v,\bG,t)\bv\\
&=&\sum_{\alpha,\bm,\bm'} \left[\varpi_{\rho\bv}\right]_\bm (t)Y_{\bm}(\bO)
\end{eqnarray}
where
\begin{equation}
 \left[\varpi_{\rho\bv}\right]_\bm (t)=\sum_\alpha
 (3\sigma^2\mu_\alpha+\mu_\alpha^3){\bf T_\bm}(t),
\end{equation}
and
\begin{eqnarray}
{\bf
T_\bm}(t)&=&\sqrt{\frac{2\pi}{3}}(c_{\alpha,1,-1}^{\bm}(t)-c_{\alpha,1,1}^{\bm}(t))\be_\theta\nonumber\\
&&+\mathi\sqrt{\frac{2\pi}{3}}(-c_{\alpha,1,-1}^{\bm}(t)-c_{\alpha,1,1}^{\bm}(t))\be_\phi\nonumber\\
&&+2\sqrt{\frac{\pi}{3}}c_{\alpha,1,0}^{\bm}(t)\be_r.
\end{eqnarray}

%\onecolumn
%%%%%%%%%%%%%%%%%%%%%%%%%%%%%%%%%%%%%%%%%%%%%%%%%%%%%%%%%%%%
\subsection{From coefficients to advected velocity dispersion}
\label{s:appen_disp}
The distribution of advected velocity dispersion is given by~:
\begin{equation}
\varpi_{\rho\sigma_i\sigma_j}(\bO,v,\bG,t)=s^e(\bO,v,\bG,t)(\bv-\bV(\bO,t))_i(\bv-\bV(\bO,t))_j,
\label{e:defrhosigma}
\end{equation}
where the subscripts $i$ and $j$ stand for $r,\theta,\phi$ and where
\begin{equation}
\bV_i(\bO,t)\equiv\frac{\int\md v \md \bG v^2 s^e(\bO,v,\bG,t) v_i}{\int\md
v \md \bG v^2 s^e(\bO,v,\bG,t)}=\frac{\varpi_{\rho v_i}(\bO,t)}{\varpi_{\rho}(\bO,t)}.
\label{e:defVavg}
\end{equation}
Using \eq{defrhosigma} and \eq{defVavg}, we  find:
\begin{eqnarray}
&&\hskip -1cm \varpi_{\rho\sigma_i\sigma_j}(\bO,t)+\frac{\varpi_{\rho
v_i}(\bO,t)\varpi_{\rho v_j}(\bO,t)}{\varpi_{\rho}(\bO,t)}=\\
 &&
\int\md v \md \bG
v^2 s^e(\bO,v,\bG,t)v_i v_j \nonumber \\
&=&
\sum_\bm \left[{\bf q}_{ij}(t)\right]_\bm Y_\bm(\bO).
\end{eqnarray}
The six independent elements of the symmetric tensor {\bf q}(t) can be
computed from $ c_{\alpha \bm '}^{\bm}(t)$ coefficients using equation
\eq{vharm} and recalling that:
\begin{eqnarray}
\int \md \bO Y_{\ell_1,m_1}Y_{\ell_2,m_2}Y_{\ell_3,m_3}=\sqrt{\frac{(2 \ell_1+1)(2
\ell_2+1)(2 \ell_3+1)}{4\pi}} \times 
\nonumber\\
\left(\begin{array}{ccc}
\ell_1&\ell_2&\ell_3\\
0&0&0\\
\end{array}
\right)
\left(\begin{array}{ccc}
\ell_1&\ell_2&\ell_3\\
m_1&m_2&m_3\\
\end{array}
\right),
\end{eqnarray}
where $\left(\begin{array}{ccc}
\ell_1&\ell_2&\ell_3\\
m_1&m_2&m_3\\
\end{array}
\right)=W_{m_1,m_2,m_3}^{\ell_1,\ell_2,\ell_3}$ is the Wigner 3-j symbol.
One can find:
\begin{eqnarray}
\left[{\bf q}_{rr}(t)\right]_{\bm}\!\! &=&\!\!\sum_{\alpha\bm'}
%c_{\alpha \bm'}^{\bm}(t) 
{\cal H}_{\alpha \bm'}^{\bm}(t) 2
%\sqrt{4\pi(2\ell'+1)}(6\sigma^2\mu_\alpha^2+\mu_\alpha^4)
W_{000}^{11\ell'}W_{00m'}^{11\ell'} \\
\left[{\bf q}_{r\phi}(t)\right]_{\bm}\!\! &=&\!\!\sum_{\alpha\bm'}
%c_{\alpha \bm'}^{\bm}(t)
 {\cal H}_{\alpha \bm'}^{\bm}(t) \mathi \sqrt{2}
%\sqrt{2\pi(2\ell'+1)}(6\sigma^2\mu_\alpha^2+\mu_\alpha^4)
W_{000}^{11\ell'}(W_{1-1m'}^{11\ell'}+W_{11m'}^{11\ell'})\nonumber \\
\left[{\bf q}_{r\theta}(t)\right]_{\bm}\!\! &=&\!\!\sum_{\alpha\bm'}
%c_{\alpha \bm'}^{\bm}(t)
 {\cal H}_{\alpha \bm'}^{\bm}(t)  \sqrt{2}
%\sqrt{2\pi(2\ell'+1)}(6\sigma^2\mu_\alpha^2+\mu_\alpha^4)
W_{000}^{11\ell'}(W_{1-1m'}^{11\ell'}-W_{11m'}^{11\ell'})\nonumber \\
\left[{\bf q}_{\phi\phi}(t)\right]_{\bm}\!\! &=&\!\!\sum_{\alpha\bm'}
%c_{\alpha \bm'}^{\bm}(t)
 {\cal H}_{\alpha \bm'}^{\bm}(t)  (-1)
%(-\sqrt{\pi(2\ell'+1)})(6\sigma^2\mu_\alpha^2+\mu_\alpha^4)
W_{000}^{11\ell'}(W_{11m'}^{11\ell'}+2W_{1-1m'}^{11\ell'}+W_{-1-1m'}^{11\ell'})\nonumber \\
\left[{\bf q}_{\phi\theta}(t)\right]_{\bm}\!\! &=&\!\!\sum_{\alpha\bm'}
%c_{\alpha \bm'}^{\bm}(t)
 {\cal H}_{\alpha \bm'}^{\bm}(t)  \mathi 
%\mathi\sqrt{\pi(2\ell'+1)}(6\sigma^2\mu_\alpha^2+\mu_\alpha^4)
W_{000}^{11\ell'}(W_{-1-1m'}^{11\ell'}-W_{11m'}^{11\ell'})\nonumber \\
\left[{\bf q}_{\theta\theta}(t)\right]_{\bm}\!\! &=&\!\!\!\! \sum_{\alpha\bm'}
%c_{\alpha \bm'}^{\bm}(t)
  {\cal H}_{\alpha \bm'}^{\bm}(t) 
%\sqrt{\pi(2\ell'+1)}(6\sigma^2\mu_\alpha^2+\mu_\alpha^4)
W_{000}^{11\ell'}(W_{11m'}^{11\ell'}-W_{-1-1m'}^{11\ell'}-2W_{1-1m'}^{11\ell'}).\nonumber
\end{eqnarray}

where ${\cal H}_{\alpha \bm'}^{\bm}(t)=\sqrt{4\pi(2\ell'+1)}(6\sigma^2\mu_\alpha^2+\mu_\alpha^4) c_{\alpha \bm'}^{\bm}(t)
$ 

\section{Notations}
\label{s:notations}
\begin{table*}
\centering
\begin{minipage}{140mm}
\begin{tabular}{|l|l|}
\hline Symbol & Meaning\\\hline
$(\br,\bv)$ & position and velocity\\
$ t,\tau$ & lookback time variables\\
$R_{200}$ & virial radius measured at z=0\\
$V_{c}$ & circular velocity measured at $R_{200}$\\
F & the phase space distribution function of the halo\\ 
$\psi$& the self-gravitating potential\\
$\psi^e$ & the external potential, induced by external perturbations\\
$s^e$ & the source function in phase space\\
$\varpi_x$ & the flux density of $x$ \\
$\varpi_\rho$ & the flux density of mass \\
$\varpi_{\rho\bv}$ & the flux density of momentum \\
$\varpi_{\rho\sigma\sigma}$ & the flux density of velocity dispersion \\
$\psi^{[n]}(\br)$ & 3D projection basis of the potential\\
$\phi^{[n]}(\br)$ & 6D projection basis of the source term\\
${\bf a}(t)$ & expansion coefficients of the potential/density response\\
${\bf b}(t)$ & expansion coefficients of the external potential perturbation\\
${\bf c}(t)$ & expansion coefficients of the source perturbation\\
$\bO$ & angular position on the virial sphere (2 angles)\\
$\bG$ & angular orientation of the velocity vectore on the virial sphere (2
angles)\\
$v$ & velocity's amplitude\\
$\bar X$ & angular average of $X$\\
$\underbar X$ & temporal average of $X$\\
$\langle X\rangle$ & statistical expectation (or average value) of $X$\\
$\llangle X\rrangle$ & most probable value (or mode) of $X$\\
$\delta_{[X]}(\bO)$ & contrast density of $X$ measured on the virial sphere\\
$\bm=(\ell,m)$ & harmonic coefficients related to $\b0$\\
$\bm'=(\ell',m')$ & harmonic coefficients related to $\bG$\\
$a_{\ell,m}(t)$ & harmonic expansion coefficients of $\delta_{[\varpi_\rho]}$\\
$b_{\ell,m}(t)$ & harmonic expansion coefficients of $\delta_{[\psi^e]}$\\
$C_\ell(t)$ & angular power spectrum \\
$T_\ell(t,t+\Delta t)$ & angular-temporal power spectrum \\
$w(\theta,t,t+\Delta t)$ & angulo-temporal correlation function measured on the
sphere for an angulo-temporal separation $\theta$ and $\Delta t$\\
$\Phi^M(t)$ & accretion rate measured at the virial radius (averaged over all directions)\\
$\vartheta(\Gamma_1,t)$ &  PDF of the velocity's incidence angle\\
b & impact parameter\\
$\varphi(v,t)$ & PDF of the velocity's amplitude\\
$\wp(v,t)$  & Joint PDF of the velocity's incidence angle and amplitude\\
\hline
\end{tabular}
\caption{A summary of the notations used throughout the paper.}
\end{minipage}
 \label{t:notations}
\end{table*}

\end{document}